\documentclass[12pt]{article}
\pdfoutput=1
\usepackage{jheppub}
\usepackage{comment}
\newlength{\abstractwidth}
\usepackage{amsthm}
\usepackage{amsmath}
\usepackage{amsfonts}
\usepackage{amssymb}
\usepackage{latexsym}
\usepackage{dsfont}
\usepackage{bbm}
\usepackage{enumitem}

\usepackage{epsf}
\usepackage{color}
\usepackage{graphicx}
\usepackage{tikz}
\usepackage{dsfont}

\usetikzlibrary{arrows,shapes,positioning}
\usetikzlibrary{decorations.markings}
\usepackage[rightcaption]{sidecap}
\tikzstyle arrowstyle=[scale=1]
\tikzstyle directed=[postaction={decorate,decoration={markings,
    mark=at position .65 with {\arrow[arrowstyle]{stealth}}}}]
\tikzstyle reverse directed=[postaction={decorate,decoration={markings,
    mark=at position .65 with {\arrowreversed[arrowstyle]{stealth};}}}]
\usetikzlibrary{positioning}

    \usepackage{hyperref}
\definecolor{darkred}{rgb}{0.8,0.1,0.1}
\hypersetup{colorlinks=true, linkcolor=darkred, citecolor=blue, linktoc=page}

\def\ni{\noindent}
\def\ceff{c_{\rm eff}}
\def\A{\mathcal{A}}
\def\ttb{\tau, \bar\tau}

\renewcommand{\thanks}[1]{\footnote{#1}}

\newcommand{\bea}{\begin{eqnarray}}
\newcommand{\eea}{\end{eqnarray}}
\newcommand{\be}{\begin{eqnarray}}
\newcommand{\ee}{\end{eqnarray}}
\newcommand{\bma}{\begin{matrix}}
\newcommand{\ema}{\cr\end{matrix}}
\newcommand{\<}{\langle}
\renewcommand{\>}{\rangle}

\newtheorem{thm}{Theorem}[section]
\newtheorem{lem}[thm]{Lemma}

\newtheorem{cor}[thm]{Corollary}
\newtheorem{conj}[thm]{Conjecture}

\newtheorem{res}[thm]{Result}
\newtheorem{claim}[thm]{Claim}

\def\cA{{\cal A}}

\def\cC{{\cal C}}

\def\cE{{\cal E}}
\def\cF{{\cal F}}

\def\cH{{\cal H}}
\def\cI{{\cal I}}

\def\cL{{\cal L}}
\def\cM{{\cal M}}
\def\cN{{\cal N}}
\def\cO{{\cal O}}
\def\cP{{\cal P}}

\def\cS{{\cal S}}

\def\ZZ{{\mathbb Z}}
\def\RR{{\mathbb R}}
\def\NN{{\mathbb N}}
\def\CC{{\mathbb C}}
\def\HH{{\mathbb H}}
\def\QQ{{\mathbb Q}}

\def\Tr{{\rm Tr}}
\def\det{{\rm det \,}}

\def\half{{1\over 2}}
\def\p{\partial}

\def\a{\alpha}
\def\m{\mu}

\def\g{\gamma}

\def\l{\lambda}
\def\eps{\epsilon}

\def\no{\nonumber}

\def\qqb{q,\qb}
\def\({\left(}
\def\){\right)}
\def\[{\left[}
\def\]{\right]}

\def\<{\langle}
\def\>{\rangle}

\def\Vir{\text{Vir}}
\def\HMR{{\rm HMR}}
\def\Gl{G_{\ell}}
\def\id{\mathbbm{1}}

\def\wh{\widehat}

\def\min{{\rm min}}

\def\qb{\overline q}

\def\hb{\overline h}

\def\min{{\rm min}}

\def\Z{\mathbb{Z}}
\def\R{\mathbb{R}}

\def\C{\mathbb{C}}

\def\qb{\bar q}
\def\tb{\bar\tau}

\def\x{\times}

\def\eps{\epsilon}

\def\bul{$\bullet$~}

\def\hb{\overline h}
\def\M{\mathcal{M}}

\def\1{{\rm 1-loop}}

\def\h{{1\o2}}
\def\Tr{{\rm Tr}}

\def\c{\cite}

\def\cA{\mathcal{A}}

\def\cM{\mathcal{M}}
\def\cH{\mathcal{H}}
\def\zb{\bar z}

\def\cN{\mathcal{N}}
\def\cA{\mathcal{A}}

\def\C{\mathcal{C}}

\def\vs{\vskip .1 in}

\def\p{\partial}
\def\o{\over}
\def\g{\gamma}
\def\D{\Delta}
\def\rar{\rightarrow}

\def\eqr{\eqref}

\def\O{{\cal O}}

\def\ssec{\subsection}
\def\sssec{\subsubsection}
\def\sec{\section}
\def\i{\infty}
\def\foot{\footnote}
\newcommand{\es}[2] {\begin{equation} \label{#1} \begin{split} #2 \end{split} \end{equation}}
\newcommand{\e}[2] {\begin{equation} \label{#1} #2 \end{equation}}

\def\t{\tau}

\makeatletter
\def\@fpheader{\ }
\makeatother

\title{Discreteness and Integrality in \\ Conformal Field Theory}
\author[1,2]{Justin Kaidi}
\emailAdd{jkaidi@physics.ucla.edu}

\author[3]{and Eric Perlmutter} 
\emailAdd{perl@caltech.edu}

\affiliation[1]{ Simons Center for Geometry and Physics, Stony Brook University \\
Stony Brook, NY 11794\\[-4mm]}
\affiliation[2]{Mani L.\ Bhaumik Institute for Theoretical Physics\\
Department of Physics and Astronomy, UCLA, CA 90095\\[-4mm]}

\affiliation[3]{Walter Burke Institute for Theoretical Physics\\
Caltech, Pasadena, CA 91125}

\abstract{Various observables in compact CFTs are required to obey positivity, discreteness, and integrality. Positivity forms the crux of the conformal bootstrap, but understanding of the abstract implications of discreteness and integrality for the space of CFTs is lacking. We systematically study these constraints in two-dimensional, non-holomorphic CFTs, making use of two main mathematical results. First, we prove a theorem constraining the behavior near the cusp of integral, vector-valued modular functions. Second, we explicitly construct non-factorizable, non-holomorphic cuspidal functions satisfying discreteness and integrality, and prove the non-existence of such functions once positivity is added.  Application of these results yields several bootstrap-type bounds on OPE data of both rational and irrational CFTs, including some powerful bounds for theories with conformal manifolds, as well as insights into questions of spectral determinacy. We prove that in rational CFT, the spectrum of operator twists $t\geq {c \over 12}$ is uniquely determined by its complement. Likewise, we argue that in generic CFTs, the spectrum of operator dimensions $\Delta > {c-1\o 12}$ is uniquely determined by its complement, absent fine-tuning in a sense we articulate. Finally, we discuss implications for black hole physics and the (non-)uniqueness of a possible ensemble interpretation of AdS$_3$ gravity.
}

\preprint{CALT-TH 2020-033}

\begin{document}

\maketitle

\section{Introduction}

One of the overarching goals in the study of conformal field theory (CFT) is to obtain a complete understanding of the space of consistent theories. Despite success in the classification of various two-dimensional rational CFTs (RCFTs), it is safe to say that a full classification remains far out of reach. However, a number of techniques have been developed to work gradually towards this end. Collectively, these techniques make up the conformal bootstrap approach, the most famous instantiation of which is the imposition of crossing symmetry and unitarity on four-point correlators. A powerful element in the two-dimensional bootstrap arsenal is the modular bootstrap \c{Hellerman:2009bu,Friedan:2013cba,Qualls:2013eha,Keller:2014xba,Collier:2016cls,Kraus:2016nwo,Collier:2017shs,Bae:2017kcl,Anous:2018hjh,Bae:2018qym,Afkhami-Jeddi:2019zci,Ganguly:2019ksp,Mukhametzhanov:2019pzy,Benjamin:2019stq,Hartman:2019pcd,Gliozzi:2019ewk,Brehm:2019pcx,Collier:2019weq,Mukhametzhanov:2020swe,Benjamin:2020swg,Benjamin:2020zbs,Pal:2020wwd,Afkhami-Jeddi:2020ezh,Afkhami-Jeddi:2020hde}, which bounds the space of consistent theories by demanding mutual compatibility of unitarity and modular invariance of the torus partition function. 

Typically, one is interested in compact CFTs, which will henceforth be our focus. In two dimensions, the torus partition function of such a theory, $Z(\qqb)$, can be written in terms of conformal characters (whose explicit form will be recalled later) as
\bea
\label{eq:partfunctdef}
Z(\qqb) = |\chi_{\rm vac}(q)|^2 + \sum_{h,\hb} d_{h, \bar h} \,\chi_{h}(q) {\bar \chi}_{\bar h} (\qb)
\eea
where $q:=e^{2\pi i \t}$ and $\qb := e^{-2\pi i \tb}$. Here $|\chi_{\rm vac}(q)|^2$ counts the normalizable vacuum operator and its descendants under the chiral algebra of the CFT, while the remaining terms count non-vacuum modules. The full sum must be invariant under the modular group $SL(2,\Z)$. In unitary, compact CFTs, the degeneracies $d_{h, \bar h}$ must satisfy the following three properties:

\begin{enumerate}
\item {\bf Positivity:} \hspace{0.69 in} $d_{h, \bar h} \geq 0$
\item {\bf Discreteness:} \hspace{0.45 in} $\p_\t d_{h, \bar h}= \p_{\tb} d_{h, \bar h} = 0$
\item {\bf Integrality:} \hspace{0.62 in} $d_{h, \bar h} \in \ZZ $
\end{enumerate}

\ni The first of these, together with modularity, forms the backbone of the modular bootstrap program. Discreteness and integrality are sometimes observed as outputs of increasingly high-precision numerics in the modular bootstrap \c{Collier:2016cls, Bae:2017kcl, Bae:2018qym, Afkhami-Jeddi:2020hde, Afkhami-Jeddi:2020ezh}. In addition, the existence of a normalizable vacuum state is sometimes assumed in both analytics and numerics. But discreteness of the remainder of the spectrum, much less the finer condition of integrality, has not yet been systematically utilized as an {\it input} in the conformal bootstrap; without the ability to impose all of these conditions from the outset, we are surely missing something. 

Part of the challenge -- and indeed, the interest -- is purely mathematical. Torus partition functions and local observables in irrational CFTs exhibit non-holomorphic dependence on the modular parameter $\t$. Relative to the bevy of classical results on holomorphic and meromorphic modular forms, knowledge of genuine non-holomorphic modular forms is much more limited. There are, for example, no known dimension formulas or asymptotic scaling bounds that apply universally. This is reflected not just in the great simplifications present in holomorphic CFTs, but also, perhaps, in the paucity of explicitly known irrational CFTs \c{Yin:2017yyn}. Due in part to this mathematical gap, answers to basic questions about the space of irrational CFTs rest squarely in the realm of opinion. But, if nothing else, the modular bootstrap has taught us that despite this gap, not just anything goes.

We propose an alternative to the traditional modular bootstrap which aims to harness the power of discreteness and integrality. Our CFT results will follow from the derivation of some simple mathematical results on modular forms, both holomorphic and non-holomorphic. A central moral of what follows is that modularity, discreteness, and integrality -- even without positivity -- are powerful enough to produce novel bounds on the data of rational and irrational CFTs. Some of our results run counter to conventional wisdom about non-holomorphic CFT, making them seem closer to holomorphic CFT than one might expect. Of course, all of our results will have implications for quantum gravity in AdS$_3$. The approach taken here is orthogonal to the modular bootstrap, but we hope that the current work generates ideas for a merger that algorithmically incorporates all of the constraints on compact CFT.

Before delving into a detailed summary of our results, we establish a bit of notation. The conformal dimension $\D$, and spin $J$, of an operator are given in terms of the chiral and anti-chiral conformal weights $h$ and $\hb$ as
\es{}{\D = h+\hb\,,\quad J = |h-\hb|~.}
We denote the twist $t$ as\footnote{We use $t$ for the twist to distinguish it from the modular parameter $\t$. For reference, $t  = t_{\text{\cite{Collier:2016cls}}} = t_{\text{\c{Anous:2018hjh}}} = 2t_{\text{\c{Benjamin:2019stq}}} = 2t_{\text{\c{Benjamin:2020mfz}}}+{c-1\o 12}$.}
\e{}{t:=\D-J = 2\, {\rm min}(h, \bar h)~,}
and the central charge as $c$. 

Our physical results naturally split into two classes -- bounds on the twist spectrum, and bounds on the full dimension spectrum -- associated to which are two classes of mathematical results. The first pertains to vector-valued modular functions, and the second to non-holomorphic cuspidal functions. We now discuss these in turn. 

 \ssec*{I. Twist spectrum and vector-valued modular forms}

In a compact CFT on a torus, Hilbert space traces with operator insertions can be decomposed into a sum of products of chiral and anti-chiral ``characters,''
\bea
\label{eq:genericmodfunc}
F(q, \bar q) = \sum_{i,j} N_{ij} v_i(q) \bar v_j(\bar q)~.
\eea
We call $N_{ij}$ the ``gluing matrix''. Let $d$ be the number of characters appearing in this decomposition -- for the moment we will assume that $d$ is finite, as in RCFTs, though we will consider relaxing this restriction later. Each of the components admits a Fourier expansion as
\bea
\label{eq:chargenFourierexp}
v_i(q) = q^{m_0^{(i)}}\sum_{m\geq 0} c_m^{(i)} q^{m/ b^{(i)}}~,\hspace{0.8 in}m_0^{(i)} \in \RR,\,\,\,\, b^{(i)} \in \NN
\eea
where, by discreteness, the Fourier coefficients $c_m^{(i)}$ are $\tau$-independent. The components $v_i(q)$ together form a {\it vector-valued modular form (vvmf)}, $\vec{v}(q)$, transforming in some representation $\rho:SL(2, \ZZ) \rightarrow GL(d, \CC)$ as, 
\bea
\vec{v}(q|_\g)= \rho(\g) \vec{v}(q)~,\hspace{0.8 in} \g \in SL(2, \ZZ)~,
\eea
where $q|_\g :=q(\g\t)$. The modular $T$-matrix can be taken to be diagonal, with $T= {\rm diag}(e^{2 \pi im_0^{(1)}}, \dots, e^{2 \pi i m_0^{(d)}})$. 

The study of vvmfs has a long history in both physics (e.g. \cite{mathur1988classification,Mathur:1988gt, Kiritsis:1988kq,Naculich:1988xv,Gaberdiel:2008ma,Hampapura:2015cea,Gaberdiel:2016zke, Chandra:2018pjq, Chandra:2018ezv,Harvey:2018rdc,Mukhi:2020gnj,Cheng:2020srs}) and mathematics (e.g. \cite{kaneko2002modular,knopp2004,Bantay:2005vk,bantay2007vector,marks2010structure,marks2012fourier,Marks:2010fm,99dc2d57859749fcabfc151b9730f994,franc2014fourier,franc2016hypergeometric,KANEKO2017332,franc2018constructions, mason2018vertex, Franc2020, Cheng:2020srs}); for concise recent reviews, see e.g. \cite{gannon2014theory} and \c{Mukhi:2019xjy}. Note that to each vvmf is associated a number 
\bea
\label{eq:m0def}
m_0 := \mathrm{min}\left(m_0^{(1)}, \dots, m_0^{(d)} \right)
\eea
which is smallest of the leading exponents in the $q$-expansion of each component. 

Though discreteness is already incorporated, the generic vvmf is neither positive nor integral. Depending on the CFT quantity which we are calculating, positivity and integrality may not be physically necessary. However, one or both of these conditions {\it is} necessary for a certain class of observables. This includes the partition function and, as we will show, one-point functions of marginal operators. Imposing integrality in these cases will prove to be particularly useful, as we now sketch. 

Leaving subtleties to Section \ref{sec:intratsubsec}, let us for the moment refer to a representation $\rho$ as ``integral" if there exists a finite $M\in \ZZ$ such that $M c_{m}^{(i)}\in \ZZ$ for all $m$ and $i$. In other words, a representation is integral if all components of the vvmf have Fourier coefficients which are rational numbers with bounded denominator.\footnote{We note the related concept of ``admissibility'' defined in \cite{Bantay:2005vk,bantay2007vector,Gaberdiel:2008ma}, which is a necessary condition for integrality.} A strongly supported conjecture in the math literature states that a representation is integral only if the components $v_i(q)$ are modular functions for a principal congruence subgroup $\Gamma(N) \subset SL(2, \ZZ)$. By using results from the theory of modular forms of congruence subgroups, we will prove in Section \ref{sec:mainthm} that \textit{every weight-0 vvmf in an integral representation has $m_0 <0$}. 

The physical implications of this simple fact are plentiful. First, let us interpret $F(q, \bar q) = Z(q, \bar q)$ as an RCFT partition function. Then the $v_i(q)$ are conformal characters for the extended chiral algebra, and their exponents $m_0^{(i)}$ encode the chiral dimensions of primaries, 
\e{}{m_0^{(i)} = h_i - {c \over 24}~.}
Our bound $m_0 < 0$ then implies that \textit{all RCFTs satisfy $\ceff:=c - 24 h_{\rm min}>0$, where $h_{\rm min}$ is the dimension of the lightest operator in the theory.} This is non-trivial for non-unitary CFTs. 

Next let us interpret $F(q, \bar q) = Z_1(q, \bar q)-Z_2(q, \bar q)$ as a difference of partition functions of two theories CFT$_1$ and CFT$_2$ with the same central charge. Let us furthermore assume that CFT$_1$ and CFT$_2$ have spectra which are identical up to twist $t = {c \over 12}$. Then upon taking the difference of partition functions, the $t< {c \over 12}$ portions of the spectrum cancel out, and we conclude that $F(q, \bar q) \sim q^{m_0}\bar q^{\bar m_0}+ \dots $ with $m_0 \geq 0$. Since we claimed above that all non-trivial $F(q, \bar q)$ must have $m_0 <0$, we conclude  $F(q, \bar q)=0$. In other words, \textit{in RCFT, the twist spectrum with $t \geq {c \over 12}$ is uniquely determined by its complement, $t < {c \over 12}$.} This striking result shows that RCFTs are more like holomorphic CFT than previously believed. 

Finally, let us interpret $F(q, \bar q) =\eta^{-2h_\cO}\bar \eta^{-2 \bar h_\cO} \langle \cO \rangle$ as the torus one-point function of some operator $\cO$, dressed with $\eta$-functions to absorb modular weight. This admits an OPE,
\e{cbexp}{\langle \cO \rangle \sim \sum_{\phi} C_{\phi^\dagger \cO \phi} \,q^{h_\phi - {c \over 24}} \bar q^{\bar h_\phi - {c \over 24}}~,}
which can be further rearranged into a conformal block expansion. The vvmf components are ($\eta$-functions times) torus one-point blocks. Generically such blocks are not integral, though in certain interesting cases to be described momentarily, they are. In such cases, our bound implies that {\it there exists an operator $\phi_*$ obeying }
 \e{eq:chiraldimcond1}{h_{*} \leq {c+2h_\O\o 24}~,\quad C_{\phi_*^\dagger \O\phi_*}\neq 0~.}
We emphasize that because $C_{\phi^\dagger \cO \phi}=0$ for any holomorphic $\phi$, assuming $\O$ is primary, this bound is not trivially satisfied by the identity or any chiral algebra descendant (though it may be satisfied by the primary $\cO$ itself). This is an inherent advantage of a bootstrap using torus one-point functions. Moreover, taking $F(q,\qb)$ to be a difference of two one-point functions, we conclude that {\it terms in \eqr{cbexp} with $h_\phi \geq {c+2h_\O\o 24}$ are uniquely determined by their complement.} This fixes all OPE coefficients above the aforementioned threshold (or, in the presence of degeneracy, the sum of OPE coefficients at a given level $h_\phi$): in other words, there is a form of ``OPE determinacy'' in RCFT. 

As for when these bounds hold, in Section \ref{sec:twistgapboot} we invoke a remarkable fact: the torus one-point blocks of operators with $h_\O=1$ are \textit{always} integral. This follows from a simple conformal invariance argument which we recall, and applies universally to blocks of any chiral algebra at any central charge, and for any internal field $\phi$. If $\O$ has $\D_\O=2$ and is {\it exactly} marginal, the CFT has a conformal manifold. Consistency of first-order conformal perturbation theory \cite{Cardy:1987vr} implies $C_{\phi^\dagger\cO\phi} = 0 $ for all marginal $\phi$.\footnote{For more on conformal perturbation theory, see e.g. \cite{Komargodski:2016auf,Bashmakov:2017rko,Behan:2017mwi, Sen:2017gfr}.} This in particular means that $C_{\cO^\dagger\cO\cO}=0$, and hence that the field $\phi_*\neq \O$. Moreover, it can be shown that any $\phi$ with $C_{\phi^\dagger \cO \phi}\neq 0$ must pick up an anomalous dimension to first order in conformal perturbation theory, $\gamma^{(1)}_\phi\propto C_{\phi^\dagger \cO \phi}$ . Hence this is a bound on the spectrum of non-protected (in the supersymmetric case, non-BPS) operators of the theory. So to summarize, \textit{any RCFT with an exactly marginal primary operator $\O$ with non-trivial one-point function must have a non-protected primary $\phi_*$ satisfying $t_{*} < {c+2 \over 12}$ and $C_{\phi_*^\dagger\O\phi_*}\neq 0$. Moreover, contributions to $\<\O\>$ with $t_\phi\geq{c+2\o 12}$ are uniquely determined by their complement.} 

\ssec*{II. Dimension spectrum and non-holomorphic cuspidal functions}

Having obtained a number of constraints on the chiral content of CFTs, we proceed to constraining the full spectrum, no longer restricting to the realm of vvmfs. To do so, we return to our generic modular function (\ref{eq:genericmodfunc}) and restrict to the imaginary axis. It then admits a Fourier expansion of the form
\e{eq:qeqqbexp}{F(q,q) = q^{x_0} \sum_{x\geq 0} a_x q^x}
where 
\e{eq:x0def1}{x_0 := \mathrm{min}_{\cI}(m_0^{(i)} + \bar m_0^{(j)})~,\hspace{0.5 in} \cI  = \{(i,j) \,\,| \,\, N_{ij} \neq 0\}~.}
In cases of physical interest, the quantity $x_0$ can be related to the dimension $\Delta$ of some low-lying operator by $x_0 = \Delta - {c \over 12}$, and hence it is $x_0$ which we aim to bound. In particular, one might ask whether, as for $m_0$, one can prove that $x_0 <0$ whenever discreteness and integrality are imposed. 

This question is of independent mathematical interest. Mathematically, a modular function $F(q,\bar q)$ which has leading exponent $x_0>0$ in the $q=\bar q \rightarrow 0$ limit is known as a ``non-holomorphic cuspidal function." We are thus led to ask whether there exist any non-holomorphic cuspidal functions satisfying discreteness and integrality. In Section \ref{sec:nonholocusps}, we begin by proving that if one requires in addition positivity of Fourier coefficients, then no such forms exist for any non-negative modular weight. However, if one relaxes the constraint of positivity, such forms are not forbidden, though to the best of our knowledge no such examples have appeared explicitly in the literature (which we survey in Section \ref{app:literaturesearch}). In Section \ref{sec:existenceofcuspfunc}, we remedy this by providing a number of simple constructions of such cuspidal functions. 

Having done so, we then proceed to the bootstrap. First interpreting $F(q, \bar q)$ as the difference of two partition functions, this leads immediately to the formal conclusion that, by modularity {\it alone}, the dimension spectrum with $\Delta > {c \over 12}$ is not uniquely determined by its complement, $\Delta \leq {c \over 12}$. This is to be contrasted with the analogous result for the twist spectrum. However, we stress that discrete, non-holomorphic cusp forms necessarily contain an infinite number of negative degeneracies. Consequently, we argue in Section \ref{sec:finetuning} that \textit{absent fine-tuning in the pairing between the space of unitary CFT primary partition functions and the space of weight-$\(\h,\h\)$ non-holomorphic cusp forms, the primary spectrum with $\D>{c-1\o 12}$ {\textbf{is}} uniquely determined by its complement, $\D\leq {c-1\o 12}$, in a compact, unitary, irrational CFT.}

As was done for the twist bounds, we may obtain bounds on conformal dimensions $\D$ by taking $F(q,\bar q)$ to be a torus one-point function. However, as opposed to the twist bounds which depended only on the chiral vector $\vec{v}(q)$, we see from (\ref{eq:x0def1}) that dimension bounds will depend also on the gluing matrix $N_{ij}$. To begin, let us consider the diagonal invariant $N_{ij} = \delta_{ij}$. In this case it is clear that $x_0 = 2m_0$, with $m_0$ defined in (\ref{eq:m0def}). Hence by our previous results for vvmfs, in this case we actually \textit{do} have $x_0<0$ for any quantity built from integral vvmfs. Then precisely the reasoning outlined earlier for the twist bounds yields dimension bounds. 
 
 In the case of non-diagonal invariants, it is generically no longer true that $x_0<0$. However, in Section \ref{sec:howcuspidal} we will show that for any quantity which can be built from $d$-dimensional vvmfs, one has the weaker bound
\e{}{x_0 \leq {d-1 \over 6}~.}
We highlight one particular bootstrap bound following from these observations: {\it any RCFT with an exactly marginal operator $\O$ with non-trivial one-point function must have a primary $\phi_*\neq \O$ satisfying}
\e{margintro}{\D_{*} \leq {c+2+\eps \over 12}~, \qquad C_{\phi_*^\dagger\O\phi_*}\neq 0~.}
{\it If $\<\O\>$ is diagonal, then $\eps = 0$; otherwise $\eps=2(d_\O-1)$, where $d_\O$ is the number of primaries $\phi$ such that $\cO\subset \phi^\dagger \times \phi$.} Following the interpretation of the exactly marginal bounds given below \eqr{eq:chiraldimcond1}, $d_\O$ is the rank of the one-loop dilatation matrix, whose eigenvalues for tree-level dimensions above the bound \eqr{margintro} are fixed by those obeying the bound. A number of other results will be discussed in Section \ref{sec:dimboundssubsec}. 

With an eye towards AdS$_3$/CFT$_2$, we make two brief comments.\foot{The realm of 3d gravity, pure and otherwise, with RCFT duals has been discussed in e.g. \c{Gaberdiel:2010pz, Castro:2011zq, Jian:2019ubz}. Pure higher-spin gravity also appears in connection with ensemble averaging over Narain moduli space \c{Maloney:2020nni,Afkhami-Jeddi:2020ezh}.} First, at large $c$, the bound \eqr{margintro} asymptotes to $\Delta_* \lesssim {c \over 12}$. Though our bounds were obtained assuming rationality, it is significant that $\phi_*$ must be \textit{non-holomorphic}. In particular, these CFTs are {\it irrational} with respect to Virasoro, and the bound is necessarily independent of the extra conserved currents. In this sense, our bounds lie ``halfway" between exclusively RCFT bounds, and the sought-after asymptotic bound $\D_*\lesssim {c\o 12}$ in generic irrational CFT. Second, though we will prove that there are no non-holomorphic cuspidal functions satisfying discreteness, integrality, and positivity, if we relax the condition of discreteness this proof no longer holds. It is thus conceivable that continuous, integral, positive cuspidal functions \textit{do} exist. This has implications for tentative ensemble average interpretations of pure gravity in AdS$_3$, in which the partition function $Z_{\rm grav}$ is continuous with a gap above a normalizable vacuum. In particular, one is faced with the same ambiguity in the spectrum of dimensions $\Delta > {c \over 12}$ as in the discrete case, but now without any of the subtleties of fine-tuning. This and some implications for the black hole spectrum are discussed further in Section \ref{sec:ensembleavg}.
 
\subsection*{Outline}

This main sections of this paper alternate between physical and mathematical focus. The more mathematically-oriented reader may wish to first direct their attention to Sections \ref{sec:mainvvmfsec} and \ref{sec:nonholocusps}, before turning towards the physical implications in Sections \ref{sec:twistgapboot}, \ref{sec:dimboundssubsec} and \ref{sec:3dgravity}. Conversely, the physically-oriented reader may wish to first direct their attention to Sections \ref{sec:twistgapboot}, \ref{sec:dimboundssubsec} and \ref{sec:3dgravity}, taking the mathematical results developed in Sections \ref{sec:mainvvmfsec} and \ref{sec:nonholocusps} on faith.

Concretely, the paper is organized as follows. We begin in Section \ref{sec:preliminaries} with some preliminaries, which includes reviewing relevant math terminology and rephrasing our physical questions in a mathematical language. Section \ref{sec:mainvvmfsec} has a more detailed review of vvmfs and conditions for their integrality. In particular, in Section \ref{sec:mainthm} we prove our main result concerning the sign of $m_0$, see Theorem \ref{thm:conj1}. In Section \ref{sec:twistgapboot} we apply these results to obtain bounds on the twist gap in various situations, reobtaining the results sketched in the introduction and more. We also mention a connection to Schur indices of 4d $\cN=2$ theories in Section \ref{sec:Schur}, and discuss possible extensions to irrational CFT in Section \ref{secirr}.  
 
Sections \ref{sec:nonholocusps} and \ref{sec:dimboundssubsec} discuss non-holomorphic cuspidal functions and their relation to bootstrap bounds. In Section \ref{sec:noposintconst} it is shown that no non-factorizable, non-holomorphic cuspidal functions satisfying positivity, discreteness, and integrality can exist. However, if one sacrifices the condition of positivity then it is possible to construct such functions. Rather surprisingly, to our knowledge no such non-holomorphic cuspidal functions have appeared in the literature, as we review in Section \ref{app:literaturesearch}. We fill in this gap by outlining some simple constructions in Section \ref{sec:existenceofcuspfunc}. We discuss the physical implications of all of these results on the $\D$ spectrum in Section \ref{sec:dimboundssubsec}.

In Section \ref{sec:3dgravity} we ask what our 2d CFT results tell us about 3d gravity. Section \ref{sec:ensembleavg} begins with a discussion of implications for ensemble average interpretations of 3d gravity, and ends with comments about the quantum-corrected black hole threshold. We then proceed to a brief discussion of enigmatic black holes in Section \ref{sec:enigmaticBH}.

Some extra results are collected in the appendices. The first three appendices give more details on integral vvmfs. In Appendix \ref{sec:congtestsubsec} we review simple techniques to determine when a vvmf is integral. In Appendix \ref{app:2dproof} we use these techniques to give alternative proofs of Theorem \ref{thm:conj1} in the case of two-dimensional vvmfs. In Appendix \ref{sec:twistexampsubsec}, we give numerous examples of integral vvmfs that appear in familiar RCFTs. 

In Appendix \ref{app:cuspfuncts} we give many novel constructions of non-factorizable, non-holomorphic cuspidal functions satisfying both integrality and discreteness. Some of these constructions make use of the data of $\wh{su(N)}_k$ WZW models, which we collect in Appendix \ref{appwzw} for the reader's convenience. 

Finally, Appendix \ref{sec:modsystems} is reserved for readers interested in a concrete building of modular systems.

\sec{Preliminaries}
\label{sec:preliminaries}

Modular forms $F_{(w,\bar w)}(\ttb)$ of weight-$(w,\bar w)$ transform under $SL(2,\Z)$ transformations  as 
\e{}{F_{(w,\bar w)}(\gamma \t, \gamma \tb) = (c\t+d)^w(c\tb+d)^{\bar w}F_{(w,\bar w)}(\ttb)\,,\quad \g\in SL(2,\Z)~.}
We will mostly be concerned with modular functions, $F(\ttb) := F_{(0,0)}(\ttb)$. These admit a Fourier expansion\footnote{We will alternate freely between writing  the arguments of modular functions as $(\t, \tb)$ and $(q, \qb)$.}
\es{Fqexp}{F(\qqb) &=q^{m_0} \bar q^{\bar m_0} \sum_{m,\bar m \geq 0}a_{m,\bar m}(\t_2) \,q^m \bar q^{\bar m}~,} 
 where $q:=e^{2\pi i \t}$, $\qb := e^{-2\pi i \tb}$, and $\t:=\t_1+i\t_2$. We say that $F$ is {\it discrete} if $a_{m,\bar m}(\t_2)$ are constants, independent of $\t_2$. We say that $F$ is {\it integral} when $a_{m,\bar m}\in\Z$. If $F$ has definite parity under $\t_1 \rar -\t_1$, i.e. definite spacetime parity in CFT language, then $a_{m,\bar m} (\t_2)= \pm a_{\bar m, m}(\t_2)$. 

When $F$ is {\it cuspidal}, we will denote it by $\C$. ``Cuspidal'' means that $\C$ vanishes near the cusp at $\qb=q \rar 0$. Specializing to the diagonal and writing
\bea
\label{eq:cC2}
\cC(q, q) = q^{m_0+\bar m_0} \sum_{x\geq \tilde x} a_x q^x\,, \hspace{0.8in} a_x = \sum_{ m+\bar m=x}a_{m,\bar m}~,
\eea
we see that $\cC$ is cuspidal if 
\e{cuspineq}{x_0 := m_0+\bar m_0 + \tilde x\geq0~,}
where $\tilde x$ is defined such that the first non-zero Fourier coefficient is $a_{\tilde x} \neq 0$. This inequality can be satisfied even if $m_0, \bar m_0 \leq 0$, as long as $\tilde x$ is sufficiently large. 

We will sometimes use the alternative Fourier parameterization
\bea
\label{cuspfunc}
F(q, \bar q) = \sum_{N=0}^\infty \left( a_N(\tau_2) \,e^{2 \pi i N \tau_1} + \tilde{a}_N(\tau_2)\, e^{-2 \pi i N \tau_1}\right)~,
\eea
with coefficients
\bea
a_{N}(\tau_2) = \sum_{\Delta} c_N(\Delta)\, (q \bar q)^\Delta e^{-2 \pi N \tau_2}~,
\no\\
\tilde{a}_{N}(\tau_2) = \sum_{\Delta} \tilde{c}_N(\Delta)\, (q \bar q)^\Delta e^{-2 \pi N \tau_2}~.
\eea
The sums are over some set of real numbers $\Delta$, not necessarily integer. In this parameterization, definite parity means $\tilde a_N(\t_2) = \pm a_N(\t_2)$, discreteness means $c_N(\Delta)$ is independent of $\t_2$, integrality means $c_N(\D)\in\Z$, and cuspidality means $\D>-{N\o 2}$. 

Due to Virasoro symmetry, a compact CFT partition function $F(q,\qb)=Z(q,\qb)$ with a normalizable vacuum may be written as a sum over conformal characters as in \eqr{eq:partfunctdef}, 
where
\e{chars}{\chi_{h}(q) := {q^{h-{c-1\o 24}}\o \eta(q)}~,\quad \chi_{\rm vac}(q) := (1-q)\chi_0(q)~.}
The vacuum character $\chi_{\rm vac}(q)$ accounts for the null state at level one in the vacuum module. One may wish to define a ``primary'' partition function\foot{One may equally work with the ``reduced'' partition function $\hat Z := \sqrt{\t_2}\, Z_p$, which is modular-invariant. For our purposes, this choice slightly obscures questions of discreteness.}
\e{}{Z_p(\qqb) := |\eta(q)|^2Z(\qqb)}
of modular weight $\(\h,\h\)$, the Fourier expansion of which counts primary degeneracies. However, we will work mostly with $Z(\qqb)$.

\ssec{Spectral determinacy and $c$ versus $c-1$}
\label{cc1}

With our notation in place, we can relate the following two physical questions
\es{}{&\textit{{\bf Q1:}  Does the spectrum with $t \leq {c\o 12}$ uniquely specify the spectrum with $t>{c\o 12}$?}\nonumber\\
&\textit{{\bf Q2:}  Does the spectrum with $\D \leq {c\o 12}$ uniquely specify the spectrum with $\D>{c\o 12}$?}\nonumber}
 to sharp questions about modular functions:
\es{}{&\textit{{\bf Q1:} Do there exist modular functions (\ref{Fqexp}) with $a_{m,\bar m}\in\mathbb{Z}$ and $m_0,\bar m_0>0$?}\quad\quad~\nonumber\\
&\textit{{\bf Q2:} Do there exist cuspidal functions (\ref{eq:cC2}) with $a_{m,\bar m}\in\mathbb{Z}$ and $x_0>0$?}\nonumber}

Let us pause to explain why, from the physics point of view, we choose to use a threshold of $c\o 12$ when addressing questions of spectral determinacy, as opposed to ${c-1\o 12}$. Indeed, the latter might naively seem more physically relevant: ${c-1\o 12}$ is the threshold for polarity in the primary partition function $Z_p$, and for the support of the modular crossing kernel \c{zamps}. However, the questions we have posed are trivial to answer when phrased in terms of a threshold ${c-1\o 12}$. To see this, first note that any compact CFT partition function with central charge $c'=1$, viewed as additive to the partition function of an auxiliary CFT with central charge $c$, modifies only states with $(h,\hb) \geq {c-1\o 24}$ and preserves compactness, as was pointed out in \cite{Keller:2014xba}. This is because the vacuum of the $c'=1$ CFT ``looks like'' a primary state with $(h,\hb) = {c-1\o 24}$. This alone allows us to conclude that the twist spectrum with $t\geq {c-1\o 12}$ is \textit{not} fixed by its complement. 
Moving on, we note that by the same logic, adding any $c'<1$ CFT partition function modifies only the states with $(h,\hb)>{c-1\o 24}$.  
Hence we conclude that the twist spectrum with $t> {c-1\o 12}$ is {again} not fixed by its complement, and similarly for the dimension spectrum.  With this, as well as the mathematical interest in cuspidal functions, in mind, we seek stronger statements using a threshold of $c\o 12$. 

\ssec{Invitation: A twist bound from Lorentzian modular forms}
\label{sec:Lorentzmodform}
As a warmup to our main investigation, we first prove the following theorem for non-holomorphic modular functions, 

\begin{thm}
\label{lorthm}
Consider a non-holomorphic modular function $F(\qqb)$, bounded in the interior of the fundamental domain, with Fourier expansion
\es{Zqexp}{F(\qqb) &=q^{m_0} \bar q^{\bar m_0} \sum_{m,\bar m \geq 0}a_{m,\bar m} \,q^m \bar q^{\bar m}~.} 
If $m_0,m\in\Z/2$, then $m_0\leq 0$. The same holds for $\bar m_0, \bar m$.
\end{thm}

We begin with the following fact: given a modular form $F(\qqb)$ with $\tb=\t^*$, one can extend it to a domain in $\mathbb{C}^2$ where $\t,\tb$ are independent complex variables, $\tb\neq \t^*$, while preserving modular-invariance (see e.g. footnote 3 of \cite{Hartman:2014oaa}). With this in mind, we consider 
\e{}{\mathcal{F}(\t) := F(\tau,-i)~.}
This has an expansion 
\e{}{\mathcal{F}(\t)  =  q^{m_0}\sum_{m\geq 0} c_mq^m \quad \text{where}\quad c_m := \sum_{\bar m\geq 0} e^{-2\pi (\bar m_0+\bar m)}a_{m,\bar m}~.}
We make the mild assumption that $c_m\neq 0$ for at least some $m$. Under $S$ transformation of $\t$ and $\tb$, one has
\e{}{F(-1/\tau,-i) = F(\tau,-i)~.}
By construction, we can regard this as $S$-invariance of the weakly holomorphic function $\mathcal{F}(\t) = \mathcal{F}\(-{1\o\t}\)$, i.e. invariance under a transformation of $\t$ alone. On the other hand, under $T$ transformation of $\t$ and $\tb$, one has
\e{}{F(\t+1,-i+1) = F(\t,-i)~.}
But now let us ask how this behaves under $T$-transformation of $\t$ alone. $\mathcal{F}\(\t\)$ is $T$-invariant if and only if $m_0\in\Z$ and $m\in\Z$. Assuming this, then $\mathcal{F}\(-{1\o\t}\)$ is a weakly holomorphic modular function. Such functions obey the bound  
\e{mmin}{m_0\leq 0~.}
More precisely, this bound holds if $\mathcal{F}(\t)$ is bounded away from the cusp at $i\i$, so rational functions of $J(q)$ with non-trivial denominator are not allowed. This boundedness is a basic finiteness requirement of all CFT observables, equivalent to the statement that they should not diverge at finite temperature.  Imposing this concludes the proof for the case $m_0,m\in\Z$. 

If we instead assume that $m_0,m\in\Z/2$, then $\mathcal{F}(\t)$ is invariant under $S$ and $T^2$. In other words, it is a weakly holomorphic function for $\Gamma_\theta\subset SL(2,\Z)$. Noting that $\Gamma_\theta$ is congruent to $\Gamma_0(2)$, and that there are no cuspidal functions for $\Gamma_0(s)$ for any $s$ (see e.g. Section 4.4 of \cite{apostol2012modular}), this concludes the proof for the case $m_0,m\in\Z/2$. 

More generally, if $\mathcal{F}(\t)$ is invariant under any subgroup $\Gamma\subset SL(2,\Z)$ which contains $S$ as a generator, then $m_0\leq 0$ if $\Gamma$ admits no cuspidal functions.\foot{\label{footnote:infindex}We point out that if $\mathcal{F}(\t)$ is invariant under $S$ and $T^p$ for $p>2$, there is no straightforward statement to make because the group generated by $S$ and $T^{p>2}$ does not have finite index in $SL(2,\Z)$ \c{conrad}; we are not aware of dimension formulas for such groups.} Rather than pushing further in this direction, we will relax the requirement of $S$-invariance and instead study an important class of $S$-covariant objects, namely, vector-valued modular forms. 

\section{Integrality and vector-valued modular forms}
\label{sec:mainvvmfsec}

In this section we prove the following theorem:
\begin{thm}
\label{thm:conj1}
Consider any weakly holomorphic $SL(2,\Z)$-invariant function $F(q, \bar q)$ admitting a decomposition into vvmfs of dimension $d>1$, with Fourier expansion of the form
\bea
F(\qqb) = q^{m_0} \bar q^{\bar m_0} \sum_{m,\bar m \geq 0}a_{m,\bar m} \,q^m \bar q^{\bar m} ~.
\eea
Then imposing integrality of the Fourier coefficients, i.e. $a_{m, \bar m} \in \ZZ$, implies 
\bea
\label{eq:conj1}
m_0,\bar m_0 < 0~.
\eea
 \end{thm} 
\noindent This theorem will follow from two slightly narrower results about vvmfs, which we give as Theorem \ref{thm:conj1b} and Corollary \ref{thm:conj1a}, and is predicated upon a well-established conjecture. As with most of the  results in this paper, this theorem can be extended to apply  to modular {forms} $F_{(w, \bar w)}$ with non-zero weight $(w, \bar w)$ under $SL(2,\ZZ)$. To do so, we simply consider the modular function $F = \eta^{-2 w} \bar  \eta^{-2 \bar w} F_{(w, \bar w)}$ and apply our results there.\foot{When $w-\bar w\neq$ 0 mod 12, this transforms with a non-trivial unitary multiplier. This will not be relevant for our purposes.} We will make use of this fact implicitly later on.

Before proving Theorem \ref{thm:conj1}, we begin with a brief review of vvmfs, paying special attention to the effects of imposing integrality. 

\subsection{Basics}
\label{sec:intratm0}
As reviewed in the introduction, we will be concerned with modular functions which can be decomposed into a (not necessarily finite) sum of products of chiral and anti-chiral ``characters'', 
\bea
F(q, \bar q) = \sum_{i,j} N_{ij} v_i(q) \bar v_j(\bar q)~.
\eea
We focus for now on the chiral ``characters'' $v_i(q)$, of which there are $d$. They admit a Fourier expansion as in (\ref{eq:chargenFourierexp}). Together, the $v_i(q)$ form a vvmf $\vec{v}(q)$ transforming in some representation $\rho:SL(2, \ZZ) \rightarrow GL(d, \CC)$ as,
\bea
\vec{v}(q|_\g)= \rho(\g) \vec{v}(q)~,\hspace{0.8 in} \g \in SL(2, \ZZ)~.
\eea
On the other hand, one generically does not expect the individual components of the vvmf to possess nice modularity properties.

It is a general fact that a $d$-dimensional vvmf satisfies a $d$-dimensional modular differential equation (MDE), with the $d$ solutions to the MDE furnishing the $d$ components of the vvmf, see e.g. \cite{Harvey:2018rdc,Cheng:2020srs}. The equation can be written as
\e{}{\[D^{(d)} + \sum_{r=1}^d f_{r,\ell}(q) D^{(d-r)}\]v_i(q) = 0}
where $f_{r,\ell}(q)$ is a meromorphic modular form of weight $2r$. The label $\ell$ is known as the ``Wronskian index" \c{Mukhi:2019xjy}, and denotes the maximum order of poles allowed for the coefficient functions.\footnote{More precisely, $\ell$ is defined as 
\bea
\ell := 6 \left(\half \mathrm{ord}_i(f) + {1 \over 3} \mathrm{ord}_\rho(f) + \sum_{\substack{p \in \HH /SL(2, \ZZ) \\ p \neq i, \rho}} \mathrm{ord}_p(f) \right) ~,
\eea
where $\rho := e^{2 \pi i /3}$. The origin of the term ``Wronskian index" is that $\ell/6$ is the number of zeroes of the Wronskian determinant of the independent solutions to the MDE.} It is generically allowed to take values $\ell = 0,2,3,4,\dots$. The Serre derivatives are defined as
\e{}{D^{(d)} := \prod_{s=1}^d D_{2s-2}\,,\qquad D_a := q {d\o dq} -{a\o 12} E_2(q)}
with $E_2(q)$ the modular connection. By convention, $D^{(0)}$ is taken to be constant.

When $\ell = 0$, the coefficient functions $f_{r,0}(q)$ in the MDE are holomorphic, giving rise to the case of monic MDEs well-studied in the literature (we reserve the term monic for MDEs with unit leading coefficient and  holomorphic coefficient functions). As an example, the generic monic second-order MDE takes the form
\e{eq:2ndorderMDE}{ \left[D_2 D_0 + \g E_4(q) \right] v_i(q) = 0}
with $E_4(q)$ the normalized weight-4 holomorphic Eisenstein series and $\g$ a free parameter. Features of the Fourier expansions of the solutions $v_i(q)$ can be understood by considering the indicial equation 
\bea
\label{eq:indiceq}
(m_0^{(i)})^2 - {m_0^{(i)} \over 6} + \g = 0~,
\eea
where $m_0^{(i)}$ is the leading exponent of $v_i(q)$, as per \eqr{eq:chargenFourierexp}. There are generically two distinct solutions $m_0^{(1)}, m_0^{(2)}$ to this quadratic equation, which satisfy\footnote{\label{footnote:indicial}If the two solutions to the indicial condition are identical, then only one solution to the MDE can be obtained via the method of Frobenius, i.e. can be written as a power series in $q$. The other solution will contain terms of the form $\log q$.  Such solutions are not compatible with our requirements of discreteness. Say instead that the two solutions to the indicial equation are distinct. If the two solutions differ by a non-integer, then two solutions can be found via the method of Frobenius. If however the two solutions differ by an integer, the larger root will be a solution, but the smaller may or may not be. This subtlety will for the most part be irrelevant for us, except in Section \ref{sec:existenceofcuspfunc}.}
\bea
\label{eq:MDEsum}
m_0^{(1)} + m_0^{(2)}= {1\over 6}~.
\eea
The general non-degenerate solutions to (\ref{eq:2ndorderMDE}) can in fact be obtained by recasting the MDE as a hypergeometric equation \cite{franc2014fourier}. From this one obtains solutions
\bea
\label{eq:2dexplicitsols}
v_1(q) &=& j(q)^{-{(1+\mu) \over 12}} {}_2 F_1\left({1 + \mu \over 12}, {5 + \mu \over 12}\,;\, 1 + {\mu \over 6}\,; \, 1728 j(q)^{-1} \right)~,
\no\\
v_2(q) &=& j(q)^{-{(1-\mu) \over 12}} {}_2 F_1\left({1 - \mu \over 12}, {5 - \mu \over 12}\,;\, 1 - {\mu \over 6}\,; \, 1728 j(q)^{-1} \right)~,
\eea
where we have defined $\mu := \sqrt{1 - 144 \g}$ and $j$ is the modular $j$-function defined in \eqr{jdef}. 

With the explicit solutions in hand, one can now verify Theorem \ref{thm:conj1} for this class of two-dimensional vvmfs on a case-by-case basis. For example, for $\mu < 1$ one sees that both of the solutions have $m_0^{(i)}>0$, and hence Theorem \ref{thm:conj1} would imply that in these cases the vvmf cannot have integer coefficients. In Appendix \ref{app:2dproof}, we give two different proofs of Theorem \ref{thm:conj1} for the case of vvmfs satisfying a second-order monic MDE. However, in what follows we will furnish a streamlined proof that applies equally well to both the non-monic case and higher-dimensional vvmfs.

For $d$-dimensional vvmfs or non-monic MDEs, a closed form expression for the solutions is generically not known. However, one quantity which is independent of all free parameters of the MDE is the sum over indicial roots \c{mathur1988classification},\footnote{To derive this, note that $-\sum_{i=1}^d m_0^{(i)}$ is the coefficient of the $s^{d-1}$ term in the indicial equation $\prod_{i=1}^d (s-m_0^{(i)})=0$. This comes solely from the $D^{(d)}$ term of the MDE. One can then use $\sum_r r = d(d-1)$, where the sum runs from $r=0,2,4,\dots,2d-2$.} 
 \bea
 \label{eq:MDEconstd}
 \sum_{i=1}^d m_0^{(i)} = {d(d-1) \over 12} -{\ell \over 6}~.
 \eea
In other words, the sum over leading exponents of each component of a vvmf is a completely universal quantity, depending only on the dimension and Wronskian index of the vvmf. This result will prove useful for us shortly.

One immediate implication of \eqr{eq:MDEconstd} is that if $\ell > {d(d-1) \over 2}$, we automatically have $m_0<0$ and Theorem \ref{thm:conj1} is trivially satisfied, even without imposing integrality. For this reason, in the rest of this section we assume that $\ell \leq {d(d-1) \over 2}$. Note that in the case of equality, we may safely neglect the degenerate case in which all roots individually vanish.

\subsection{Integrality and rationality}
\label{sec:intratsubsec}
 We now aim to impose integrality on our vvmfs. Though the full vvmfs transform (by definition) in a representation of $SL(2, \ZZ)$, the individual components generically do not have nice modularity properties. However, when the Fourier coefficients are algebraic integers (which in particular is the case for rational integers $c_n^{(i)} \in \ZZ$), one can invoke the following conjecture \cite{atkin1971modular}:
\begin{conj}[Integrality conjecture]
\label{thm:intconj}
If $m_0^{(i)}\in \QQ$ and all Fourier coefficients $c_n^{(i)}$ are algebraic integers, then $v_i(q)$ is a modular function for a principal congruence subgroup $\Gamma(N) \subset SL(2, \ZZ)$ for some $N$.
\end{conj}
\noindent In particular, if there exists an $M\in\Z$ such that all entries of $M\vec v$ have integral Fourier coefficients, then $\mathrm{ker}\, \rho \supseteq\Gamma(N)$ and all components $v_i(q)$ are modular functions for $\Gamma(N)$, defined as 
  \bea
  \label{eq:GammaNdef}
  \Gamma(N) = \left\{\left(\begin{matrix}a& b \\ c& d\end{matrix} \right) \in SL(2, \ZZ) \,\Big | \, \left(\begin{matrix}a& b \\ c& d\end{matrix} \right)  = \left(\begin{matrix}1& 0 \\ 0& 1\end{matrix} \right)\,\,\,\mathrm{mod}\,\,N\right\}~.
  \eea
  Such modular functions are well-studied in the math literature, and satisfy a number of useful properties. Copious evidence for, and low-dimensional proofs of, Conjecture \ref{thm:intconj} can be found in \c{mason2010fourier,marks2010structure,Marks:2010fm,marks2012fourier,franc2014fourier,franc2016hypergeometric,franc2018constructions,Cheng:2020srs,FRANC2016186} and references therein; the conjecture is widely believed to be true, and we will assume so here.\foot{The conjecture is often stated in the math literature as the ``unbounded denominator conjecture'', which refers to the contrapositive: if $v_i(q)$ are modular forms for noncongruence subgroups of $SL(2,\Z)$, then the denominators of Fourier coefficients are unbounded in the sense above. The conjecture was originally formulated for scalar modular forms rather than vvmfs.}

We pause to note that there is a simple diagnostic for determining when $\mathrm{ker}\, \rho \supseteq\Gamma(N)$, discussed in Appendix \ref{sec:congtestsubsec}. This turns out to be extremely useful in a variety of situations. In particular, it can be used to give a beautiful arithmetic proof of Theorem \ref{thm:conj1} in the case of $d=2$ vvmfs, as we show in Appendix \ref{app:2dproof}. However, the proof we give in the next subsection is simpler, and valid for any $d$. 

Before proving Theorem \ref{thm:conj1}, we emphasize the importance of integrality in the claim. Indeed, if we allow for non-integer coefficients then one may easily construct counterexamples using the explicit solutions (\ref{eq:2dexplicitsols}). One such counterexample is when $\mu = {6\o 7}$, in which case one finds\foot{This case yields the torus blocks for the one-point function of the $(r,s) = (1,4)$ operator in the non-unitary Virasoro minimal model $\M(7,2)$ \cite{Gaberdiel:2008ma}. $v_1(q)$ corresponds to the fusion $(1,2) \x (1,2) = (1,4)$, and $v_2(q)$ to the fusion $(1,4) \x (1,4) = (1,4)$.} 
\bea
\label{eq:g27g37}
v_{1}(q) &=& q^{13 \over 84}\left(1- {13 \over 14}q - {13 \over 49}q^2 + {299 \over 686}q^3 -{2674 \over 7725}q^4+ \dots \right)~,
\no\\ 
v_{2}(q) &=& q^{1 \over 84}\left(1- {4 \over 7}q - {267 \over 637}q^2 + {8 \over 343}q^3 -{2236 \over 7203}q^4+ \dots \right)~.
\eea
Both of these has positive leading exponent, which would be in contradiction to Theorem \ref{thm:conj1}. However, the vvmf $\vec{v}(q)$ suffers from a conspicuous lack of integrality. Indeed, the denominators in both components grow unboundedly at higher orders in $q$, so no choice of basis can have integer coefficients. 

  \subsection{Proof of Theorem \ref{thm:conj1}}
  \label{sec:mainthm}

We now come to the main point of this section, which is the proof of Theorem \ref{thm:conj1}. We will actually first prove Theorem \ref{thm:conj1b} and Corollary \ref{thm:conj1a}, stated below, which are phrased purely in terms of weakly holomorphic vvmfs for $SL(2,\Z)$. From these, Theorem \ref{thm:conj1} follows.  

\begin{thm}
\label{thm:conj1b}
Consider a $d$-dimensional weakly holomorphic vvmf $\vec v(q)$ with components $v_i(q)$. Define
\e{eq:m0def2}{m_0 := \mathrm{min}\left(m_0^{(1)}, \dots, m_0^{(d)} \right)}
where $v_i(q) \sim q^{m_0^{(i)}}$ near the cusp at $i\i$. If there is at least one component $v_i(q)$ which is a modular function for $\Gamma(N)$ for some $N$ and has $m_0^{(i)}\neq 0$, then $m_0<0$.
\end{thm}

\ni We begin by noting that modular functions for $\Gamma(N)$ obey a valence formula, 
 \bea
 \label{eq:valenceformula}
 \sum_{p\,\in \,\HH/\Gamma(N)} \mathrm{ord}_p(v_i) = 0
 \eea
 where $\HH$ is the upper half-plane, and $ \mathrm{ord}_p(v_i)$ denotes the order of a zero (counted positively) or a pole (counted negatively) of $v_i(q)$ at $\tau = p$ (see e.g. \cite{Diamond2005}). By the assumption of weak holomorphicity, our modular functions do not have any poles in the interior of $\HH/\Gamma(N)$. Note that for $\Gamma(N)$, the order of a pole at any cusp $\t_*$ is measured in the local variable $q_*^{1/N}$, where $q_* := \exp\({2\pi i \a^{-1}(\t_*)}\)$ and $\a^{-1}$ is the $SL(2, \ZZ)$ matrix that maps the cusp to infinity as $\t_* = \a(i\i)$ (see e.g. Chapter 4 of \cite{schultz2015notes}).

 Now consider the component $v_i(q)$ that is modular for $\Gamma(N)$, where $N$ is a positive integer. If $m_0^{(i)}<0$, then by definition $m_0<0$ and the proof is complete. Let us instead assume that $m_0^{(i)}>0$. This means that $\mathrm{ord}_{i \infty}(v_i) = Nm_0^{(i)}>0$. By the valence formula, this implies that $v_i(q)$ must have at least one pole. Since we are assuming no poles in the interior of $\HH/\Gamma(N)$, this pole must be at a cusp, say at $\tau=\t_*$.
 
Next, consider the diagonal modular invariant constructed from our vvmf,
\e{}{F(\qqb) = \vec{v}(q)\cdot \vec{\bar v}(\qb)~.}
Since the component $|v_i(q)|^2$ has a pole at $\t_*$, the full expression $F(q, \bar q)$ does as well. Now, it is always possible to map a cusp of $\HH/\Gamma(N)$ to $i \infty$ via an appropriate $SL(2,\ZZ)$ transformation. Since $F(\qqb)$ is $SL(2, \ZZ)$ invariant, this means that $F(\qqb)$ must have a pole at $i \infty$ as well. In other words, there must be a different component, say $v_j(q)$, of the vvmf with $m_0^{(j)}<0$. This is precisely what we wanted to prove!
 
We infer an immediate corollary, predicated upon the integrality conjecture:

\begin{cor}
\label{thm:conj1a}
Consider a $d$-dimensional vvmf $\vec v(q)$ with components $v_i(q)$. Define $m_0^{(i)}$ and $m_0$ as in (\ref{eq:chargenFourierexp}) and (\ref{eq:m0def2}), respectively. If $m_0^{(i)}\in \QQ$ and all Fourier coefficients $c_n^{(i)}$ are algebraic integers for all $i$, then $m_0<0$.
\end{cor}
\ni The proof is brief. Since we are assuming that the $m_0^{(i)}$ are rational, we can directly apply Conjecture \ref{thm:intconj} to conclude that all components $v_i(q)$ are modular functions of a congruence subgroup $\Gamma(N)$. Due to \eqr{eq:MDEconstd} and comments thereafter, we must have $m_0^{(i)}>0$ for at least one component. Applying Theorem \ref{thm:conj1b} then concludes the proof. We reemphasize that integrality of the Fourier coefficients plays a crucial role here, since this restricts the components of the vvmf to be modular functions of $\Gamma(N)$ for some $N$. 

Theorem \ref{thm:conj1b} is more powerful than Corollary \ref{thm:conj1a}, in that it does not require $m_0^{(i)}\in\QQ$, does not rely on the integrality conjecture, and requires an assumption about only one component of the vvmf. For instance, Theorem \ref{thm:conj1b} applies to ``logarithmic vvmfs,'' in which at least one component has $\log q$ appearing in the Fourier expansion \c{knopp2009logarithmic}. On the other hand, Corollary \ref{thm:conj1a} provides a useful bridge towards the physical requirement of integrality.\foot{We are not aware of a proof that, for general representations, if one component of a vvmf is modular for $\Gamma(N)$ then all components are. Indeed, in \cite{FRANC2016186} significant effort was expended in proving this statement for a narrow class of $d=3$ representations. The integrality conjecture is usually stated in the math literature with the assumption that all components of $\vec v(q)$ have bounded denominator. It may be the case that if a single component $v_i(q)$ has bounded denominator, then $v_i(q)$ is always modular for $\Gamma(N)$, but we are not aware of this question being addressed elsewhere. Finally, we note that Bantay proved the following theorem in \cite{bantay2013dimension}: if $\rho$ is an irreducible representation of finite image, then $m_0< 0$. This has overlap with Corollary \ref{thm:conj1a} because the integrality conjecture typically assumes that $\rho$ has finite image. }

\sec{Bootstrap bounds I: Twist spectrum}
\label{sec:twistgapboot}
Having proven Theorem \ref{thm:conj1}, we now discuss their physical implications, some of which were already outlined in the introduction. Our basic strategy is to construct modular functions $F(\qqb)$ from vvmfs, and then to think of $F(\qqb)$ as representing different CFT observables. This generates several new bounds. These bounds apply to any CFT observables comprised of tensor products of vvmfs, for any $d$ (not necessarily finite). While this mainly includes partition functions and local correlation functions in RCFTs with arbitrary chiral algebra $\A\times\overline\A$, it also extends to observables involving degenerate representations in irrational CFTs. Our bounds also apply to any observables in irrational CFTs obeying the assumptions in Section \ref{sec:Lorentzmodform}, as well as any vvmfs with sufficiently large Wronskian index. For concreteness, we will phrase all of the physical implications in terms of compact RCFTs, with the other aforementioned cases implied. Finally, we assume that all vvmfs in this section have $d>1$ unless otherwise noted

\ssec{Positivity of effective central charge}
\label{ceffsec}

Take $F$ to be a partition function,
\e{}{F(q, \bar q) = Z(q, \bar q)\,,}
with central charge $c$. The leading behavior of the partition function near the cusp is
\e{Zceff}{Z(\qqb) \sim q^{-{\ceff\o 24}}\qb^{-{\bar{c}_{\rm eff}\o 24}}\,,\quad \text{where}\quad \ceff : = c-24h_{\rm min}}
and $h_{\min}$ is the chiral dimension of the lightest operator in the theory (likewise for the anti-chiral piece). Then the requirement that $m_0 < 0$ implies the following,
\begin{res}\label{ceff}
All rational CFTs have $\ceff>0$.
\end{res}
%
\ni In a unitary theory, $h_\min=0$ and $\ceff$ is trivially positive. However, this result is non-trivial in a non-unitary theory. In particular, if $c<0$, then we conclude that there must exist an operator with $h<0$.\footnote{As a simple example, note that for generic Virasoro minimal models $\M(p,q)$ one has  $\ceff(p,q)=1-{6\o pq}$ \c{DiFrancesco:1997nk}, which for any $p\geq 2, q\geq 2$, and $(p,q)=1$ is indeed positive. (The case $p=2,q=3$ is a trivial theory.)}  

\ssec{Uniqueness of $t>{c\o 12}$ spectrum}
\label{twistuniqueness}
Take $F$ to be a difference of partition functions, 
\e{}{F(q, \bar q) = Z_1(q, \bar q)-Z_2(q, \bar q)\,,}
where CFT$_1$ and CFT$_2$ are taken to have the same central charge $c$. Let us furthermore assume that CFT$_1$ and CFT$_2$ have spectra which are identical up to twist $t  = {c \over 12}$. Then upon taking the difference of partition functions, the $t< {c \over 12}$ portions of the spectrum cancel out, and we conclude that $F(q, \bar q) \sim q^{m_0}q^{\bar m_0}+ \dots $ with $m_0 =  h -{c \over 24}\geq 0$. Since all non-trivial\foot{The previous section's results assumed $d>1$. For $d=1$, the basis of vvmf is given in Appendix \ref{appd1}. Except for the trivial form -- the constant -- all $d=1$ vvmf have $m_0<0$. It is consistent with modular invariance to add an arbitrary constant to a non-holomorphic CFT partition function. However, one should view this operation as adding a multiple of $1 = {|\eta(q)|^2\o |\eta(q)|^2}$, which contributes both positive and negative Virasoro {\it primary} degeneracies at {\it all} levels $h,\hb={c\o 24}+\ZZ_{\geq 0}$. We put this formal possibility aside.} $F(q,\qb)$ must have $m_0 <0$, we conclude that $F(q, \bar q)=0$. Hence the two theories are identical. In other words, 
\begin{res}\label{res42}
 The twist spectrum with $t \geq {c \over 12}$ is uniquely determined by its complement.
 \end{res} 

\noindent This immediately implies the following corollary:
\begin{res}\label{res43}
There is at most one RCFT partition function with a given extended chiral algebra, of central charge $c$ and twist gap $t_*\geq{c\o 12}$.
\end{res}
\noindent These statements about non-holomorphic RCFT may be viewed as extensions of the same statements in holomorphic CFT, where they are true owing to basic aspects of meromorphic modular forms. We emphasize that these results are exact in $c$.

\ssec{Spectral and OPE bounds from torus one-point functions} 
\label{torusoneptsec1}

Take $F$ to be the torus one-point function of a local operator $\cO$ with conformal weights $(h_\O,\hb_\O)$. More precisely, since $\<\O\>$ has modular weight $(h_\O,\hb_\O)$, take $F$ to be the non-holomorphic modular function
\bea\label{Ftorus}
{F(q, \bar q) =\eta(q)^{-2h_\cO}\bar \eta(\qb)^{-2 \bar h_\cO} \langle \cO \rangle~.}
\eea
This allows us to address the following bootstrap-type question: {\it Given some operator $\O\subset \phi^\dagger \x\phi$, what is the upper bound on the twist gap to the lowest-twist operator $\phi=\phi_*$?} This is the converse of the prototypical bootstrap OPE question, which fixes $\phi$ and bounds the conformal data of $\O$.

Let us define the following:
\es{torusdefs}{\A_\O:&\quad \text{chiral algebra with respect to which $\O$ is primary}\\
\cF_\phi(q):&\quad \text{torus one-point blocks for $\A_\O$ with internal $\phi$}\\
d_\O:&\quad \text{number of $\A_\O$-primaries $\phi$ such that $\cO\subset\phi^\dagger \times \phi$}}
We take $F(\qqb)$ to admit an expansion in vvmfs of dimension $d_\O$,
\e{oneptexp}{F(\qqb) = \sum_{i,j=1}^{d_\O} N_{ij}v_i(q)  \bar v_j(\qb)~.}
Rearranging the Fourier expansion in terms of one-point conformal blocks gives
\bea\label{blockexp}
\langle\cO\rangle= \sum_\phi C_{\phi^\dagger\cO\phi} |\cF_\phi(q)|^2~,
\eea
where $\phi$ is $\A_\O$-primary, and we normalize the blocks as $\cF_\phi(q) \sim q^{h_\phi-{c\o 24}}(1+O(q))$. Then the components $v_i(q)$ can be taken to be
\e{veta}{v_i(q) = \eta(q)^{-2h_\cO}\cF_{\phi_i}(q) ~,\quad \bar v_i(q) = \eta(\qb)^{-2\hb_\cO}\bar\cF_{\phi_i}(\qb)~.}
The leading term near the cusp is given by the operator of smallest conformal weight for which $C_{\phi^\dagger\O\phi}\neq 0$, call it $\phi_*$. Henceforth in discussions of torus one-point functions, ``primary'' means primary with respect to $\A_\O$.

In order to extract bounds from torus one-point functions using Theorem \ref{thm:conj1}, we need to ask when the one-point functions are integral.\foot{The same method employed in this section may be applied to sphere four-point functions \c{Headrick:2015gba,Maldacena:2015iua, Cheng:2020srs}. This was recently fully systematized and extended in \c{Cheng:2020srs}. When comprised of at least three identical operators, suitably defined sphere four-point blocks are vvmfs for $SL(2,\Z)$. As with torus one-point blocks, these are sometimes, but not generically, integral. Such cases can be read off from \c{Poghossian:2009mk}. In addition, in the ``sphere-torus correspondence'' of \c{Cheng:2020srs}, certain four-point conformal blocks are identified with conformal characters, and are thus integral. It would be worthwhile to use our bounds to infer OPE and spectral constraints from those and other cases.} This is a slight abuse of nomenclature: more precisely, we ask this of the torus one-point {\it blocks}, which are the components of the vmmf via \eqr{veta}, and not of the full one-point functions, which contain OPE coefficients. Before addressing the space of integral one-point functions, let us first state the bounds which apply to them. 
\begin{res}
\label{eq:chiraldimcond2}
If $\<\O\>$ is non-zero and integral, then there exists an operator $\phi_*$ obeying 
\bea
\label{eq:hphibound} h_{*} <  {c+2h_\O\o 24}~,\quad C_{\phi_*^\dagger \O\phi_*}\neq 0~.
\eea
\end{res}
\noindent 
Note that in RCFT, unlike in irrational CFT \c{Collier:2016cls,Afkhami-Jeddi:2017idc}, there is no accumulation of $\cA$-primaries below $h={c\o 24}$ required by modular invariance, so this bound is non-trivial. If $\O$ is scalar, then we can phrase this neatly in terms of the twist:
\begin{res}
\label{eq:firsttwistbound}
If $\<\O\>$ is non-zero and integral and $J_\O=0$, then there exists an operator $\phi_*$ obeying $$t_{*} < {c+t_\O\o 12}~,\quad C_{\phi_*^\dagger \O\phi_*}\neq 0~.$$
\end{res}
\ni These are non-trivial bounds on the $\phi^\dagger\x\phi$ OPE. If moreover $t_\O \geq {c\o11}$, then $t_{*}<t_\O$, and hence this is also a non-trivial bound on the twist spectrum. 
To achieve the strongest bound within a given theory, take $\O$ to be the lightest non-vacuum operator with a non-zero integral one-point function; then there must exist another primary operator $\phi_*$ obeying (\ref{eq:hphibound}), which moreover has $C_{\phi^\dagger \O\phi}\neq 0$.

Finally, by the same logic as in Section \ref{twistuniqueness}, taking $F$ to be a {\it difference} of one-point functions, each of which admits a conformal block expansion, leads to the following:
\begin{res}
\label{eq:thirdtwistbound}
If $\<\O\>$ is non-zero and integral, then contributions to $\<\O\>$ with $h_\phi<{c+2h_\O\o 24}$ uniquely determine those with $h_\phi\geq {c+2h_\O\o 24}$. 
\end{res} 
\ni We call this \textit{OPE determinacy}: all OPE coefficients $C_{\phi^\dagger \O\phi}$ at and above the aforementioned threshold -- or, in the presence of degeneracy, the sum $\sum_\phi C_{\phi^\dagger \O\phi}$ at fixed level $h_\phi$ -- are fixed by those below. 

\vs

Now, {\it when} are torus one-point functions so constrained as to obey Results \ref{eq:chiraldimcond2} -- \ref{eq:thirdtwistbound}? In other words, when are they integral?  A number of examples in known RCFTs are collected in Section \ref{sec:twistexampsubsec}, including infinite classes of one-point functions in Virasoro minimal models. These provide non-trivial checks of our bounds. Here we want to take the abstract approach, presenting a general class of torus one-point functions which are always integral in any CFT.

\sssec{$h_\O=1$ and marginal operators} 
\label{margsec1}

It is a striking fact that one-point blocks $\cF_\phi(q)$ are \textit{always integral} when $h_\O=1$. In this case, the blocks enjoy the simple relation
\e{blockchar}{\cF_\phi(q) =\chi_\phi(q)~,}
where $\chi_\phi(q)$ is the conformal character for the module $\phi$, defined in \eqr{chars}. The relation \eqr{blockchar} follows from conformal invariance of the $h_\O=1$ chiral field integrated over a circle, as recalled in \c{Kraus:2016nwo}. As such, it holds for both rational and irrational blocks alike, defined for any chiral algebra $\A$. Clearly the components \eqr{veta} have integral Fourier coefficients for any $\phi_i$ because both $\eta(q)$ and $\chi_{\phi_i}(q)$ do. Thanks to this integrality, we may use Theorem \ref{thm:conj1} to conclude that Results \ref{eq:chiraldimcond2} -- \ref{eq:thirdtwistbound} hold here with $h_\cO=1$. 

A particularly interesting sub-case is when $\O$ is a marginal scalar, 
\e{}{h_\O=\hb_\O=1~.}
Suppose $\O$ is exactly marginal. By standard first-order conformal perturbation theory, $C_{\O^\dagger\O\O}=0$. This implies that 
\begin{res}
\label{res:margtwistbound}
Any RCFT with an exactly marginal operator $\O$ with $\<\O\>\neq 0$ must have a primary $\phi_*\neq \O$ satisfying $$t_{*} < {c+2 \over 12}~,\quad C_{\phi_*^\dagger\O\phi_*}\neq 0~.$$ Contributions to $\<\O\>$ with $t_\phi<{c+2\o 12}$ uniquely determine those with $t_\phi\geq {c+2\o 12}$.
\end{res}

\noindent This is a constraint on any rational point on a conformal manifold.\footnote{We remind the reader that Virasoro RCFTs cannot have conformal manifolds:  deforming away from the rational point gives a theory with the same central charge due to the $c$-theorem, but which is necessarily irrational since the OPE data depends on the coupling. Instead, the presence of a conformal manifold is only allowed in the presence of an enhanced chiral algebra $\cA\supset \Vir$, whereupon moving away from the rational point breaks $\cA$ down to $\cA' \supseteq \Vir$.} There are several other nice aspects of the exactly marginal case. We defer their discussion to Section \ref{margsec}, where we derive analogous bounds on the dimension $\D_{*}$, rather than the twist $t_{*}$.

\sssec{Two asides}
It seems highly likely that $h_\O=1$ is the {\it only} non-zero value for which $\cF_\phi(q)$ is integral for all $h_\phi$.\foot{The torus one-point Virasoro block $\cF_\phi(q)$ is a specialization of the sphere four-point Virasoro block, as proven in \c{Poghossian:2009mk,Hadasz:2009db}. The latter is well-known to obey Zamolodchikov's recursion relations \c{Zamolodchikov:1985ie,zamo2}, which generate increasingly complicated Fourier coefficients that are rational functions of parameters; the first several coefficients of $\cF_\phi(q)$ may be found in the Appendix of \c{Alkalaev:2016fok}. Experience strongly suggests that integrality will not hold for all $h_\phi$ unless $h_\O = 0,1$.} Indeed, we believe that an even stronger result holds for non-degenerate torus one-point Virasoro blocks $\cF_\phi(q)$: for $h_\O \neq 0,1$, there do not exist triplets $(h_\O,h_\phi, c)$ for which $\cF_\phi(q)$ is integral for {\it any} fixed values of these parameters. We do not try to prove these statements here, though it would be of independent interest to do so.

We also note that similar spectral bounds are easily derived from the modular differential equation for {\it characters} of any RCFT. For an RCFT with $d$ characters, the indicial equation for chiral characters is $\sum_{i=1}^d \(h_i - {c\o 24}\) = {d(d-1)\o 12}- {\ell \over 6}$ \c{mathur1988classification}. Rearranging, and dropping the vacuum operator, one sees that there must be at least one non-vacuum operator obeying
\e{eq:charbound}{h\leq {d\o d-1}{c+2(d-1)-4 \ell/d\o 24}~.}
Over wide ranges of parameters $d, c$, and $\ell \leq {d(d-1)\over 2}$,  the optimal one-point function bound beats (\ref{eq:charbound}). In contrast to the one-point function bound, the character bound has nothing to say about the OPE.


\subsection{Application: Schur indices of 4d SCFTs}
\label{sec:Schur}

One application of our results beyond two-dimensional conformal field theory is to 4d $\cN=2$ theories, in the context of the vertex operator algebra (VOA) framework of \cite{Beem:2013sza}.
In this context, the 4d Schur index $\mathcal{I}_{\rm Schur}$ is identified with the vacuum character of a 2d VOA, which satisfies a finite-order MDE \cite{Beem:2017ooy}. The other solutions of the MDE furnish non-vacuum modules of the VOA, some of which may be interpreted as indices in the presence of $(2,2)$ surface defects in the 4d theory. We now address the question of whether \textit{all} of the non-vacuum modules may be associated to $(2,2)$ defects. 

The requirement of integrality of the Schur index means that at least one solution to the MDE has an integral Fourier expansion. Suppose the MDE is $d$-dimensional. If the other solutions are indeed to correspond to sensible indices in the presence of defects, their expansions, too, should be discrete and integral. Accordingly, if $\mathcal{I}_{\rm defect}(q)$ is the entry of a $d$-vector $\vec v(q)$ with smallest leading exponent, and writing
\e{}{\mathcal{I}_{\rm defect}(q) \sim q^{-{c_{\rm eff}\o 24}}~,}
where $\ceff$ is the effective central charge of the 2d VOA, then Corollary \ref{thm:conj1a} implies the following: \textit{if all $d$ entries of $\vec v(q)$ are (defect) Schur indices, then $c_{\rm eff}\geq 0$.} Conversely, if one were to identify a case in which $c_{\rm eff} < 0$, then one would conclude that not all characters of the VOA admit an interpretation as indices in the presence of defects. Of course, for finite-dimensional VOAs this is disallowed by Result \ref{ceff}. 

It is interesting to see how $c_{\rm eff}>0$ is realized from the 4d perspective. When $c_{4d} > a_{4d}$, one has the following expression for $\ceff$ in terms of the central charges of the 4d theory \cite{Beem:2017ooy}, 
\bea
 c_{\rm eff} =48(c_{4d}-a_{4d})>0~.
 \eea
On the other hand, when $a_{4d} > c_{4d}$ it is expected that this relation gets modified according to Section 3.3 of \cite{Ardehali:2015bla}. Schematically, the proposal there is that 
\bea
c_{\rm eff} \propto (c_{4d}-a_{4d})_{\rm shifted} :=c_{4d}-a_{4d}- {3 \over 4}\cL_{\rm min}
\eea
where $\cL_{{\rm min}}$ is the minimization of the so-called Rains function, and the proportionality constant is positive. What is relevant for our purposes is that in all examples studied there, it was shown that $(c_{4d}-a_{4d})_{\rm shifted}>0$, and hence again $c_{\rm eff}>0$.\footnote{The following open question was posed in \cite{Ardehali:2015bla}: Is there a general correlation between the sign of ${\rm Tr}R$ in a SUSY gauge theory with a semi-simple group, and the sign of the theory's $\cL_{\rm min}$? Since ${\rm Tr}R = - 16 (c_{4d}-a_{4d})$, our results suggest that for ${\rm Tr}R>0$ one must have $\cL_{\rm min}<0$ (at least for a finite-dimensional VOA).}

However, it should be noted that it is not necessarily the case that the modules of the VOA have finite degeneracies; in particular, they may furnish {\it logarithmic} vvmfs, in which $v_i(q) \sim q^\#\log q$ for at least one component. Moreover, there are known cases where an index becomes logarithmic upon adding \c{beem} or removing \c{Bianchi:2019sxz} flavor fugacity. Accordingly, the physical significance of whether the Schur index enumerates a finite-dimensional state space at fixed level is not clear. It would be interesting to sharply determine what the discreteness and integrality criteria on the 2d VOA vvmfs tell us about the set of surface defects of 4d SCFTs.

\ssec{vvmfs at large $d$ and irrational CFT}
\label{secirr}

The bounds obtained in the previous subsections hold for any $d$, including $d\rar\i$.  One might then ask if there are any special contexts in which such large $d$ vvmfs arise. In particular, would bounds obtained at large $d$ apply to generic irrational CFTs?

First consider an infinite sequence of RCFTs in which the number $d$ of irreducible modules goes to infinity in an asymptotic limit. For example, any sequence of unitary minimal models admits such a limit; for the Virasoro minimal models, one has $d\approx (c-1)^{-1}$ as $c\rar 1$. In these cases all observables are described by vvmfs at every point in the sequence, and the limit theory sits ``on the border'' of rationality and irrationality (e.g. \c{Runkel:2001ng,Ribault:2014hia,Benjamin:2018kre}). 

However, for bona fide irrational CFTs it is unclear whether vvmfs at $d\rar \i$ play any role. One argument against this comes from considering the form of the modular $S$-matrix for characters. Define $c_{\rm currents}$ as
\e{ccurrents}{\log \chi_{\rm vac}(q\rar 0)\approx {i\pi\t^{-1}\o 12}c_{\rm currents}}
Then the modular $S$-matrix has support on $h\geq c-c_{\rm currents}$, for both vacuum and non-vacuum characters. In RCFT, $c=c_{\rm currents}$, whereas in irrational CFTs, $c>c_{\rm currents}$. Thus, an essential distinction between RCFT and irrational CFT is that in the former case, the modular $S$-matrix has support on the entire spectrum with $h \geq 0$, whereas in the latter case it does not. This suggests that the latter case cannot be brought under the umbrella of vvmfs. 

That being said, we now make a few more optimistic comments. First, in Theorem \ref{lorthm} we proved that $m_0\leq 0$ in any non-holomorphic function -- not necessarily formed from vvmfs -- obeying the (admittedly rather stringent) conditions set forth there. That proof could possibly be generalized to cases where the Fourier expansion proceeds in more generic powers (though see footnote \ref{footnote:infindex} for a comment on the difficulty of establishing this). Second, insofar as Theorem \ref{thm:conj1b} is concerned the indicial exponents $m_0^{(i)}$ of vvmf components $v_i(q)$ are not constrained to be rational, and hence can accomodate the operator dimensions/central charges appearing in an irrational theory. For both of these reasons, one might consider contemplating the possibility of conjecturing that even in irrational CFTs, the twist spectrum with $t \leq {c \over 12}$ uniquely determines that with $t > {c \over 12}$. But we will neither consider nor contemplate that possibility here. We leave the question of whether generic CFTs respect $m_0<0$, and hence the bootstrap bounds laid out in this section, for future investigation.

\section{Non-holomorphic cusp forms}
\label{sec:nonholocusps}
In Sections \ref{sec:mainvvmfsec} and \ref{sec:twistgapboot} we proved Theorem \ref{thm:conj1} and discussed a number of physical implications. These implications were all restricted to statements about the twist spectrum of the theory. Indeed, we have not yet been able to say anything about the spectra of dimensions. To make progress on this front, it is useful to restrict momentarily to the imaginary axis, where we can write our generic non-holomorphic function as
\e{eq:qeqqbexp2}{F(q,q) = q^{x_0} \sum_{x\geq 0} a_x q^x}
with
\bea
\label{eq:x0def2}
x_0 := \mathrm{min}_{\cI}(m_0^{(i)} + \bar m_0^{(j)})~,\hspace{0.5 in} \cI  = \{(i,j) \,\,| \,\, N_{ij} \neq 0\}~.
\eea
The minimization above is over pairs $(i,j)$ such that there is non-trivial gluing between chiral and anti-chiral characters in the character decomposition of $F(q,\bar q)$. When $F(\qqb)=Z(\qqb)$ is a partition function, $x_0$ is related to the dimension $\D$ of a low-lying operator by $x_0 = \Delta - {c \over 12}$. In order to obtain bounds on dimensions, it is thus fruitful to contemplate bounds on $x_0$. 

Obtaining general bounds on $x_0$ is a more difficult task than for its chiral analog $m_0$, since the former is sensitive to the choice of gluing matrix $N_{ij}$ (the full classification of which remains unknown). Nevertheless, in analogy with $m_0$ let us ask whether a bound $x_0 <0$ for all integral functions $F(q,\bar q)$ could in principle hold. Conversely, we can ask the following question: {\it do discrete, integral, and positive non-holomorphic functions $F(q, \bar q)$ with $x_0>0$ exist?} In Section \ref{sec:noposintconst}, we use a simple argument to answer this question negatively -- no such function can exist. 

However, for some of the physical applications encountered below, positivity will not be necessary. In this case, we can ask a broader question:  {\it do discrete and integral (but not necessarily positive) non-holomorphic functions $F(q, \bar q)$ with $x_0>0$ exist?} In Section \ref{sec:existenceofcuspfunc} we answer this in the affirmative by explicitly constructing such functions. We may summarize the answers to these two questions with the following theorem, 

\begin{thm}
\label{thm:cuspthm}
There do not exist non-holomorphic cusp forms satisfying positivity, discreteness, and integrality. On the other hand, there \textit{do} exist non-holomorphic cusp forms satisfying discreteness and integrality (but not positivity). 
\end{thm}

\subsection{No cusp forms with discreteness, integrality, and positivity}
\label{sec:noposintconst}

We begin with a proof of the first half of the theorem, namely of the statement that there do not exist non-holomorphic cusp forms whose Fourier coefficients are discrete, integer, and positive. We begin with the parameterization \eqr{cuspfunc}, which we repeat here for a cusp form $\C(\qqb)$:
\bea
\label{cuspfunc2}
\cC(q, \bar q) = \sum_{N=0}^\infty \left( a_N(\tau_2) \,e^{2 \pi i N \tau_1} + \tilde{a}_N(\tau_2)\, e^{-2 \pi i N \tau_1}\right)
\eea
with
\es{}{a_{N}(\tau_2) &= \sum_{\Delta>-{N\o 2}} c_N(\Delta)\, (q \bar q)^\Delta e^{-2 \pi N \tau_2}~,\\
\tilde{a}_{N}(\tau_2) &= \sum_{\Delta>-{N\o 2}} \tilde{c}_N(\Delta)\, (q \bar q)^\Delta e^{-2 \pi N \tau_2}~.}
For simplicity we have suppressed the modular weights of $\C$, which may be arbitrary. Clearly we have 
\bea
\left|a_N(\tau_2)\right| = \left| \int_0^1 d\tau_1\, e^{-2 \pi i N \tau_1}\, \cC(q, \bar q)\right| \leq \int_0^1 d\tau_1 \left|\cC(q, \bar q)\right|
\eea
by the triangle inequality for integrals. Being cuspidal, $\cC(q, \bar q)$ is bounded in the upper-half plane $\HH$, and thus there exists some finite $\Lambda \in \RR_{>0}$ such that $|\cC(q, \bar q)| \leq \Lambda$ for all $\tau\in \HH$. We thus conclude that 
\bea
\label{importantbound}
\Bigg| \sum_{\Delta>-{N\o 2}} c_N(\Delta)\, (q \bar q)^\Delta e^{-2 \pi N \tau_2}\Bigg| \leq \Lambda~.
\eea
The above inequality must hold for all $\tau_2\geq 0$, and in particular for $\tau_2 = 0$, thereby giving 
\bea 
\label{sumBN}
\Bigg| \sum_{\Delta>-{N\o 2}} c_N(\Delta)\Bigg| \leq \Lambda~.
\eea
If we impose the requirement of positivity of the coefficients $c_N(\Delta)$, we conclude that for any given $\Delta$ we have 
\bea
\label{eq:Fouriercoeffbound}
c_N(\Delta) \leq \Lambda~.
\eea
We now further assume integrality of the Fourier coefficients, i.e. $c_N(\Delta) \in \ZZ_{\geq0}$ for all $N$ and $\Delta$. This then implies that only a finite number of Fourier coefficients are non-vanishing, i.e. there exist finite $N_0$ and $\Delta_0$ such that
\bea
c_N(\Delta) = 0~, \hspace{1 in} \forall\, (N>N_0 )\vee (\Delta > \Delta_0)~.
\eea
But it is not possible to have a non-trivial weight-0 cuspidal function with finite Fourier series, as this cannot be invariant under $\tau \rightarrow -1/\tau$. Hence no cuspidal functions satisfying our physical requirements exist.

Note that this depended crucially on discreteness. Indeed, if we instead had something of the form 
\bea\label{contpos}
a_{N}(\tau_2) = \sum_{\Delta>-{N\o 2}} c_N(\Delta)\, \tau_2^\a \,(q \bar q)^\Delta e^{-2 \pi N \tau_2}
\eea
for some $\a= \a(\Delta, N)$, then setting $\tau_2=0$ in the analog of (\ref{importantbound}) would not give rise to the bound (\ref{sumBN}). Instead, the simplest (and perhaps strictest, for generic $N, \Delta$) non-trivial bound is obtained by taking $\tau_2 \propto 1/(N+2 \Delta)$, for which we conclude that 
\bea
c_N(\Delta) \leq \Lambda (N+2 \Delta)^\a~.
\eea 
This does not allow us to conclude that the cuspidal function is trivial.

Which of the three conditions assumed must be sacrificed to obtain non-trivial cusp forms? It is easy to see that positivity in particular can never be compatible with cuspidality. One way to argue this is as follows. Returning to our previous parameterization of the Fourier expansion and restricting to the imaginary axis as in (\ref{eq:qeqqbexp2}), we may consider the limit $\tau=-\bar \tau\rightarrow i \infty$. Cuspidality and $S$-invariance imply that 
\e{ax}{\sum_{x\geq 0} a_x = 0~,}
from which we deduce that positivity must be sacrificed. This holds for modular forms of arbitrary weight $(w,\bar w)$: since $\C$ decays exponentially at the cusp, the powers of $\log q$ coming from $S$-covariance make no difference. 

To summarize, {\it there are no discrete, positive, integral non-holomorphic cusp forms.}

\ssec{Known constructions are not discrete}
\label{app:literaturesearch}

In the next subsection, we will prove the second half of Theorem \ref{thm:cuspthm} -- namely that if we relax the condition of positivity, there \textit{do} exist non-holomorphic cusp forms satisfying integrality and discreteness. Before doing so, we point out that, to the best of our knowledge, no examples of such functions have appeared previously in the math or physics literature. In this subsection we survey all systematic constructions of cuspidal functions known to us, showing that none of these functions is discrete. The reader uninterested in this literature review can safely skip to Section \ref{sec:existenceofcuspfunc}.

\sssec{Maass forms}
\label{sec:Maassforms}

By far the most well studied class of non-holomorphic cuspidal functions are so-called Maass cusp forms, which are defined to satisfy a Laplace eigenvalue equation,
\bea
\Delta \cC = \lambda\, \cC~,\hspace{1 in}\lambda \neq 0
\eea
where $\Delta = \tau_2^2 (\p_{\tau_1}^2 + \p_{\tau_2}^2)$. We assume that $\lambda$ is non-zero since a harmonic cusp form is exactly zero by the maximum principle for harmonic functions.

It is easy to show that no Maass forms can satisfy discreteness. To see this, we note that the Fourier coefficients of $\cC(q, \bar q)$ can be shown to satisfy a differential equation as follows,
\bea
\label{Fouriercoeffs}
a_N(\tau_2) &=& \int_0^1 d\tau_1\, e^{-2 \pi i N \tau_1}\, \cC(q, \bar q)
\no\\
&=&{1 \over \lambda} \int_0^1 d\tau_1 \,e^{-2 \pi i N \tau_1 }\,\Delta \cC(q, \bar q)
\no\\
&=& {\tau_2^2 \over \lambda} \left[(2 \pi i N)^2 a_N(\tau_2) + \p_{\tau_2}^2 a_N(\tau_2) \right]~,
\eea
or in other words
\bea
\tau_2^2 \p_{\tau_2}^2 a_N(\tau_2) = \left(\lambda + (2 \pi N)^2 \tau_2^2\right) a_N(\tau_2)
\eea
which one recognizes as a modified Bessel equation. We should discard solutions of the type $I_{n}$ since they have incorrect behavior at the cusp, and we are left with 
\bea
\label{Besselexpr}
a_N(\tau_2) = c_N \sqrt{\tau_2} \,K_{\half \sqrt{1+ 4 \lambda}}\left(2 \pi N\tau_2\right)~,
\eea
where $c_N$ are $\tau_2$-independent constants. It is then clear that we cannot get rid of terms with explicit $\tau_2$ dependence. For example, if $\tau_2 \rightarrow \infty$, then (\ref{Besselexpr}) becomes 
\bea
a_N(\tau_2) = {c_N \over 2 \sqrt{N}} \left(1 + {\lambda \over 4 \pi N \tau_2} + {\lambda (\lambda-2) \over 8 (2 \pi N \tau_2)^2} + \dots \right)e^{-2 \pi N \tau_2}~.
\eea
The same argument goes through for $\tilde{a}_N(\tau_2)$.

\sssec{Construction of \cite{Brown:2017}}

In \cite{Brown:2017}, the following construction of an infinite family of non-holomorphic cusp forms was outlined. Begin by choosing an arbitrary weakly holomorphic modular form $f$ of weight $h+2$. This admits a Fourier expansion
\bea
f= \sum_{n \geq -N} a_n q^n~, \hspace{1 in} a_n \in \CC
\eea
where $N$ is finite. Given this $f$, one then constructs 
\bea
f^{(k)} = \sum_{n \in \ZZ\backslash 0} {a_n \over(2n)^k}q^n
\eea
for any $k \geq 0$, as well as the functions 
\bea
R_{r,s}(f) = (-1)^r \binom{h}{r} \sum_{k=s}^h \binom{r}{k-s} {k! \over (2 \pi \tau_2)^k}f^{(k+1)}
\eea
for any $r,s \geq 0$ such that $r+s=h$. In terms of these ingredients, one obtains a set of weight-$(r,s)$ non-holomorphic modular forms $\cH(f)_{r,s}$ defined via
\bea 
\cH(f)_{r,s} = -{2 \pi a_0 \over h+1} \tau_2 + \a_f(-1)^r \binom{h}{r} (-2 \pi \tau_2)^{-h} + R_{r,s}(f) + \overline{R_{s,r}(\mathbf{s}(f))}
\eea
for some constant $\a_f$ and some Hecke-equivariant map $\mathbf{s}$, the details of which may be found in \cite{Brown:2017}. 

To have $\cH(f)_{r,s}$ vanish at the cusp, we must set $a_0 = 0$, i.e. choose the seed function $f$ to be a cusp form. If we furthermore want $\cH(f)_{r,s}$ to be discrete, we must avoid  factors of $\tau_2$ arising from $R_{r,s}(f) + \overline{R_{s,r}(\mathbf{s}(f))}$, which can only be done by choosing $h=0=r=s$. Hence our seed function $f$ must be a weight-two holomorphic cusp form; however, no such cusp forms -- indeed, no such modular forms -- exist.

More generally, we can construct the following infinite family of modular-invariant functions:
\e{brownfns}{\mathcal{H}(\vec f)_{\vec r, \vec s}:= \t_2^{t}\prod_{i}\mathcal{H}(f_i)_{r_i,s_i} \,,\quad \text{where}\quad t: = \sum_i r_i = \sum_i s_i}
and the functions $f_i$ are weakly holomorphic modular forms of weight $r_i+s_i+2$. In order to eliminate the naked factors of $\t_2$ for this function, we require 
\e{t2cond}{\prod_{i}\mathcal{H}(f_i)_{r_i,s_i}  \propto \t_2^{-t}~.}
In order to achieve \eqr{t2cond} without elaborate cancellations among different elements of the product \eqr{brownfns}, each $\mathcal{H}(f_i)_{r_i,s_i}$ should be proportional to a fixed power of $\t_2$. By the same logic as above, we need $h_i=0=r_i=s_i$. Again, this is not possible. Thus, barring cancellations, it is not possible to eliminate naked factors of $\t_2$ from $\mathcal{H}(\vec f)_{\vec r, \vec s}$, even without imposing cuspidality.

\sssec{Construction of \cite{DHoker:2019txf}}

In \cite{DHoker:2019txf}, a number of explicit constructions of non-holomorphic cuspidal functions were given. These involve the application of differential operators to certain combinations of modular graph functions, with the results being broadly organized into odd and even cuspidal functions (this distinction refers to parity under $\tau_1 \rightarrow - \tau_1$). For introductions to and applications of modular graph functions, see e.g. \cite{DHoker:2015wxz,DHoker:2016mwo,DHoker:2016quv,DHoker:2017zhq,Gerken:2018zcy,Gerken:2018jrq,Gerken:2019cxz}.

The simplest example of odd cuspidal functions are obtained by applying the Cauchy-Riemann operator to non-holomorphic Eisenstein series (i.e. one-loop modular graph functions). In particular, we recall the definition of the weight-$w$ non-holomorphic Eisenstein series,
\bea
E_w(\tau, \bar \tau) = \sum_{\substack{m,n \in \ZZ \\ (m,n) \neq (0,0)}}{\tau_2^w \over \pi^2 |m\tau+n|^{2w}}
\eea
which admits the following Fourier expansion, 
\bea
E_w(\tau, \bar \tau)&=& - {B_{2w} \over (2w)!}(-4 \pi \tau_2)^w + {4 (2 w - 3)! \zeta(2 w - 1) \over (w-2)! (w-1)! (4 \pi \tau_2)^{w-1}}
\no\\
&\vphantom{.}&\hspace{0.5 in}+ {2 \over (w-1)!} \sum_{N=1}^\infty N^{w-1}\sigma_{1-2w}(N) P_w(4 N \pi \tau_2) \left(q^N + \bar q^N \right)
\eea
where $B_{2w}$ are the Bernoulli numbers, $\sigma_{1-2w}(N)$ is the divisor function, and $P_w(x)$ is a polynomial in $1/x$ given by 
\bea
P_w(x) = \sum_{m=0}^{w-1} {(w+m-1)! \over m! (w-m-1)! x^m}~.
\eea
Then using the Cauchy-Riemann operator $\nabla = 2 i \tau_2^2 \p_\tau$ one can construct the combinations 
\bea
\cP_k(w_1, w_2) = \tau_2^{-2k}\left( \nabla^k E_{w_1} \overline{\nabla}^k E_{w_2} - \nabla^k E_{w_2} \overline{\nabla}^k E_{w_1}  \right)
\eea
for any $k \geq 1$. These are clearly odd under $\tau_1 \rightarrow -\tau_1$ and thus cannot have constant terms (i.e. terms without any powers of $q$ or $\qb$) in their Fourier expansion, since such terms would necessarily be even. Hence these are examples of odd cuspidal functions. Though the non-holomorphic Eisenstein series are Maass forms, the combinations $\cP_k(w_1, w_2)$ are not, and as such the arguments of Section \ref{sec:Maassforms} do not rule them out. However, it is easy to check explicitly that for any choices of $k, w_1,$ and $w_2$ there exist naked factors of $\tau_2$ in the $q$-expansion. Hence this class of non-holomorphic cuspidal functions is not discrete. 

Beyond the simple example given above, the space of all odd cuspidal functions built from one- and two-loop modular graph functions was obtained in \cite{DHoker:2019txf}. We have checked that for all ``weights" $w \leq 12$, none of them is discrete.\footnote{Note that this use of the word ``weight," common in the modular graph function literature, refers not to ``modular weight" -- which in the current case is zero -- but rather to the sum of the entries in the first line of the $2 \times 2$ matrix labelling the one- or two-loop modular graph functions.}

Several classes of even cuspidal functions were also constructed in \cite{DHoker:2019txf}. The simplest of these are constructed by first defining the following differential operator, built out of the Laplacian, 
\bea
\Delta_w = \prod_{\ell =1}^w \left(\Delta - \ell (\ell-1) \right)~.
\eea
Then the functions $\Delta_w\left(E_{w_1} E_{w_2} \right)$ with $w_1+w_2\in\Z$ and $w_1 + w_2 \leq w$ are even cuspidal functions: as shown in \cite{DHoker:2019txf}, $\D_w$ annihilates annihilates any Laurent polynomial with degree $(w,1-w)$. Indeed, these are the lowest of an infinite class of even cuspidal functions
\e{}{E(\vec w|w) := \Delta_w\Bigg(\prod_i E_{w_i} \Bigg)\,,\quad \text{where}\quad \sum_i w_i \leq w}
and we demand $\sum_i w_i\in\Z$. Again, by construction $\D_w$ annihilates the constant term. Similarly, the functions $\D_w\C_{u,v;w}$ are even cuspidal functions, where $\C_{u,v;w}$ are two-loop modular graph functions defined in \cite{DHoker:2019txf}. One can again explicitly check that for all of the aforementioned functions, there are naked factors of $\tau_2$. 

\subsection{Building cuspidal functions with discreteness and integrality}
\label{sec:existenceofcuspfunc}

\sssec{Generalities}

In this subsection, we will explicitly construct examples of non-factorizable, non-holomorphic cuspidal functions satisfying discreteness and integrality. Our strategy will be to build cuspidal functions from the integral vvmfs studied in previous sections. It is important to realize that Theorem \ref{thm:conj1} does not rule out such cuspidal functions. Indeed, it is possible for a modular function $\cC(q, \bar q)$ with $m_0, \bar m_0 \leq 0$ to still be cuspidal, as per the comments after \eqr{cuspineq}. 

Let us begin by considering the case of an indecomposable two-dimensional vvmf with $\vec{v} = (v_1, v_2)$, i.e. one which cannot be decomposed into a pair of subrepresentations. If we consider the diagonal invariant, 
\bea
\cC(q, \bar q) = |v_1(q)|^2 + |v_2(q)|^2~,
\eea
then we have $x_0 = 2 m_0$, which is negative by Theorem \ref{thm:conj1}. Hence this is not cuspidal. On the other hand, we may try to consider the non-diagonal invariant 
\bea
\cC(\qqb) =v_1(q) \bar v_2(\bar q) + v_2(q) \bar v_1(\bar q)
\eea
which would have $x_0 = m_0^{(1)} + m_0^{(2)} = {1 \over 6}$ and therefore would seemingly be cuspidal. However, in this case $T$-invariance would require $m_0^{(1)} - m_0^{(2)}\in \ZZ$. In fact, this is not allowed for an indecomposable two-dimensional vvmf. This can be seen from the explicit form of two-dimensional solutions given in (\ref{eq:2dexplicitsols}). Indeed, in the notation used there we have $m_0^{(1)} - m_0^{(2)} = {\m\over 6}$, and when $\m \in 6 \ZZ$ we see that the third entry of one of the two hypergeometric functions becomes a non-positive integer, and is thus divergent. This means that the second solution to the MDE should actually be logarithmic, and indeed it is known to be given in terms of the Meijer G-function \cite{Beem:2017ooy}, 
 \bea
 g_{\rm log}(q) = G_{2,2}^{2,0}\left(\begin{matrix}{2 \over 3} & 1 \\ {1- \m \over 12} &{1+ \m \over 12}  \end{matrix} \,\,\Big | \,\,1728 j(q)^{-1}\right) ~,
 \eea
c.f. footnote \ref{footnote:indicial}. As a result, this case does not satisfy our desired discreteness property.  We conclude that we cannot obtain any of the desired cuspidal functions using \textit{indecomposable} two-dimensional vvmfs.  

Similar comments hold for indecomposable three- and four-dimensional vvmfs. In those cases we again cannot have $m_0^{(i)} - m_0^{(j)}\in \ZZ$ for any $i,j$, and thus only the diagonal invariant is allowed, giving $x_0 = 2 m_0<0$.   For $d=5$ there can be a non-diagonal invariant, as in the case of the $\widehat{su(2)}_4$ WZW model partition function. However, in that case the modular invariant is still \textit{block} diagonal (and hence diagonal under the maximally extended chiral algebra), which again implies that there is a term with leading exponent $x_0 = 2 m_0<0$. 

This leads us to suspect that it is not possible to obtain a non-holomorphic cuspidal function by adjoining \textit{indecomposable} admissible characters. For modular functions $\cC(q, \bar q)$ corresponding to e.g. partition functions, the physical intuition behind this suggestion is that one always has the negative contribution from the vacuum block $|\chi_{\rm vac}(q)|^2 \sim (q\qb)^{-{c \over 24}}$. The validity of this suspicion will actually not be of much importance to us in this work, so we do not attempt to prove it here. What is more important is showing the existence of \textit{any} discrete and integral cuspidal function at all. We will thus simply take this as motivation to begin our search by considering cuspidal functions built from \textit{decomposable} vvmfs. 

\sssec{Explicit constructions}
\label{sec:constructionsoutline}

Allowing vvmfs to be in decomposable representations immediately gives rise to constructions of discrete, integral cuspidal functions. We first give a simple example, then describe several general algorithms.

The simplest example is obtained by starting with the $\widehat{su(2)}_4$ WZW model, whose characters transform in an indecomposable 5-dimensional representation $R_5$. In the usual conventions, we label the characters by $\chi_0, \dots, \chi_4$ with $m_0^{(i)} = h_i - {c \over 24} $, where we have  $c=2$ and 
\bea
h_{i=0,\dots,4} = 0\,,\, {1 \over 8}\,,\, {1 \over 3}\,,\, {5 \over 8}\,,\, 1
\eea 
(c.f. Appendix \ref{appwzw}.) It is easy to check that (\ref{eq:MDEconstd}) is satisfied. The diagonal partition function is given by 
\bea
Z^{\widehat{su(2)}_4}_{\rm diag}(\qqb) = \sum_{i=0}^4|\chi_i(q)|^2~.
\eea

Note that the five-dimensional representation $R_5$ is \textit{indecomposable}, but not \textit{irreducible}. What this means is that it contains a lower-dimensional subrepresentation, whose complement is not itself a representation.\footnote{If the complement to this subrepresentation were {also} a subrepresentation, then it would be decomposable. See e.g. the example in Section \ref{sec:d4example}.}
Indeed, by inspecting the $S$-matrix,
\bea
S = {1 \over 2\sqrt{3}}\left(\begin{matrix} 
1 & \sqrt{3} & 2 & \sqrt{3} & 1 \\
\sqrt{3} & \sqrt{3} & 0 & - \sqrt{3} & - \sqrt{3} \\
2 & 0 & -2 & 0 & 2\\
\sqrt{3} & - \sqrt{3} & 0 & \sqrt{3} & - \sqrt{3} \\
1 & - \sqrt{3} & 2 & - \sqrt{3} & 1
\end{matrix} \right)~,
\eea
and by diagonality of the $T$-matrix, we see that $\{ \chi_{0}+\chi_4, \chi_2\}$ decouple from $\{\chi_{1}, \chi_3\}$ and form their own two-dimensional subrepresentation $R_2$. It is easy to check that  $m_0^{(2)}+ \mathrm{min}(m_0^{(0)}+m_0^{(4)}) ={1 \over 6}$, as required for a two-dimensional representation, c.f. (\ref{eq:MDEsum}). The modular function 
\bea
Z^{\widehat{su(2)}_4}_{\rm non-diag}(\qqb) = |\chi_0(q) + \chi_4(q)|^2 + 2 |\chi_2(q)|^2
\eea
is interpreted as the non-diagonal partition function for the $\widehat{su(2)}_4$ WZW model, or alternatively as the diagonal invariant for the $\widehat{su(3)}_1$ WZW model. As we have mentioned earlier, neither $Z^{\widehat{su(2)}_4}_{\rm diag}$ nor $Z^{\widehat{su(2)}_4}_{\rm non-diag}$ is cuspidal.

Now we consider the following modular invariant built from characters transforming in the decomposable representation $R_5 \oplus R_2$:
\bea
\label{eq:SU24cuspidal}
\C(\qqb)  := Z^{\widehat{su(2)}_4}_{\rm non-diag}(\qqb) - Z^{\widehat{su(2)}_4}_{\rm diag}(\qqb) ~.
\eea
This in fact \textit{is} cuspidal, with $x_0 ={c \over 24}= {1 \over 12}$! Physically, all we have done is subtract the diagonal and non-diagonal partition functions to get rid of the vacuum module. As required by the arguments in Section \ref{sec:noposintconst}, we have necessarily sacrificed positivity.

To the best of our knowledge, (\ref{eq:SU24cuspidal})  is the first concrete example of a non-factorizable, non-holomorphic cuspidal function satisfying integrality and discreteness. 

While the number of cuspidal functions that can be obtained in the way outlined above is surprisingly limited -- indeed, among $\widehat{su(2)}_k$ and $\widehat{su(3)}_k$ WZW models no other diagonal/non-diagonal subtractions work -- we have found several morally similar constructions. These may be categorized as follows, with explicit examples of each construction given in Appendix \ref{app:cuspfuncts}:

\sssec*{\bul Differences of partition functions:}  Consider two CFTs with equal central charge, and define
\e{cdiff}{\C(\qqb) = Z_{1}(\qqb) - Z_{2}(\qqb)~.}
The entire vacuum module cancels. Suppose both CFTs have a gap to $\D={c\o 12}$. This is consistent with constraints of modularity, for all values of $c$ \c{Friedan:2013cba}. Then by construction, we have $x_0>0$. 

More generally, we can allow primaries with $\D\leq{c\o 12}$ and retain cuspidality, as long as $Z_1$ and $Z_2$ have these states with the same multiplicity. A sub-case of this is the situation described above, in which $Z_1$ and $Z_2$ are diagonal and non-diagonal invariants of an RCFT. In Appendix \ref{sec:identicalc}, we give examples of the form (\ref{cdiff}) involving differences of WZW partition functions with sufficient gaps and identical central charges. 

\sssec*{\bul Differences of powers of partition functions:} Consider two CFTs with central charges $c_1$ and $c_2$, where $c_1 = n c_2 < 24$ with $n\in\Z_{\geq 2}$, and define
\e{}{\C(\qqb)  = Z_1(\qqb) - Z_2(\qqb)^n~.}
This is, of course, just a version of \eqr{cdiff} in which the second CFT is an $n$-fold tensor product. This has an integral Fourier expansion. Specializing to $\qb=q$, we have
\e{}{\C(q,q) \sim q^{\D_1-{c_1\o 12}} +(2+2n)q^{2-{c_1\o 12}} - 2nq^{\D_2-{c_1\o 12}}+\ldots}
where $\ldots$ indicates terms of higher powers. The second term represents the holomorphic and anti-holomorphic stress tensors from $Z_1$ and $Z_2$. If both CFTs have gaps to $c_1 \over 12$ and if $c_1 <24$, then $\C(q,q)$ will be cuspidal. 

One can generalize this construction to greater values of $c_1>24$ in multiple ways. First, one can take more complicated linear combinations. In Appendix \ref{sec:SpinIsingcusp}, we give examples of this sort involving the Ising model and WZW models. Second, one can use extremal CFT partition functions -- i.e. the modular $J$-function and Hecke operators -- to ``soak up'' the difference in central charge. We do so in Appendix \ref{sec:Jfuncdiff}. 

\sssec*{\bul Orbifolds:} Consider the difference of the partition function of a CFT with that of its orbifold, and define
\bea
\cC(q, \bar q) = Z_{\rm CFT}(q, \bar q)-Z_{{\rm CFT}/H}(q, \bar q)~.
\eea  
The simplest example of this is again the cuspidal function obtained by applying the non-diagonal algorithm to the $\widehat{su(2)}_4$ WZW model. Indeed, it is known that the non-diagonal ($D$-type) invariant is obtained by gauging a $\ZZ_2$ symmetry in the diagonal ($A$-type) theory. A different example is that of the compact boson of radius $R$ and its $\Z_2$ orbifold. A simple calculation given in Appendix \ref{sec:cusporbifold} shows that, in a convention for $T$-duality where $R^2\geq 2$,
\e{}{m_0=-{1\o 24}\,,\quad x_0 = -{1\o 12} + \text{min}\left({1\o R^2},{1\o 8}\right)~.}
This is cuspidal as long as $R^2< 12$. Unlike some of our other examples, this has support on {\it irrational} powers of $q$ and $\qb$, whenever $R^2$ is irrational.

\ssec{How cuspidal?}
\label{sec:howcuspidal}

With an eye toward bootstrap applications, it is natural to ask the following question: {\it Given a cuspidal function $\C$ with some fixed $m_0, \bar{m}_0$, how cuspidal can $\C$ be? In other words, how large can $x_0$ be?}

Let us consider in particular cuspidal functions formed by gluing vvmfs together. Note that the constraint (\ref{eq:MDEconstd}) implies that
\e{eq:sumeq}{\sum_{i,j=1}^d  (m_0^{(i)}+ \bar{m}_0^{(j)}) = {d(d-1)\o 6} - {\ell + \bar \ell \over 6}~,}
where $\ell$ and $\bar \ell$ are the respective Wronskian indices for the chiral and anti-chiral vvmfs. If we distribute the right-hand side evenly among all $d$ pairs on the left, one obtains $x_0:= \mathrm{min}(m_0^{(i)}+ \bar{m}_0^{(j)}) = {d-1\o 6}- {\ell + \bar \ell \over 6 d}$. If we instead increase one pair, (\ref{eq:sumeq}) implies that we must decrease another, which lowers $x_0$. Hence for any $N_{ij}$, $m_0$, and $\bar m_0$, 
\e{eq:weakerdbound}{x_0 \leq {d-1\o 6}- {\ell + \bar \ell \over 6 d}~.}

We expect that the bound (\ref{eq:weakerdbound}) is very sub-optimal. In particular, it is agnostic about the constraints on possible gluing matrices $N_{ij}$, and it does not invoke integrality of Fourier coefficients. 
Nevertheless, note that for sufficiently large Wronskian index, $x_0$ is forced to be negative. For example, taking $d=2$ and recalling that $\ell=1$ is not allowed, we see that if either $\ell>0$ or $\bar\ell>0$ then $x_0 \leq 0$.

\sec{Bootstrap bounds II: Dimension spectrum}
\label{sec:dimboundssubsec}
We now explore some physical implications of the existence of non-holomorphic cuspidal functions satisfying discreteness and integrality.

\ssec{On uniqueness of the $\D>{c\o 12}$ spectrum}
\label{sec:finetuning}

Because non-holomorphic cuspidal functions $\cC(q, \bar q)$ are non-positive, we should not interpret them as partition functions of any single (unitary) theory. Instead we begin by interpreting  $\C$ as the difference of partition functions for two distinct theories CFT$_1$ and CFT$_2$ at identical central charge,
\e{}{\cC(q, \bar q) = Z_1(\qqb) - Z_2(\qqb)~.}
If $\cC(q, \bar q)$ is cuspidal, the two theories have identical operator spectra up to $\Delta \leq {c \over 12}$, but can differ above $\Delta > {c \over 12}$. Hence the existence of non-holomorphic cuspidal functions established in the previous section gives us the following, 
\begin{res}\label{heavy} On the basis of modularity alone, the spectrum with $\Delta > {c\over 12}$ is \textbf{not} determined by its complement, $\Delta \leq {c\over 12}$. 
\end{res}

As a formal statement about modular invariance, this is definitive. However, even if such a function were positive, discrete, and integral, it would not be guaranteed to be the partition function of a CFT: there may exist some yet-unaccounted-for constraints which rule out such a candidate partition function or theory. A more pressing issue is that non-holomorphic cuspidal functions obeying discreteness and integrality must have negative degeneracies. Therefore, given a unitary irrational CFT, in order to preserve unitarity while tweaking the $\D>{c\o 12}$ states, there must exist cuspidal functions whose negative degeneracies are all sufficiently supported on the spectrum of the theory. In light of that, we here advance the perspective that the $\D>{c\o 12}$ spectrum -- indeed, the $\D>{c-1\o 12}$ spectrum -- {\it is} in fact uniquely determined if one imposes no fine-tuning.

To phrase this more rigorously, we define the following two spaces:\foot{ We are using the primary partition functions because this allows us to establish stronger fine-tuning criteria, with respect to a lower threshold ${c-1\o 12}$. }

\begin{quote}
 $\mathsf{Z:}$ the space of primary partition functions $Z_p(\qqb)$ of unitary, compact, irrational CFTs with a normalizable vacuum. 

 $\mathsf{C:}$ the space of discrete, integral weight-$(\h,\h)$ cusp forms $\C_p(\qqb)$.

\end{quote}
\noindent Recall that $Z_p$ is weight-$\(\h,\h\)$. For every element $Z_p \in \mathsf{Z}$, denote the Fourier coefficients along $\qb=q$ as $a_{x}$. Likewise for every element $\C_{p}\in\mathsf{C}$, denote the Fourier coefficients along $\qb=q$ as $c_{x}$. Finally, define $\lbrace x_i(\C_p)\rbrace$ as the set of dimensions for which $c_{x_i(\C_p)}<0$. Then 
\e{finecond}{Z_p+\C_p \in\mathsf{Z}~~~\text{iff}~~~\text{$a_{x_i(\C_p)} + c_{x_i(\C_p)} > 0 ~\quad \forall~i$}~.}
The question of fine-tuning is that of {\it typicality} of pairs $(Z_p,\C_p)$ obeying this set of conditions. Let us emphasize: this is an infinite number of conditions, since $\lbrace x_i(\C_p)\rbrace$ necessarily contains an infinite number of elements due to the integrality condition, c.f. (\ref{ax}). 

A ``typical'' CFT may be defined by many measures. At the level of OPE data, a conservative statement is that a typical CFT has irrational central charge and a sporadic spectrum of (mostly) irrational operator dimensions.\foot{Many dynamical phenomena indicative of typicality may be boiled down to these properties of the OPE data. For instance, an irrational spectrum may be invoked in explaining the absence of periodic dynamics in correlators \c{Datta:2019jeo} or spectral form factors \c{Dyer:2016pou}, forbidden singularities \c{Maldacena:2015iua,Fitzpatrick:2016ive}, and the presence of chaos. } Thus the question of fine-tuning is three-fold: Do there exist discrete cusp forms $\C_p$ with support on an infinite set of sporadic, irrational values of $h,\hb$? And if so, can they be paired with CFT partition functions $Z_p$ such that \eqr{finecond} is obeyed? And if so, with finite measure on either space? 

Given the lack of classification of either space $\mathsf{Z}$ or $\mathsf{C}$, these questions cannot be answered definitively. But it is reasonable to believe that the answer to at least one of these questions is negative. This justifies the following claim.
\begin{claim}
\label{prop:finetuning}
Absent fine-tuning, the primary spectrum with $\D>{c-1\o 12}$ is uniquely determined by its complement, $\D\leq {c-1\o 12}$, in a compact, unitary, irrational CFT. 
\end{claim}
\ni We have stated this for CFTs with Virasoro symmetry alone. For the analogous claim for CFTs with extended chiral algebras, take $c-1 \rar c-c_{\rm currents}$. 

\sssec{On Rademacher expansions}
\label{sec:Rademacher}
Result \ref{heavy} can be phrased as a negative result on the formal existence of Rademacher expansions for non-holomorphic CFT. This topic was recently addressed in \c{Alday:2019vdr}. It was proposed that ``ambiguities'' in Rademacher expansions are spanned by a basis of Poincar\'e series which are $SL(2,\Z)$ sums over seed characters $\chi_{h}(q)\chi_{\hb}(\qb)$ with conformal weights $h,\hb > {c-1\o 24}$. Given that cusp forms of weight $\(\h,\h\)$ are the definition of an ambiguity allowed by modularity, our results formally rule out this proposal: we have constructed cusp forms that are not Poincar\'e sums over conformal characters.\foot{Poincar\'e sums over conformal characters generate continuous functions with support on $\D\geq {c-1\o 12}$ \c{Keller:2014xba}, whereas our functions are discrete with support on $\D> {c\o 12}$. We also point out that even if we allow the cusp forms to be continuous, Theorem 3.1 of \cite{DHoker:2019txf} gives examples of cuspidal functions that are not Poincar\'e sums of conformal characters.} On the other hand, an irrational CFT {\it does} admit a convergent Rademacher expansion if it is typical in the sense explained above. 

\ssec{Spectral and OPE bounds from torus one-point functions} 

Next we interpret $\cC = \langle \cO \rangle$ as the one-point function of an operator $\cO$. Assume first that $\C$ admits an expansion in vvmfs, as in Section \ref{torusoneptsec1}. As discussed above, the bounds that we can obtain now depend on the gluing matrix $N_{ij}$. 

To begin, let us consider the diagonal matrix $N_{ij} = \delta_{ij}$. A sufficient (but not necessary) condition for the one-point function to be diagonal is that the RCFT partition function is  diagonal. In this case it is clear that 
\e{}{x_0 = 2m_0\,\,\,\qquad (N_{ij} = \delta_{ij})}
with $m_0$ as defined in (\ref{eq:m0def}). Then by Theorem \ref{thm:conj1} applied to $m_0$, in this case we actually \textit{do} have $x_0<0$ for any quantity built from integral vvmfs. By precisely the same reasoning as for the twist bounds in Section \ref{torusoneptsec1}, we thus obtain bounds on the spectrum of $\D$ analogous to Results \ref{eq:chiraldimcond2} -- \ref{eq:thirdtwistbound}.

 If $N_{ij}\neq \delta_{ij}$, it is generically no longer true that $x_0<0$. This is because the leading-twist operator may not be a scalar. However, recall that we derived a weaker, though completely general, bound in (\ref{eq:weakerdbound}). Then by identical logic as for the twist bounds, we can state the key result as follows:
\begin{res}
\label{res:dim1}
If $\<\O\>$ is non-zero and integral, then there exists an operator $\phi_*$ obeying $$\D_{*} \leq {c+\D_\O+\eps_\O\o 12}~,\qquad C_{\phi_*^\dagger \O\phi_*}\neq 0~.$$  If $\<\O\>$ is diagonal, then $\eps_\O = 0$, otherwise $\eps_\O=2(d_\O-1)$. Contributions to $\<\O\>$ with $\D_\phi\leq{c+\D_\O+\eps_\O\o 12}$ uniquely determine those with $\D_\phi> {c+\D_\O+\eps_\O\o 12}$.
\end{res}
\ni As with Result \ref{eq:firsttwistbound}, these are non-trivial bounds on the $\phi^\dagger\x\phi$ OPE. If $\D_\O > {c+\eps_\O\o11}$, then $\D_{*}<\D_\O$ and these become non-trivial bounds on the spectrum. Note that at non-zero Wronskian index, (\ref{eq:weakerdbound}) implies that we can actually push $\eps_\cO$ down to $2(d_\O-1)- {2(\ell+\bar \ell) / d_\cO}$, but we choose not to exhibit this above so that our result as stated applies to all vvmfs.

Before applying these to the especially interesting case of marginal operators, we make a few comments. First, the above bounds are, modulo possible exceptions, only useful when $\O$ is primary with respect to the extended chiral algebra, $\A_\O=\A$: if $\A_\O\neq \A$, then $\<\O\>$ is generically not diagonal, and $d_\O$ is generically infinite, on account of the infinite branching of highest-weight $\A$-modules into highest-weight $\A_\O$-modules. 

Second, we reemphasize that the bound \eqr{eq:weakerdbound} is likely sub-optimal, in which case \eqr{res:dim1} for the non-diagonal case is likewise sub-optimal. We believe it should be possible to improve \eqr{eq:weakerdbound} by incorporating $S$-invariance and, optimistically, by constraining the possible automorphisms of RCFT fusion rules \c{Moore:1988ss}. 

Finally, let us make contact with some recent analytic bootstrap bounds. The bound  \cite{Hartman:2019pcd} on Virasoro primaries is
\e{hmr}{\D_*^{\rm HMR} = {c+4\o 8}\,,\quad c\in(1,4]\cup [12,\i)~.}
Our diagonal bound in Result \ref{res:dim1} is stronger than \eqr{hmr} if
\e{stronger}{\D_\O < {c\o 2}+6~.}
Now recall from  \cite{Hartman:2019pcd} that \eqr{hmr} is in fact {\it optimal} at $c=4$.\foot{At $c=12$, it is also the optimal solution to the {\it spinless} bootstrap \cite{Hartman:2019pcd}, but not the full modular bootstrap \cite{Collier:2016cls}.} Therefore, the CFT which optimizes $\D_*^{\HMR}$ at $c=4$ must not contain operators $\O$ that satisfy the conditions required for \eqr{stronger} and Result \ref{res:dim1} to apply. This is indeed true for the saturating CFT at $c=4$, namely, eight free fermions with diagonal GSO projection.\footnote{To see this, note that this theory can be bosonized to the $\wh{\rm spin(8)}_1$ WZW model, whereby the chiral and anti-chiral fermions combine to give rise to a bosonic operator $\eps_{i \bar \j} = \psi_i \bar \psi_{\bar \j}$ of dimension $\Delta_\eps =1$, transforming in the $(\mathbf{8_v},\mathbf{8_v})$ of ${\rm Spin}(8)\times {\rm Spin}(8)$. Invariance under ${\rm Spin}(8)\times {\rm Spin}(8)$ demands that the only non-zero correlators of $\eps_{i \bar \j}$ involve products of 8 of them, giving an operator of dimension $\Delta = 8$. The $\wh{\rm spin(8)}_1$ WZW model admits two other non-trivial primaries of dimension $\Delta =1$, and similar comments apply to those as well.}

 For $c\in(1,4)$, the modular bootstrap bound for CFTs with chiral algebra $\A= U(1)^c$ \cite{Afkhami-Jeddi:2020ezh} or $\A= \Vir$ \cite{Collier:2016cls} is at least as strong as 
\e{2006}{\D_* \leq {c+2\o 6}~.}
The bound \eqr{2006} is known to be optimized at $c=1$ by the $\wh{su(2)}_1$ WZW model, and at $c=2$ by the $\wh{{su}(3)}_1$ WZW model. It would thus be inconsistent for the bound obtained from Result \ref{res:dim1} to be stronger than this. This means that in these optimal CFTs, there should not exist any operator $\cO$ with dimension $\Delta_\cO < c+4$ satisfying the assumptions of Result \ref{res:dim1}. The $\wh{su(2)}_1$ case is consistent since the only non-vacuum primary in the theory has vanishing one-point function,\foot{The proof is as follows. The $\wh{su(2)_1}$ non-vacuum primary $\O$ has $h_\O=1/4$ and, by virtue of being the lone primary, either $d_\O=0$ or $d_\O= 1$. When $d_\O=1$, the condition $\<\O\>\neq 0$ implies $h_\O = {c\o 22} +{12 \over 11}m_0$, with $m_0$ one of the values in (\ref{eq:1dm0}). Recalling that $c=1$, we see that for no choice of $m_0$ can this equation be satisfied.} and hence Result \ref{res:dim1} does not apply. The spectrum of non-vacuum primaries in the $\wh{su(3)}_1$ WZW model is $h={1\o 3}, {3\o 4}$ and obeys $\Delta_\cO < c+4$, so we learn that these one-point functions do not satisfy the criteria necessary for Result \ref{res:dim1} to apply. 
 
\ssec{Bootstrap bounds on conformal manifolds}\label{margsec}
If $\O$ is exactly marginal, then $\<\O\>$ is integral and we obtain the following specialization of Result \ref{res:dim1}: 
\begin{res}
\label{res:diagRCFTdim}
Any RCFT with an exactly marginal operator obeying $\<\O\>\neq 0$ must have another primary $\phi_*\neq \O$ satisfying 
\e{}{\D_{*} \leq {c+2+\eps \over 12}~, \qquad C_{\phi_*^\dagger\O\phi_*}\neq 0~. \nonumber}
If $\<\O\>$ is diagonal, then $\eps = 0$, otherwise $\eps=2(d_\O-1)$. Contributions to $\<\O\>$ with $\D_\phi\leq{c+2+\eps \over 12}$ uniquely determine those with $\D_\phi> {c+2+\eps \over 12}$.
\end{res}

\ni That $\phi_*\neq \O$ follows from the vanishing of the one-loop beta function.\foot{The absolute minimal spectral assumption needed for $\eps=0$ is that the lightest contribution to $\<\O\>$ be a scalar. So as to avoid contrivance, we have instead phrased the bounds in terms of diagonality.} 

There are several interesting aspects of this case. Any operator $\phi$ with $C_{\phi^\dagger\O\phi}\neq 0$ necessarily acquires an anomalous dimension to first order in conformal perturbation theory. This is seen by extracting the logarithmic part of the first-order perturbation of $\<\phi^\dagger(z,\zb)\phi(0,0)\>$, which gives (e.g. \c{Sen:2017gfr})
\e{anomdim}{\delta S = \l\int d^2w \,\O(w,\bar w) \quad \Rightarrow \quad \g^{(1)}_\phi = -\l C_{\phi^\dagger\O\phi}~.}
The fact that $C_{\phi^\dagger\O\phi}\neq 0$ implies $\g^{(1)}_{\phi}\neq 0$, and not just the converse, follows from the existence of a unique tensor structure for three-point functions of operators of arbitrary spin. Therefore, the primary $\phi_*$ that obeys the bound in Result \ref{res:diagRCFTdim} is the lightest primary with $\g^{(1)}_{\phi}\neq 0$. In a superconformal field theory, such an operator is necessarily non-BPS. The degree $d_\O$ counts the number of primary operators $\phi$ with $\g^{(1)}_{\phi}\neq 0$, i.e. the rank of the one-loop dilatation matrix. Result \ref{res:diagRCFTdim} thus relates this rank to the tree-level dimension of the lightest operator. 

In addition, the OPE determinacy of Result \ref{res:diagRCFTdim}, together with \eqr{anomdim}, implies that $\g_\phi^{(1)}$ for operators $\phi$ with tree-level dimension $\D_\phi > {c+2+\eps\o 12}$ are in fact determined by $\g_\phi^{(1)}$ for operators $\phi$ with tree-level dimension $\D_\phi \leq {c+2+\eps\o 12}$. Altogether, this is a remarkable reduction in the number of independent data needed to specify the one-loop dilatation matrix. 

We have performed some checks of Result \ref{res:diagRCFTdim}. In particular, one can confirm that they are satisfied in the unitary $W_3$ minimal models \c{Iles:2014gra}, as well as $\wh{su(N)_k}$ WZW models, using the known explicit spectra.

A limitation of our bound is that it does not apply to conformal manifolds generated by current-current deformations $\O=J\bar J$. This rules out application to supersymmetric conformal manifolds in which $\O=J\bar J$ where $J$ is an $R$-current.

\sssec{Interpretation}
\label{sec:interpretation}
Result \ref{res:diagRCFTdim} is a strong constraint on rational points of conformal manifolds. We have made only the minimal assumption necessary for deriving such a bootstrap bound -- namely, the existence of an exactly marginal $\D_\O=2$ operator. Perhaps the most obvious feature of our bound is that
\e{c12}{\D_* \approx {c\o 12}\,,\quad c\gg 1~.}
The reader might worry about taking a large $c$ limit when our results hold only for rational theories. But recalling that rationality requires $c=c_{\rm currents}$, we may take $c\gg 1$ while preserving rationality and finite $d_\O$ as long as we increase the ``size'' of the chiral algebra $\A\x\bar\A$. As a result, in the context of AdS$_3$/CFT$_2$, the holographic interpretation of this bound is that of a massless scalar coupled not to Einstein gravity, but rather to a higher-spin gravity with asymptotic symmetry algebra $\A\x\bar \A$ containing an infinite number of higher-spin gauge fields. Thus interpreted, the bounds obtained coincide with the classical threshold for BTZ black holes in theories of pure gravity in AdS$_3$. Result \ref{res:diagRCFTdim} bounds the non-gauge sector of the bulk theory.

A crucial point of interpretation of \eqr{c12} is the following. Although rationality has played an important role in obtaining our bounds, the operator $\phi_*$ obeying the bounds is necessarily non-holomorphic. Thus, {\it while rationality is required to establish Result \ref{res:diagRCFTdim}, the bound constrains the non-holomorphic spectrum of CFTs which are irrational with respect to Virasoro!} 

There is also a gravity interpretation of the condition $C_{\phi_*^\dagger\O\phi_*}\neq 0$. We may view $\O$ as a light bulk field in AdS$_3$ and $\phi_*$, in view of the spectral bound, as a near-extremal black hole microstate. Then $C_{\phi_*^\dagger\O\phi_*}$ controls the amplitude for  Hawking radiation, i.e. the emission of the light field from the black hole. Results \ref{res:dim1} and \ref{res:diagRCFTdim} may be read as the statement that black holes Hawking radiate all the way down to threshold. However, a major caveat in this interpretation is that these bounds only apply to integral operators $\O$, and not to all light operators. It would be nice to interpret this from the gravity side.\foot{The bound \eqr{eq:hphibound} also bears some similarity to ``heavy-heavy-light'' OPE bounds in irrational CFTs at large central charge (c.f. \c{Kraus:2016nwo} and references thereto). Those works study the quantity $\overline{C_{HHL}}$, averaged over heavy operator dimensions $\D_H\approx c$. Taking $c\gg1$ in \eqr{eq:hphibound}, the bound tells us that given a ``light'' operator $\O$, there must be a ``heavy'' operator $\phi_*$ that obeys the bounds as stated. Unlike the HHL story, our bound applies to individual, non-averaged OPE coefficients, but we do not derive OPE asymptotics.}

Understanding Result \ref{res:diagRCFTdim} from other perspectives, and optimizing the $d_\O$-dependence by improving \eqr{eq:weakerdbound}, would be logical avenues for future work.

\section{3d gravity}
\label{sec:3dgravity}

In this final section, we discuss connections between aspects of non-holomorphic modular forms and AdS$_3$/CFT$_2$. 

\ssec{Ensemble averages and the black hole threshold}
\label{sec:ensembleavg}

As discussed around \eqr{contpos}, our previous arguments do not forbid positive cuspidal functions if they are continuous. Only upon imposing discreteness do we forbid them. How are we to interpret this distinction in the context of 3d gravity? As we now argue, this has implications for ensemble averages over CFTs and their possible duality to theories of 3d gravity with asymptotically AdS boundary conditions \c{Saad:2019lba,Cotler:2020ugk,Maloney:2020nni,Afkhami-Jeddi:2020ezh,Perez:2020klz,Belin:2020hea,Cotler:2020hgz}.

The above distinction suggests that the possible {\it uniqueness} of a positive partition function with a normalizable vacuum and a Virasoro primary gap to $\D \approx {c\o 12}$ depends sensitively on whether the partition function has a discrete or continuous spectrum. In particular, let us define the semiclassical partition function of a theory of ``pure 3d gravity'' in AdS$_3$ as
\e{Zgrav}{Z_{\rm grav}(q,\qb) \approx |\chi_{\rm vac}(q)|^2 + \int_{{c\o 24}}^\i dh\int_{{c\o 24}}^\i d\hb\, \rho(h,\hb)\chi_h(q) \chi_{\bar h}(q)~.}
We assume that such a theory exists and is modular-invariant and unitary, $\rho(h,\hb)\geq 0$. The spectrum comprises only gravitons up to a threshold 
\e{thresh1}{\D_{*}^{\rm grav}\approx {c\o 12} +\O(1)~.}
While the discussion to follow does not rely on the absence of extra currents, the canonical study of pure 3d gravity assumes that $\mathcal{A}=$ Virasoro.

Let us first suppose that $\rho(h,\hb)$ is continuous. Then $Z_{\rm grav}(q,\qb)$ may not be unique: it could be modified above threshold without spoiling unitarity by adding a positive cuspidal function. A continuous spectrum with a gap above a normalizable vacuum is a feature of ensemble averaged systems. Therefore, if one wishes to view pure 3d gravity as dual to an ensemble average over unitary theories, the possible existence of continuous, positive cuspidal functions would suggest that {\it there is not a unique ensemble average dual to pure 3d gravity.}

This dovetails with the following expectations. To establish such a duality, one must first decide what ensemble of CFTs to average over, and with respect to what measure, and how. A natural guess is that pure 3d gravity is dual to an ensemble average over {\it all} CFTs.\foot{This suggestion was made independently by Alex Maloney \c{alex}.} But there may be many choices that yield partition functions of the form \eqr{Zgrav}. In order to quantify that freedom, two algorithmic approaches come to mind. First, one could attempt to understand the (sub)spaces of CFTs over which one can rigorously average (and how to do so), as well as the available measures on these subspaces. This is, to put it conservatively, not presently practical.  Instead, one could parameterize the ambiguities in the partition functions that could possibly {\it result} from averaging. Our assertion is that the space of continuous, positive cuspidal functions reflects this ambiguity. We stress that while modularity does not forbid such functions, we have not yet identified one. Doing so is an interesting avenue for future work.\foot{Preliminary inspections of the cuspidal functions described in Section \ref{app:literaturesearch} do not turn up any examples with a positive density of states.}

Now let us suppose that $\rho(h,\hb)$ is instead discrete and integral. In this case the $\O(1)$ correction in \eqr{thresh1} is especially important: in particular, there is a direct relation between the {\it uniqueness} of pure 3d gravity and the {\it quantum-corrected black hole threshold}. For concreteness let us take $\mathcal{A}=$ Virasoro and work with the primary partition function,
\e{}{Z_{p,\,\rm grav}(q,\qb) = |\eta(q)|^2 Z_{\rm grav}(q,\qb)~.}
This has modular weight $(\h,\h)$. As proven in Section \ref{sec:noposintconst}, there are no discrete, positive cusp forms. In view of this, let us rewrite the black hole threshold as
\e{thresh}{\D_{*}^{\rm grav}\approx {c-1\o 12} +\delta_{*}~.}
Now we see that the sign of $\delta_{*}$ is important. If $\delta_{*}>0$, then by the absence of discrete, positive cusp forms of weight $(\h,\h)$, $Z_{p,\,\rm grav}$ is unique modulo fine-tuning in the sense of Section \ref{sec:finetuning}. If $\delta_{*}\leq 0$, then $Z_{p,\,\rm grav}$ is {\it not} unique: we may modify it by adding positive non-cuspidal forms while preserving unitarity and the threshold condition \eqr{thresh}.

This then revives the question: What is the quantum-corrected black hole threshold of pure 3d gravity? A first guess is that, in fact, $\delta_{*}=0$. The $c\rar c-1$ shift is computed directly from a {\it perturbative} quantum correction to small black hole entropy \c{Benjamin:2016aww}, and is suggested indirectly by the form of the modular crossing kernel. What we seek is a non-perturbative proof. We can attempt to approach this from a different perspective following recent work \c{Maxfield:2020ale}. Let us define the threshold for spin-$J$ black holes, 
\e{}{\D_{*}^{\rm grav}(J):={c-1\o 12} + J + \delta_{*}(J)~.}
An elegant argument \c{Maxfield:2020ale} asserts that in the large-spin, ``near-extremal'' regime,
\e{}{J\gg c\,,\quad \D-J-{c-1\o12} \ll1~,}
the black hole threshold sits at 
\e{MTtraj}{\delta_{*}(J) \approx -{1\o (2\pi)^2}(-1)^J e^{-{1\o2}S_0(J)}\,,\quad \text{where}\quad S_0(J) := 2\pi\sqrt{{c-1\o 6}J}~.}
This gives an exponentially small asymptotic correction to the universal large-spin, near-extremal spectrum in any irrational Virasoro CFT with a twist gap above the vacuum. The determination of $\delta_{*}$ would follow from extending these results down to finite $J$: specifically, the question is whether, at $c\gg1$,
\e{}{\delta_{*} = \text{min\big($J+\delta_{*}(J)\big)\stackrel{?}{>}0$}~.}
%
Note that extrapolation of \eqr{MTtraj} down to $J=0$ is not justified even at $c\gg1$: even ignoring other possible obstructions, $S_0(0)=0$. 

Here is a concrete CFT-based approach that could aid in determining the location of the black hole threshold in pure 3d gravity. First, rigorously define Virasoro Regge trajectories; then constrain the shape of the leading trajectory, on which the lightest black hole necessarily sits, given the asymptotic behavior \eqr{MTtraj}. Analyticity results have been established for leading Regge trajectories in higher-dimensional CFTs, which are monotonic, and convex functions of $J$ \c{Komargodski:2012ek,Fitzpatrick:2012yx}. Initial steps toward formalizing Virasoro Regge trajectories were taken in \c{Collier:2018exn}. We hope to pursue this avenue in future work.

\ssec{Comment on enigmatic black holes}
\label{sec:enigmaticBH}
Holography instructs us to interpret CFT Rademacher expansions as sums over asymptotically AdS black hole geometries \c{Dijkgraaf:2000fq}. Thus the aforementioned formal absence of a convergent Rademacher expansion in compact non-holomorphic 2d CFTs has a bulk dual interpretation as a non-universality of microcanonical black hole entropy.\foot{In other words, Farey tails can have multiple endings.} Microstate counting deep in the Cardy regime, $\D\gg c$, is unaffected because cuspidal degeneracies are subleading to the Cardy entropy (indeed, Cardy behavior comes from the $S$-transform of $|\chi_{\rm vac}|^2$, which is absent for cuspidal functions). For sparse CFTs of large central charge, dual to prototypical theories of weakly coupled gravity, the Cardy formula holds down to $\D\approx {c\o 6}$ \c{Hartman:2014oaa}. Thus, to leading order in large $c$, the cuspidal degeneracies are only visible in the range ${c\o 12} \lesssim \D\lesssim {c\o 6}$. This range is precisely the ``enigmatic'' range \c{Hartman:2014oaa}, in which there can exist non-BTZ black holes that dominate BTZ in the microcanonical ensemble \c{Denef:2007vg, deBoer:2008fk,Bena:2011zw}. The spectrum and thermodynamics of enigmatic black holes is known not to be universal. On the other hand, the notion of fine-tuning in Proposition \ref{prop:finetuning} leads to the following suggestion: {\it Enigmatic black holes exist only in theories containing ``matter'' degrees of freedom obeying $\D\lesssim {c-1\o 12}$.} This is indeed an empirical feature of all known theories containing enigmatic black holes: the light states are string or perturbative fields of compactified string/M-theory.

\section*{Acknowledgements}
We wish to thank Chris Beem, Ying-Hsuan Lin and Arnav Tripathy for helpful discussions, and Alex Belin, Nathan Benjamin and Sunil Mukhi for helpful comments on an earlier version. JK thanks the Mani L. Bhaumik Institute for generous support. EP is supported by
Simons Foundation grant 488657 (Simons Collaboration on
the Nonperturbative Bootstrap) and by the U.S. Department of Energy, Office of Science, Office of High Energy Physics, under Award Number DE-SC0011632.

\appendix 
\section{Integrality conditions on vvmf}
\label{sec:congtestsubsec}

In this appendix we give additional details on the necessary and sufficient conditions for a vvmf to have integral Fourier coefficients. This material complements Section \ref{sec:mainvvmfsec}. 

First, we note that Conjecture \ref{thm:intconj} implies that a vvmf transforming in a representation $\rho$ is integral (i.e. each element of the vector has integer Fourier coefficients) only if $\Gamma(N) \subset \mathrm{ker}\, \rho$, where $\Gamma(N)$ was defined in \eqr{eq:GammaNdef}. It is useful to have a simple diagnostic for determining when this is the case. Such a diagnostic can be obtained by analyzing the eigenvalues of the modular $T$-matrix \cite{bantay2007vector,Gaberdiel:2008ma},

\begin{thm}[Congruence test]
\label{thm:congtest}
Denote by $\cE(T)$ the unordered set of eigenvalues of the modular $T$-matrix in the representation $\rho$, with degeneracies included. Assume that $T$ is of finite order $N$, i.e. $T^N = 1$. Then one has $\Gamma(N) \subset \mathrm{ker}\, \rho$ only if for every $\ell$ coprime to $N$ one has $\cE(T^{\ell^2}) = \cE(T)$. 
\end{thm}
\noindent Following \cite{Gaberdiel:2008ma}, we will refer to representations passing the congruence test as ``admissible" representations. More precisely, admissible representations passing the congruence test have the properties that $\Gamma(N) \subset \mathrm{ker}\, \rho$, $T$ is diagonal, and $S^2$ is a permutation matrix. The congruence test turns out to be extremely useful in a variety of situations, and can in particular be used to give a fun arithmetic proof of Theorem \ref{thm:conj1} in the case of monic two-dimensional vvmfs, as in Appendix \ref{app:2dproof}.

Second, we note that though \textit{necessary}, $\Gamma(N) \subset \mathrm{ker}\, \rho$ is not \textit{sufficient} to ensure integrality of a vvmf. As it turns out, one must also impose rationality -- i.e. the condition that the coefficients in the Fourier expansion of the characters are rational \cite{bantay2007vector}. The condition $\Gamma(N) \subset \mathrm{ker}\, \rho$ then ensures that the denominators of these coefficients do not grow unboundedly, so that some integer multiple of the original character indeed has integer coefficients. This subtlety will arise in the example of Appendix \ref{sec:d4example}.

The rationality of a modular form of $\Gamma(N)$ can be adressed via Galois theory. For any $\ell\in \ZZ_N$ coprime to $N$, let $\sigma_\ell \in \mathrm{Gal}\left( \QQ[(1)^{1/N}]/\QQ\right)$ be the Galois automorphism sending $(1)^{1/N} \rightarrow (1)^{\ell/N}$. When acting on a matrix, $\sigma_\ell$ is taken to act entry-by-entry. Now define the following matrix 
\bea
\label{eq:Gldef}
G_\ell = S T^{\ell^{-1}} S T^\ell S T^{\ell^{-1}}
\eea
taken in the appropriate representation $\rho$. $\ell^{-1}$ is the modular inverse of $\ell$, i.e. $\ell^{-1}\in\Z$ such that $\ell\ell^{-1} = 1$ mod $N$. One then has the following test for rationality \cite{bantay2007vector},
\begin{thm}[Rationality test]
\label{thm:rationalitytest}
Let $\Gamma(N) \subset \mathrm{ker}\, \rho$ and consider a vector-valued modular form $\vec{v}(q)$ whose polar coefficients $\vec{a}_n$ for $n\leq 0$ are all rational. Then $\vec{v}(q)$ is rational if and only if for all $\ell\in \ZZ_N$ coprime to $N$, we have 
\bea
\label{eq:rationalitytest}
\sigma_\ell S = G_\ell S~.
\eea
Here $S$ is the modular $S$-matrix and $G_\ell$ is the matrix defined in (\ref{eq:Gldef}).
\end{thm}

When the modular $S$-matrix is real, the rationality test can be recast in a tidy form. Since $\sigma_\ell S = S$ when $S\in \R$ ($\ell$ is odd so $\sigma_\ell (\pm 1) = \pm 1$), we have
\e{}{(\Gl-\id)S = (\Gl-\id)S^2 = 0~.}
Using the fact that $S^2$ is a permutation matrix we conclude that
\e{}{\Gl= \id~.}
Note that $S^2$ being permutation implies 
\e{}{\det(S) = \pm 1~.}
We will use these facts to prove Theorem \ref{thm:conj1} for monic $d=2$ in Appendix \ref{app:2dproof}.

\section{Proofs of Theorem \ref{thm:conj1} for low-dimensional vvmfs}
\label{app:2dproof}
In this appendix, we provide two alternative proofs of Theorem \ref{thm:conj1} for monic vvmfs of dimension $d=2$. We also give a short summary of the simple case of $d=1$.

\subsection{Dimension $d=1$}\label{appd1}
We begin with a brief summary of the case of one-dimensional vvmfs. In other words, these are modular functions with a multiplier. We will assume that $S^2$ is a permutation ``matrix," as necessary for an admissible representation, in which case we have simply $S^2= 1$. This implies $S= \pm1$, and hence by the relation $(ST)^6= 1$ we conclude that the action of $T$ gives a sixth-root of unity. Any such $T = e^{2 \pi i t/6}$ for $t \in \ZZ_6$ is admissible since the only totative is $\ell=5$, and $T^{25} = e^{50 \pi i t/6}= e^{2 \pi i t/6} = T$, c.f. Theorem \ref{thm:congtest}. 

A basis of admissible one-dimensional vvmfs is given by \cite{bantay2007vector,Cheng:2020srs},
\bea
\label{eq:1dvvmflist}
1, {E_4 \over \Delta^{1/3}}, {E_6 \over \Delta^{1/2}}, {E_8 \over \Delta^{2/3}},{ E_{10} \over \Delta^{5/6}}, {E_{14} \over \Delta^{7/6}}~.
\eea
where $\Delta := \eta(q)^{24}$ is the modular discriminant, and the $E_k:=E_k(q)$ are normalized weight-$k$ holomorphic Eisenstein series. These have leading exponents 
\bea
\label{eq:1dm0}
m_0 =0, -{1\over3}, -\half, -{2\over 3},-{5\over 6}, -{7 \over 6}~.
\eea
All non-trivial cases have negative $m_0$, in line with Theorem \ref{thm:conj1}.

\subsection{Dimension $d=2$}

We now give two alternative proofs for Theorem \ref{thm:conj1} in the case of $d=2$ vvmfs satisfying a monic MDE. These are useful for building familiarity with various charming number theoretic aspects of vvmfs. 

\sssec{Proof 1} We begin with the indicial roots of (\ref{eq:indiceq}). Let us rewrite them as
\bea
m_0^{(i)} = {t_i \over N}~, \hspace{0.8 in}{t_i \in \ZZ}~, \hspace{0.4 in} i=1\,,
\eea
which together with (\ref{eq:MDEsum}) requires that $N \in 6 \ZZ$. Assuming that the modular $T$-matrix is diagonal, we can write 
\bea
T = \left(\begin{matrix} e^{2 \pi i t_1 / N} & 0 \\ 0 & e^{2 \pi i t_2 / N}\end{matrix} \right)~.
\eea
Recall that for an admissible representation, one requires by the congruence test of Appendix \ref{sec:congtestsubsec} that for every $\ell\in\ZZ_N$ such that $(\ell, N) = 1$, one has $\cE(T^{\ell^2}) = \cE(T)$. We would now like to see what non-trivial constraints are implied by this admissibility criterion on the form of the leading exponents $m_0^{(i)}$. 

First, note that there are two ways to satisfy  $\cE(T^{\ell^2}) = \cE(T)$, namely
\bea
\label{eq:admisscond}
&\vphantom{.}&\mathrm{\textbf{Case}}\,\,1: \hspace{0.2 in} \ell^2 t_1 = t_1 + N m \hspace{0.5 in}\ell^2 t_2 = t_2 + N n\hspace{0.5 in}m,n\in \ZZ
\no\\
&\vphantom{.}&\mathrm{\textbf{Case}}\,\,2: \hspace{0.2 in} \ell^2 t_1 = t_2 + N m \hspace{0.5 in}\ell^2 t_2 = t_1 + N n\hspace{0.5 in}m,n\in \ZZ
\eea
For a given $N$, there are generically multiple $\ell \in \ZZ_N$ such that $(\ell, N) = 1$, and for each one we must satisfy $\cE(T^{\ell^2}) = \cE(T)$ in either of the two ways listed above.

\subsubsection*{I. $N \in 6 (2 \NN-1)$}
As we will prove below, Case 2 is only available when $N \in 12 \NN$. Hence to simplify our discussion let us begin by assuming that $N \in 6 (2 \NN-1)$. Then for every $\ell$ coprime to $N$ we must have the equations of Case 1 satisfied. These, together with $t_1 + t_2 = {N \over 6}$, can be recast as the following equations, 
\bea
\label{eq:t1t2l1}
t_1 = {N m \over \ell^2 - 1}~, \hspace{0.6 in} t_2 = {N n \over \ell^2 - 1}~, \hspace{0.6 in} \ell^2  = 1 + 6 (m+n)~.
\eea
Since $\ell^2$ is a positive number we require that $m+n >0$. In fact, we may obtain further restriction on $m,n$ by requiring that $\ell^2$ be a perfect square. This requires that
\bea
\label{eq:mnconstraint}
m + n = 4 k
\eea
where $k$ takes values in the following set 
\bea
\label{eq:kvalues}
k \in \{ 1,2,5,7,12,15,22,26,\dots\}~,
\eea
the values of which correspond to 
\bea
\ell \in \{5,7,11,13,17,19,23,25,\dots \}~.
\eea
 
 We now prove that either $t_1$ or $t_2$ is negative (due to (\ref{eq:MDEsum}), it is not possible for both to be negative). Assume without loss of generality that $t_1>0$. Since $t_1$ must be integer, we require that 
 \bea
 \label{eq:maindivcond}
  { \ell^2 -1 \over \mathrm{gcd}(N, \ell^2 - 1)} \,\, \big | \,\, m~.
 \eea
 Note that this condition must hold for \textit{any} coprime $\ell$, e.g. if we have $(\ell_{A}, N)=(\ell_{B}, N) = 1$, then we can write
\bea
t_1\,\, =\,\, {N m_A \over \ell^2_A-1} \,\,= \,\,{N m_B \over \ell^2_B-1}  \hspace{0.8 in }m_A, m_B \in \ZZ
\eea
and we must have (\ref{eq:maindivcond}) for both $(\ell_A, m_A)$ and $(\ell_B, m_B)$. 

 We now make use of the following lemma, 
\begin{lem}
\label{thm:gcdconj}
Let $N\in 6(2\NN-1)$ and let $\cS = \{ \ell \in \ZZ_N \, | \, (\ell, N) = 1\}$ be the set of mod-$N$ integers coprime to $N$. Then there always exists some $\ell_* \in \cS$ with $\mathrm{gcd}(N, \ell_*^2 - 1)=6$.
\end{lem}
\noindent 
Since $N\in 6\NN$ and $\ell^2 - 1 \in 24 \NN$, it is automatic that $\mathrm{gcd}(N, \ell^2 - 1)\geq 6$. The lemma claims that there is at least one $\ell=\ell_*$ such that this bound is saturated. For example, for $N=6$ one can take $\ell_*=5$, while for $N=18$ one can take $\ell_*=5,7,11,13$. 
\newline\newline
\noindent{\textbf{Proof:}} We prove this by construction. Write $N=6p$ for $p \in 2\NN-1$. Consider $\ell_* = {N \over 2} + 2$, which is one of the totatives of $N$,
\bea
\mathrm{gcd}\left(N, \ell_* \right) = \mathrm{gcd}\left(6p, 3p+2 \right)=1~.
\eea
The latter equality is proven by considering
\bea
f_1 := 2(3p+2)-6p = 4~.
\eea
Any common divisor $d$ with $d\, | \, 6p$ and $d\, | \, (3p+2)$ must also satisfy $d\, | \, f_1$, and hence must be $d\in\{1,2,4\}$. But since $p$ is odd, so is $3p+2$, and thus we conclude that $d=1$. 

To prove the lemma, we now simply note that
\bea
\mathrm{gcd}\left(N, \ell_*^2 - 1 \right) = \mathrm{gcd}\left(6p, 3(3p+1)(p+1)\right)= 6 ~.
\eea
To see this, we may consider 
\bea
f_2 := 6(3p+1)(p+1) - 6p (3p+4) = 6~.
\eea
Any common divisor $d$ with $d\, | \, 6p$ and $d\, | \, 3(3p+1)(p+1)$ must also satisfy $d\, | \, f_2$, and hence must be $d\in\{1,2,3,6\}$. On the other hand, recall that for any totative $\ell$ we have $\ell^2 - 1 \in 24 \ZZ$ so that $\mathrm{gcd}\left(N, \ell^2 -1 \right)\geq 6$. Thus we conclude that $d=6$ is the greatest common divisor.\hfill$\square$
\newline\newline
\noindent
Let us now return to (\ref{eq:maindivcond}) and restrict to $\ell = \ell_*$. This implies that
\bea
m_* \geq {1 \over 6} (\ell_*^2 - 1)
\eea
from which it immediately follows that 
\bea
t_2 \,\,=\,\, {N n_* \over \ell_*^2 -1} \,\,=\,\, {N \over  \ell_*^2 -1}\left( {1 \over 6} (\ell_*^2 -1)-m_*\right) \leq0~.
\eea

\subsubsection*{II. $N \in 12 \NN$}
\label{sec:d2N12}
In the case of $N\in 12 \NN$ we may also satisfy $\cE(T^{\ell^2}) = \cE(T)$ via the equations of Case 2 of (\ref{eq:admisscond}). In that case one finds 
\bea
\label{eq:t1t2l2}
t_1 = {N \over 6}{6 m+1 \over \ell^2 +1} \hspace{0.6 in}t_2 = {N \over 6}{6 n+1 \over \ell^2 +1} \hspace{0.6 in}\ell^2 = 1 + 6(m+n)
\eea
where one again has the constraint (\ref{eq:mnconstraint}) on $m,n$ with $k$ taking values in (\ref{eq:kvalues}).
We assume without loss of generality that $t_1 > 0$. For some $\ell$ we may satisfy Case 1, whereas for others we might satisfy Case 2 -- i.e. there could exist $\ell_A, \ell_B$ such that 
\bea
&\vphantom{.}&t_1 \,\,= \,\, {N m_A \over \ell^2_A-1} \,\,= \,\,{N \over 6}{6 m_B+1 \over \ell_B^2 +1}\hspace{0.6 in}m_A, m_B \in \NN
\no\\
&\vphantom{.}&t_2 \,\,= \,\, {N n_A \over \ell^2_A-1} \,\,= \,\,{N \over 6}{6 n_B+1 \over \ell_B^2 +1}\hspace{0.75 in}n_A, n_B \in \NN
\eea
For the case of $\ell_A$ we may apply the reasoning of the previous subsection to conclude that 
\bea
m_A\geq {\ell_A^2 - 1 \over \mathrm{gcd}(N,\ell_A^2 -1)}~.
\eea
However, in the current case Lemma \ref{thm:gcdconj} does not hold. Indeed for $N \in 12\ZZ$ one has $\mathrm{gcd}(N, \ell^2 -1) \geq 12$, so the best we could hope for is to saturate this bound, giving 
\bea
\label{eq:mabounds1}
m_A \geq {1 \over 12}(\ell_A^2 -1) \hspace{0.5 in}\Rightarrow \hspace{0.5 in} t_1 \geq {N \over 12} \,,\,\,\,\,\,\,\,t_2 \leq {N \over 12}~.
\eea
Next assume that $t_2$ is \textit{also} positive. Then applying the same reasoning above to $t_2$ gives
\bea
\label{eq:nabounds1}
n_A \geq {1 \over 12}(\ell_A^2 -1) \hspace{0.5 in}\Rightarrow \hspace{0.5 in} t_1 \leq {N \over 12} \,,\,\,\,\,\,\,\,t_2 \geq {N \over 12}~.
\eea
Neglecting the degenerate case $t_1 = t_2 = {N \over 12}$ (which as we explained in Footnote \ref{footnote:indicial} does not correspond to an admissible representation) and (\ref{eq:nabounds1}), we conclude by contradiction that $t_2\leq 0$.

The saturation of $\mathrm{gcd}(N,\ell^2 -1)=12$ quoted above is achieved for $N \in 12 (2 \NN-1)$, but not for $N \in 24 \NN$. Indeed, in the latter case one has $\mathrm{gcd}(N,\ell^2 -1)=24$ instead, which gives the constraints 
\bea
\label{eq:N24problemregion}
{N \over 24} \leq t_2 < t_1 \leq {N \over 8}~.
\eea
In order to rule these cases out, we may utilize the following theorem \cite{franc2014fourier},
\begin{thm}
\label{thm:francmason}
If the leading exponents $m_0^{(i)}$ are rational and satisfy $m_0^{(1)}- m_0^{(2)} = P/Q$  with $(P,Q)=1$, then the second-order MDE has two independent solutions with exactly one of the following properties true:
\begin{enumerate}
\item At least one of the solutions is non-integral.
\item $Q \leq 5$.
\end{enumerate}
\end{thm}
\noindent If $N= 24 k$, then we may without loss of generality write $t_1$ in (\ref{eq:N24problemregion}) as $t_1 = M k$ for $M \in (2,3]$. We then have that 
\bea
m_0^{(1)}- m_0^{(2)}  = 2m_0^{(1)}-{1 \over 6} = {M-2 \over 12}
\eea
and for no $M$ in the allowed range do we have $Q \leq 5$. Hence these cases are not admissible.

By exactly analogous steps, one can argue towards a similar result in the case in which Case 2 is satisfied. 

\sssec{Proof 2}
Our starting point is $G_\ell=\id$, as explained in Appendix \ref{sec:congtestsubsec}. We recall that this form of the rationality test \eqr{thm:rationalitytest} requires $S\in\R$. We begin in general $d$. It is useful to consider the case
\e{}{\ell=\ell^{-1}}
subject to the necessary constraint that $\ell\ell^{-1}=1$ mod $N$. Then the condition is
\e{}{(ST^\ell)^3 = \id~.}
This implies the existence of a similarity matrix $P$ obeying
\es{}{P(ST^\ell)P^{-1} &=\text{diag}(e^{{2\pi i\o 3} n_i})\,,\quad n_i \in \Z\\
&: = \Sigma~.}
This further implies the determinant condition $\det(ST^\ell)  =  \det(\Sigma)$, which is
\es{detcond}{\pm e^{\pi i{ \ell d(d-1)\o 6}} = e^{{2\pi i\o 3}m}\,,\quad m\in\Z }
where we have used the indicial equation $\sum_i m_i = {d(d-1)\o 12}$. There is also the trace condition $\Tr(S T^\ell) = \Tr(\Sigma)$, which is
\es{trcond}{\sum_{i=1}^d S_{ii} e^{2\pi i \ell m_{i}} = \sum_{i=1}^d e^{{2\pi i\o 3} n_i}~.}

Now consider the case 
\e{}{\ell=\ell^{-1} = N-1~.}
This holds for all $N$ because $\ell\ell^{-1} = $1 mod $N$. Using $T^{N-1} = T^{-1}$, we have
\e{}{(ST^{-1})^3 = \id}
from which we get the determinant and trace conditions \eqr{detcond} and \eqr{trcond} with $\ell=-1$. Note that ${d(d-1)\o 6} \in \Z/3$ for all $d\in\Z$, so the determinant condition can always be satisfied for some choice of $\det(S)$. But we learn that for $d=2+3\Z$, we have
\e{}{\pm e^{-{\pi i\o 3}} = e^{{2\pi i\o 3}m}\,,\quad m\in\Z ~.}
This can only be satisfied for the minus sign on the LHS. This implies that we must have
\e{}{\det(S) = -1 \qquad (d=2+3\Z)~.}
Recall that for general $d$, $\det(S)=\pm 1$. 

Now specify to $d=2$. The following relation holds in that case:
\e{}{\Tr(\Sigma)^2 = \det(\Sigma)(2+\Tr(\Sigma) \det(\Sigma))~.}
Using $\det(\Sigma) = e^{-{\pi i \o 3}}$, we obtain a quadratic equation for $\Tr(\Sigma)$. The solutions to this equation are
\e{}{\Tr(\Sigma) =  \lbrace e^{\pi i \o 3}\,, -2 e^{\pi i \o 3}\rbrace~.}
Parameterizing the indicial roots as
\e{m1m2}{(m_1,m_2) = \({1+h\o 12},{1-h\o 12}\),}
we have the trace condition
\e{d2trace}{S_{11} e^{-{\pi i h\o 6} }  + S_{22} e^{{\pi i h\o 6}} = \lbrace e^{\pi i \o 6}\,, -2 e^{\pi i \o 6}\rbrace~.}
We now ask whether this can be satisfied for $S_{11}, S_{22}\in\R$ and $h\in[0,1]$ where the latter condition imposes $m_0>0$ via \eqr{m1m2}. From \cite{bantay2007vector}, we know that in $d=2$ the only choices are\foot{For general $d$, the allowed values are $\Tr(S) = d-2n$ where $n=0,1,\ldots d$ \cite{bantay2007vector}.}
\e{}{\Tr(S) = 0,\pm 2~.}
Plugging in above, it is easy to see that $S_{11}\in\R$ if and only if $\Tr(S)=0$. Then $S_{22} = -S_{11}$, whereupon $\det(S) = -S_{11}^2 - S_{12}^2 = -1$, and \eqr{d2trace} simply yields\foot{Indeed, one can compute from the explicit hypergeometric solution (\ref{eq:2dexplicitsols}) that $S_{11} = -{1\o 2} \csc\({\pi h\o 6}\)$, using the monodromy of the hypergeometric functions (e.g. \cite{Beem:2017ooy,Cheng:2020srs}). We have recovered this here in another way.}
\e{}{S_{11} = \Big\lbrace \csc\({\pi h\o 6}\), -{1\o 2} \csc\({\pi h\o 6}\)\Big\rbrace~.}
This does not obey $S_{12}\in [0,1]$, and hence $S_{12}\in\R$, when $h\in [0,1]$. This concludes the proof.

\sec{Examples of integral vvmfs obeying $m_0<0$}
\label{sec:twistexampsubsec}
Having proven that integral vvmfs satisfy Theorem \ref{thm:conj1}, we here give some simple examples of CFT correlation functions built from integral vvmfs, and verify that indeed, $m_0<0$, and Result \ref{eq:chiraldimcond2} is satisfied. These examples have been studied previously in the literature, most notably in \cite{Gaberdiel:2008ma}. 

\ssec{Dimension $d=1$}
\paragraph{Lee-Yang one-point function}
\label{sec:LYoneptfunct}
An extremely simple, though somewhat trivial, example is that of the Lee-Yang model (i.e. the $\M(5,2)$ Virasoro minimal model) which has central charge $c=-{22 \over 5}$ and one non-trivial primary $\cO$ of dimension $h_\cO=-{1\over 5}$. The only non-trivial fusion rule is
\bea
\label{eq:LYfusion}
\cO\times \cO = \id + \cO~.
\eea
In the usual language for minimal models, $\O$ is labelled by $(r,s) = (1,2)$ and hence has a null vector at level two, which turns out to be
\bea
|N_2 \rangle = \left( L_{-2}-{5 \over 2}(L_{-1})^2  \right)  | -{1 / 5} \rangle~.
\eea
This implies a first-order MDE of the form \cite{Gaberdiel:2008ma}
\bea
\left[q {d \over dq} + {1 \over 60}E_2(q) \right]v_{-{1 \over 5}}(q)=0
\eea
for the torus blocks, which is solved by $\cF_{-{1 \over 5}}(q)= \eta^{-2/5}(q)$, a one-dimensional vvmf. Note that the inequality of Theorem \ref{thm:conj1} is saturated here, since the stripped torus block is simply $v_{-{1 \over 5}}(q)= 1$, and so $m_0 = 0$. 

It is clear that the torus one-point function given above is integral. Thus we should be able to invoke Theorem \ref{thm:conj1} and utilize the bound (\ref{eq:chiraldimcond2}). Plugging in $c$ and $h_\cO$ leads us to conclude that $\cO$ must appear in the $\phi \times \phi$ OPE of a primary $\phi$ with dimension $h_\phi \leq {c + 2 h_\cO \over 24} = -{1 \over 5}$. Indeed, $\cO$ appears in its own OPE by (\ref{eq:LYfusion}), so this is true.

\paragraph{Ising one-point function}
The next simplest example is  the one-point function of the energy operator in the Ising model. This receives contributions from a single block $\cF_\sigma(q)$ since the only non-trivial fusion rules are 
\bea
\eps \times \eps = \id~, \hspace{0.8 in} \sigma \times \sigma = \id + \eps~, \hspace{0.8 in} \sigma \times \eps = \eps \times \sigma = \sigma
\eea
and hence $C_{\sigma \sigma \eps} \neq 0$ while $C_{\eps\eps\eps} = C_{\id\id\eps} =0$. Indeed, one finds that $\langle \eps \rangle=|\eta(q)|^2$. Clearly, this is an integral vvmf, and hence we can utilize Result \ref{eq:chiraldimcond2}. Plugging in $c=\half$ and $h_\cO=\half$ leads us to conclude that $\cO$ must appear in the $\phi \times \phi$ OPE of a primary $\phi$ with dimension $h_\phi \leq {c + 2 h_\cO \over 24}={1\over 16}$. Indeed, this is the chiral dimension of the $\sigma$ operator.

\ssec{Dimension $d=2$}

\paragraph{Level 4 Virasoro minimal model one-point functions}
We now consider a less trivial class of examples -- namely, torus one-point functions of Virasoro minimal model primaries $\cO_{r,s}$ with $rs = 4$. Because such primaries have a null state at level four, in this case one can obtain a second-order MDE of the form (\ref{eq:2ndorderMDE}), with \cite{Gaberdiel:2008ma}
\bea
\g = {1 \over 720}\left({c + 8 h_{r,s} \over 2} + 3 h_{r,s} \a_{r,s} \right)~.
\eea
We recall that for Virasoro minimal models the central charge is
\bea
c = 1 - {6 (p-q)^2 \over p q}
\eea
and the operator dimensions are 
\bea
h_{r,s} = {(p r-q s)^2 - (p-q)^2 \over 4 pq}
\eea
for $1 \leq r \leq q-1$ and $1 \leq s \leq p-1$. Above we have also made use of $\a_{r,s}$, whose relevant values are 
\bea
&\vphantom{.}& (1,4): \hspace{0.2 in} \a_{(1,4)}\,\, =\,\, - {2}  \,{p^2 + 4 p q + 6 q^2 \over3  p q}~,
\no\\
&\vphantom{.}& (4,1): \hspace{0.2 in} \a_{(4,1)}\,\, =\,\, - {2}\,  {q^2 + 4 p q + 6 p^2 \over 3 p q}~,
\no\\
&\vphantom{.}& (2,2):\hspace{0.2 in} \a_{(2,2)} \,\,=\,\, -   {3 p q \over (p-q)^2}~.
\eea
Solving the indicial equation (\ref{eq:indiceq}) gives 
\bea
m_0^{(1)} = {1 \over 12} \left( 1 + \sqrt{1- 144 \g}\right)~, \hspace{0.8 in} m_0^{(2)} = {1 \over 12} \left( 1 - \sqrt{1- 144 \g}\right) 
\eea
and by Theorem \ref{thm:conj1} it follows that $m_0^{(2)}$ is negative whenever the one-point function is integral. We now ask when integrality is achieved. 

\subsubsection*{I. \,\,$(2,2)$}

Begin with the one-point function of the $(2,2)$ operator. In this case, it turns out that the one-point functions are \textit{always} integral. This can be seen by making use of the congruence test introduced in Appendix \ref{sec:congtestsubsec}. In particular if the order of $T$ is $N$, i.e. $T^N=1$, we must show that for all totatives $\ell\in \ZZ_N$ such that $(\ell, N ) =1$, we have $\cE(T^{\ell^2}) = \cE(T)$. For the $(2,2)$ one-point function, the order of the $T$-matrix can be shown to be $N=24$ for every choice of $(p,q)$, and thus by the following theorem \cite{Coste:1999yc}, 
\begin{thm}
\label{thm:defof24}
Every $\ell$ coprime to $N$ satisfies $\ell^2 = 1 \,\,{\mathrm{mod}}\,\, N$ if and only if $N$ divides 24.
\end{thm}
\noindent we conclude that $T^{\ell^2} = T$, from which $\cE(T^{\ell^2}) = \cE(T)$, and hence integrality, trivially follows.

We may now use Result \ref{eq:chiraldimcond2} to conclude that the $(2,2)$ operator with non-trivial one-point function must always appear in the $\phi \times \phi$ OPE of an operator $\phi$ with dimension 
\bea
\label{eq:22opbound}
h_{\phi} \leq { c + 2 h_{2,2}\over 24} = {1\over 48}\left( 20 - {9 p \over q} -{9 q \over p}  \right)
\eea
and \textit{a forteriori} that such an operator $\phi$ must exist in the spectrum.

If either of $p$ or $q$ is even, it turns out that the $(2,2)$ one-point function vanishes identically, whereas when $p$ and $q$ are odd it does not \cite{Gaberdiel:2008ma}. Hence (\ref{eq:22opbound}) does not give rise to constraints in the case of unitary minimal models. However, we can consider the bound for non-unitary theories with e.g. $(p,q) = (p,p-2)$. 
Taking for example the case of the $(p,q) = (5,3)$ model, assuming that the $(2,2)$ operator (which has dimension $h_{2,2} = {1 \over 5}$) has non-trivial one-point function, we conclude that there must exist a dimension $h_\phi \leq  -{1 \over 120}$ primary in the theory. Indeed, this role can be played by the $(2,1)$ operator, which has $h_{2,1} = - {1 \over 20}$. 

\subsubsection*{II. \,\,$(1,4)$}
Now consider the one-point function of the $(1,4)$ operator in the $(p,q)$ minimal model. This is only non-trivial for $p$ odd and $q$ even \c{Gaberdiel:2008ma}, to which we restrict ourselves below. 

To determine in which cases these one-point functions are integral, we again apply the congruence test of Appendix \ref{sec:congtestsubsec}. In particular, in this case the order of the $T$-matrix is $N=12p$, so we require that for any $\ell$ such that $(\ell, 12p)=1$ we have $\cE(T^{\ell^2}) = \cE(T)$. This requires  
\bea
\label{eq:14admisscond}
\ell^2 (p+3q) = p\pm 3 q \,\,\,\mathrm{mod}\,\,12p~.
\eea
It turns out that this is only possible if $p$ divides $120$. Thus we must check whether (\ref{eq:14admisscond}) is satisfied for any $1 \leq q < p \leq 120$ such that $p \,| \, 120$. It can be checked that the only solutions subject to this condition are\footnote{Another way to argue for this result is as follows. We begin by noting that $m_0^{(1)} - m_0^{(2)} = {q \over 2 p}$ and that $q$ is even. Then Theorem \ref{thm:francmason} implies that for the representation to be admissible, we need $p\leq 5$. On the other hand, the $(1,4)$ operator only exists in models with $p\geq 5$. Thus we conclude that $p=5$, and need only enumerate all even $q <p$ .}
\bea
(p,q) = (5,2), (5,4)~.
\eea
These are the Lee-Yang and tricritical Ising CFTs, respectively.

The tricritical Ising theory has $c={7 \over 10}$ and $h_{1,4} = {3 \over2}$. Our bounds predict the presence of an operator of dimension $h_\phi \leq {c+ 2 h_{1,4} \over 24}= {37 \over 240}$. Indeed, this role can be played by the $(2,2)$ primary of dimension $h_{2,2} = {3 \over 80}$, which has $C_{{3 \over 80}, {3 \over 2},{3 \over 80}}\neq 0$.

\subsubsection*{III. \,\,$(4,1)$}

Finally consider the one-point function of the $(4,1)$ operator. This is non-trivial only for $p$ even and $q$ odd. This situation is basically the same as the previous one, but with $p$ and $q$ exchanged. Hence very similar comments about integrality apply. In particular, the only integral cases are found to have $q=5$. However, since we conventionally restrict to $p>q$, we now have an infinite family of such integral cases, in theories with 
\bea
(p,q) = (p,5)~, \hspace{0.8 in} p>5~.
\eea
Our bounds can then be applied to this infinite family of theories, though we will not pursue this here.

\ssec{Dimension $d=3$}
\paragraph{Tricritical Ising model one-point functions}
\label{sec:d3tricritIsingex}
We may now consider a three-dimensional example, obtained by considering the torus one-point function of the $(r,s) = (1,3)$ operator in the tricritical Ising model ($h_{1,3} = {3\over 5}$). That this one-point function is three-dimensional follows from the fusion rules, which state that
\bea
C_{{3 \over 5},{3 \over 5},{3 \over 5}}~, \hspace{0.5 in} C_{{1 \over 10},{1 \over 10},{3 \over 5}}~, \hspace{0.5 in} C_{{3 \over 80},{3 \over 80},{3 \over 5}}
\eea
 are the relevant non-zero OPE coefficients. The first captures the three-point coupling of $(1,3)$ operators, while the second involves the $(1,2)$ operator of chiral dimension $h_{1,2} = {1 \over 10}$, and the third involves the $(2,2)$ operator of chiral dimension $h_{2,2} = {3 \over 80}$. 
 
The blocks in this case are known to be \cite{Gaberdiel:2008ma}
\bea
v_{3/80}(q) &=& {\eta(\tau) \over \eta(2 \tau)}~,
\no\\
v_{1/10}(q) &=& {\eta(\tau) \over \eta(\tau/2)} + e^{i \pi/24 } {\eta(\tau) \over \eta(\tau/2+1/2)}~,
\no\\
v_{3/5}(q) &=& {\eta(\tau) \over \eta(\tau/2)} -e^{i \pi/24 } {\eta(\tau) \over \eta(\tau/2+1/2)}~,
\eea
and, in particular, are integral. We can thus apply our bounds to conclude that there exists an operator of dimension 
 \bea
 h_{\phi} \leq  {c + 2 h_{1,3} \over 24} = {19 \over 240}~.
 \eea
This is satisfied by the $(2,2)$ operator.

\ssec{Dimension $d=4$}
\paragraph{$(12,11)$ minimal model four-point function}
\label{sec:d4example}
Finally, we consider the example of the $(r,s)=(1,4)$ four-point function in the $(p,q) = (12,11)$ unitary Virasoro minimal model. This has been studied previously in e.g. \cite{Maloney:2016kee,Cheng:2020srs}. The fusion rules take the form 
\bea
\cO_{(1,4)} \otimes\cO_{(1,4)}  = \cO_{(1,1)} \oplus\cO_{(1,3)} \oplus \cO_{(1,5)} \oplus \cO_{(1,7)} 
\eea
and $\<\cO_{(1,4)} \cO_{(1,4)} \cO_{(1,4)} \cO_{(1,4)} \>$ is comprised of a four-dimensional vvmf \c{Cheng:2020srs}. 

In order to obtain bounds, we must check that this vvmf is integral. However, one can check by explicit calculation that it is actually \textit{not} integral. This might seem to render our bounds inapplicable. But in fact, this case exemplifies how our results can have implications beyond their naive range of validity. Indeed, even though the four-dimensional vvmf is not integral, it turns out that it can be decomposed into a direct sum of two two-dimensional vvmfs, one of which \textit{is} integral.\footnote{That the four-dimensional vvmf can be decomposed in this way can be seen by noting that the $T$-matrix is diagonal, and that the $S$-matrix can be put in block-diagonal form \cite{Maloney:2016kee}, 
\bea
S' = \left(\begin{matrix}- \sqrt{2 - \sqrt{3}} &\, - \sqrt{\sqrt{3} -1} & 0 & 0 
\\
- \sqrt{\sqrt{3} -1} &\, \sqrt{2 - \sqrt{3}}  & 0 & 0 
\\
0 & 0 & - {1 \over \sqrt{2}} & {1 \over \sqrt{2}}
\\
0& 0& {1 \over \sqrt{2}} & {1 \over \sqrt{2}}
 \end{matrix} \right)
\eea
}
We can then simply apply Theorem \ref{thm:conj1} to this subrepresentation to conclude that $m_0<0$. So even though the full four-dimensional vvmf is not integral, the fact that a subrepresentation of it \textit{is} integral is enough to obtain our usual bounds. 

\section{Cuspidal function constructions}
\label{app:cuspfuncts}

As listed in Section \ref{sec:existenceofcuspfunc}, our constructions of non-factorizable, non-holomorphic cuspidal functions satisfying discreteness and integrality can be separated into three categories: 1) Differences of partition functions, 2) Differences of powers of partition functions, and 3) Differences of orbifold theories. We will discuss these in turn.

Before doing so, we first expand upon the construction given explicitly in Section \ref{sec:existenceofcuspfunc}, which involved subtracting the diagonal partition function of the $\widehat{su(2)}_4$ WZW model from the non-diagonal partition function. We refer to this construction as the ``non-diagonal algorithm." Of course, for general $k\geq 4$ the $\widehat{su(2)}_k$ WZW model again admits a non-diagonal invariant, so we may hope that the non-diagonal algorithm could be easily generalized to obtain an infinite number of integral, discrete cuspidal functions. However, this is not the case. 

We begin by recalling that non-diagonal invariants for $\wh{su(2)}_k$ admit an ADE classification \cite{DiFrancesco:1997nk}. Except the $k=4$ example above, it turns out that none of these have $x_0>0$. To see this, note that for $k>6$ the lightest state in the diagonal WZW model is below the threshold $\D={c\o 12}$ (see \eqr{hfnk}), and thus to get a cuspidal function one must subtract out this contribution. But one can check from the classification in \cite{DiFrancesco:1997nk} that for any $k$, no non-diagonal invariants of $\wh{su(2)}_k$ include the lightest primary. Hence it is impossible to subtract this contribution using a non-diagonal partition function. For $k\leq 6$, the only case with a non-diagonal invariant is $k=4$, which is the case originally studied. Thus no further cuspidal functions can be obtained via the non-diagonal algorithm for $\wh{su(2)}_k$ WZW models. 

For $\wh{su(3)}_k$ a similar ADE classification again allows us to conclude that no cuspidal functions can be constructed via the non-diagonal algorithm. More explicitly, there are no $k<4$ non-diagonal invariants, so any candidate non-diagonal invariant for $k>4$ must subtract out the fundamental $(1,0)$ field. However, one can check that none of them does. 

Finally, let us consider unitary Virasoro minimal models $\cM(p,p-1)$. In this case the non-diagonal algorithm again proves fruitless. For $p=4,5$ there are no non-diagonal invariants. For $p>5$, the primary $\phi_{(2,2)}$ obeys
\es{}{{24 h_{2,2}\o c} &= {18\o (p+2)(p-3)}<1\quad \text{for}\quad p>4~.}
Therefore a necessary condition for cuspidal $\cC(q, \bar q) = Z_{\rm non-diag}(q, \bar q)- Z_{\rm diag}(q, \bar q)$ is that the non-diagonal invariants include $|\chi_{2,2}|^2$ with multiplicity one. A quick look at Table 10.3-10.4 of \cite{DiFrancesco:1997nk}, which provides the explicit partition functions of the full ADE classification, shows that this does not happen. Therefore we cannot apply the non-diagonal algorithm for constructing possible non-holomorphic cuspidal functions from unitary Virasoro minimal models either. 

Though the non-diagonal algorithm proves difficult to generalize, we now discuss successful applications of the other constructions listed above. 

\ssec{Theories with identical $c$}
\label{sec:identicalc}
 
We begin by considering cuspidal functions obtained by simply taking differences of CFT partition functions.  
\sssec{WZW models at identical $c$}

Consider WZW models  $\wh{su(N_1)}_{k_1}$ and  $\wh{su(N_2)}_{k_2}$ with the same value of $c$, and take the difference of their respective partition functions,
\e{Cdiffapp}{\cC(q, \bar q) = Z_1(q, \bar q) - Z_2(q, \bar q) ~.}
Clearly the vacuum module cancels between the two terms. We will furthermore restrict to $N_ik_i<12$ for $i=1,2$, since these models are guaranteed not to have any primaries with $\D\leq {c\o 12}$. If such a pair of theories can be found,  $\cC(q, \bar q)$ will be cuspidal. Among diagonal WZW models, there are three pairs with the above properties. Their behavior near the cusp is as follows:
\es{pairs}{&c(2,4) = c(3,1)=2 \, , \quad x_0 = {1\o 12}\\
&c(3,3) = c(5,1)=4\, , \quad  x_0 = {1\o 9}\\
&c(4,2) = c(6,1)=5\, , \quad  x_0 = {5\o 24}}
where $c(N,k)$ is the central charge of $\wh{su(N)_k}$ defined in Appendix \ref{appwzw}. By construction, each pair involves WZW models with different $N$, and hence different numbers of currents. Therefore we have
\bea
m_0=\bar m_0=-{c\over 24} \hspace{0.5 in} \mathrm{and}\hspace{0.5 in} x_0>0~.
\eea
This construction thus gives us three examples of non-holomorphic cuspidal functions satisfying the desired properties. However, note that 
\e{}{Z^{\wh{su(2)_4}}_{\rm non-diag}(\qqb)  = Z^{\wh{su(3)_1}}_{\rm diag}(\qqb) }
and hence the first of these three examples is actually identical to the cuspidal function obtained by applying the non-diagonal algorithm to $\wh{su(2)_4}$.

\subsection{Product theories with identical $c$}
We now consider two theories whose central charges are not equivalent, but rather satisfy $c_1 = n c_2$ for some $ n \in \ZZ_{\geq 2}$. We give two examples: one involves the difference of $\wh{\mathrm{spin}(N)}_1$ WZW model partition functions with products of the Ising model partition function, and the other involves a more general construction using the $J$-invariant.

\subsubsection{$\widehat{\mathrm{spin}(N)}_1-(\mathrm{Ising})^N$ cuspidal functions}
\label{sec:SpinIsingcusp}
We begin by considering an infinite family cuspidal functions obtained by taking the difference of partition functions of the $\widehat{\mathrm{spin}(N)}_1$ WZW model and $N$ copies of the Ising model. We give the construction explicitly for $N< 48$, but the results are easily generalized to arbitrary $N$. 

Recall that the $\widehat{{\rm spin}(N)}_1$ WZW model partition function can be written as 
\bea
Z_{\widehat{{\rm spin}(N)}_1} = \left\{ \begin{matrix} &|\chi_{\omega_0}|^2 + |\chi_{\omega_1}|^2 + |\chi_{\omega_r}|^2 &~~N \,\,\, \mathrm{odd}\\& |\chi_{\omega_0}|^2 + |\chi_{\omega_1}|^2 + |\chi_{\omega_{r-1}}|^2+|\chi_{\omega_r}|^2 & ~~N \,\,\,\mathrm{even}  \end{matrix}\right.
\eea
with the characters given by certain combinations of theta functions \cite{DiFrancesco:1997nk}. A special case is that of $N=1$, which is the usual Ising theory,
\bea
Z_{\rm Ising} = |\chi_1|^2 + |\chi_\eps|^2 + |\chi_\sigma|^2 ~.
\eea
We now define the following function 
\bea
\label{eq:ZdiffSpinIsing}
\C_{N}(q,\bar q):= Z_{\widehat{{\rm spin}(N)}_1}(q, \bar q) - F_N(Z_{\rm Ising}(\qqb))
\eea
with $F_N(x)$ an appropriate degree-$N$ polynomial of $x$. For our purposes up to $N<48$, we take 
\bea
\label{eq:cusppolyF}
F_N(x) &=&  \sum_{k=0}^{\left[{N \over 3} \right]} {(-1)^k \,N \over (N-3k)!}\left[{\Gamma(N-2k) \over \Gamma(k+1)} + {2\Gamma(N-2k-8) \over \Gamma(k-7)}(N^2 - 25 N + 8 + 2k) \right]x^{N-3k}~. \qquad\quad
\eea
One can then check via explicit computation that $\C_{N}$ is cuspidal for any $1 \leq N <48$. 

To give a bit more detail, recall the definition of $x_0$ as the location of the first non-zero Fourier coefficient at $q=\qb$, and introduce a similar quantity $x_1$ defined via 
\bea
x_1 := \mathrm{min}_\cS (x)~, \hspace{0.8 in} \cS = \left\{x \,\, | \,\, a_x / a_0 <0 \right\}~;
\eea
i.e. $x_1$ is the location of the first sign change in the Fourier coefficients. It is amusing to tabulate the values of $x_0$ and $x_1$ for these functions, as shown in Table \ref{tab:SpinIsing} (we start at $N=2$ since $Z_1=0$ by definition). Indeed, in all cases the $x_0$ are positive, and hence the corresponding functions are cuspidal. Furthermore, we empirically observe the periodicity $x_{0}(N) = x_{0}(N+24)$. By definition, one always has $x_1 > x_{0}$. More non-trivially, we may observe that in all cases the following two inequalities hold,
\bea
\label{eq:x0x1ineq}
x_{0} \leq {c \over 24} + {1 \over 16}~, \hspace{1 in} x_1 \leq {c \over 24}+ \half~,
\eea
where the central charge $c={N\over 2}$. The second inequality is saturated only in the cases of $N=8,24$, i.e. for $c=4,12$. In all cases, one has $m_0 = -{c\o 24}$.

 \begin{table}[htp]
\begin{center}
\caption{Values of $x_0(Z_N)$ and $x_1(Z_N)$ for $N=2,\dots,25$}
\label{tab:SpinIsing}
\begin{tabular}{|c|c|c|c|c|c|c|c|c|c|c|c|c|}
\hline
N & 2 & 3 & 4 & 5 & 6 & 7 & 8 & 9 & 10 & 11 & 12 & 13
\\ \hline
$x_{0}$ & ${1\over 24}$ & ${1 \over 8}$ & ${1 \over 12}$ & ${1\over 6}$ & ${1 \over 8}$ &$ {5 \over 24}$ & ${1 \over 6}$ & ${1 \over 4}$ & ${5 \over 24}$ & ${7 \over 24}$ & ${1 \over 4}$ &  ${1\over 3}$
\\ \hline
$x_1$ & ${1 \over 6} $& ${1 \over 4}$ & ${1 \over 3}$ & ${5 \over 12}$ & ${1\over 2} $&$ {7 \over 12}$ &${ 2\over 3} $& ${5 \over 8}$ &${7 \over 12}$ & ${13 \over 24}$ & ${1\over 2}$ & ${11 \over 24} $\\ \hline
\hline
N &  14 & 15 & 16 & 17 & 18 & 19 & 20 & 21 & 22 & 23 & 24 & 25
\\ \hline
$x_{0}$  & ${7 \over 24}$ & ${3 \over 8}$ & ${1\over 3}$ & ${7 \over 24}$ &$ {1 \over 4}$ & ${5 \over 24}$ & ${1 \over 6}$ & ${1 \over 8}$ & ${1 \over 12}$ & ${1 \over 24}$ & ${1 \over 8}$ & ${1 \over 12}$
\\ \hline
$x_1$ & ${5 \over 12}$ & ${1 \over 2}$ & ${11 \over 24}$ & ${5\over 12} $&$ {3 \over 8}$ &${ 1\over 3} $& ${7 \over 24}$ &${1 \over 4}$ & ${5 \over 24}$ & ${1\over 6}$ & ${1}$ & ${23 \over 24}$\\ \hline
\end{tabular}
\end{center}
\end{table}%

\sssec{Differences involving the $J$-invariant} 
\label{sec:Jfuncdiff}
Consider two CFTs with central charges $c_1$ and $c_2$ where $c_2<c_1<24$. Now take
\e{Jconst}{\C(\qqb)= Z_1(\qqb) - Z_2(\qqb)(J(q)\bar J(\qb))^{{c_1-c_2\o 24}}}
where $J(q)$ is the modular $J$-function,
\e{Jdef}{J(q) = {E_4(q)^3\over \Delta(q)}-744~,\quad \Delta(q) = \eta(q)^{24}~.}
In the Fourier expansion, $J$ has no constant term,
\e{Jfourier}{J(q) = {1\over q}+196884q+21493760q^2+\ldots}
In the body of the paper we also make use of 
\e{jdef}{j(q) := J(q) + 744~.}
$\C(\qqb)$ is modular invariant for any $c_1$ and $c_2$. $\C(\qqb)$ is cuspidal if $Z_1$ and $Z_2$ have identical primary spectra (for example, a gap) up to $\D={c_1\o 12}$, and if $c_1<24$. The latter condition comes from the product of the level-2 states in the $J$-functions and the vacuum operator in $Z_2$. In order to also preserve integrality of $\C(\qqb)$, we require
\e{cc'}{{c_1-c_2= 0 \text{ mod }6}~.}
This follows from properties of Fourier coefficients of the $J$-function: first, factorize the first two non-trivial coefficients in the $q$-expansion of $J$,
\e{JFourier}{196884 = 2^2 \cdot 3^3 \cdot 1823~, \quad 21493760 = 2^{11} \cdot 5 \cdot 2099~,}
then expand $J^{{c_1-c_2\o 24}}$. That \eqr{cc'} gives integer Fourier coefficients to {\it all} orders in $q$ can presumably be proven using the binomial theorem and congruence properties of $J$, but we content ourselves with having checked a $q$-expansion up to $\O(q^{1000})$ for all values of $c_1-c_2=6n$ with $ 0\leq n \leq 100$. 

 Suppose $c_2<c_1$ with $c_1>24$. In this case we can still obtain a cuspidal function by replacing $J(q)$ with the action of Hecke operators on $J(q)$:
\e{hecke}{\C(\qqb)= Z_1(\qqb) - Z_2(\qqb) \left( T_mJ(q) T_m \bar J(\qb)\right)^{\tt x}\,,\quad {\tt x}:= {c_1-c_2\o 24m}}
$T_m$ is a Hecke operator, whose action on $J(q)$ produces another modular function of the form
\e{}{T_mJ(q) \sim q^{-m}+ \O(q)~.}
Replacing $J(q)$ and $\bar J(\qb)$ by the action of Hecke operators allows us to push their ``descendants'' to arbitrarily high level. 

Naively, \eqr{hecke} allows $c_1$ arbitrary while retaining cuspidality, as long as both CFTs have identical spectra up to $c_1/12$. However, if we assume that the CFTs have a {\it gap} to $c_1/12$, an interesting thing happens at sufficiently large central charge: enforcing integrality quantizes $c_1-c_2$, and the resulting constraint may fail to be compatible with modular bootstrap bounds. One can see this as follows. First, we impose cuspidality. By design, the power ${\tt x}$ cancels the vacuum module in the difference \eqr{hecke}. Expanding in $q$, we have the following three further conditions for cuspidality:
\es{}{i)&\quad \D_{*,1}>{c_1\o 12}\\
ii)&\quad \D_{*,2}>{c_1\o 12}\\
iii)&\quad m>{c_1\o 12}-1~.}
We assume that CFT$_1$ and CFT$_2$ satisfy the gap conditions \textit{i)} and \textit{ii)}, respectively. Approximating the best known modular bootstrap bound at large central charge as $\D_*\approx {c\o 9}$, we require that ${c_1\o 12}< \D_2 \leq {c_2\o 9}$. In other words, $c_1 - c_2 \leq {c_1\o 4}$. Therefore, ${\tt x}\leq {c_1\o 96m}$, and the bound on $m$ further implies
\e{xhecke}{{\tt x} \leq {1\o 8}~.}
Now we ask: is \eqr{xhecke} consistent with integrality of the Fourier coefficients of \eqr{hecke}? The answer is no. The strategy is to first write down a set of low-lying Fourier coefficients of $T_m J(q)$ using known formulas for the Fourier coefficients of $T_m$ acting on any holomorphic modular function (e.g. \c{iwaniecspectral}), then take the ${\tt x}$'th power and demand integrality. Doing so reveals that $(T_m J(q))^{\tt x}$ has integral Fourier coefficients if and only if the following quantization conditions are met:
\es{}{m\in\lbrace 3,4,9\rbrace:&\quad {\tt x}\in\Z/2\\
m\in6\Z:&\quad {\tt x}\in \Z/4\\
m\notin\lbrace3,4,9,6\Z\rbrace:&\quad {\tt x}\in\Z}
In particular, \eqr{xhecke} is not allowed. Therefore, while this construction works for seed CFTs of sufficient gap and finite central charge, it fails when the CFTs have arbitrarily large central charge. While somewhat involved, this is a nice example of how integrality, cuspidality and modularity are mutually constraining. 

\ssec{Orbifolds}
\label{sec:cusporbifold}
Finally, we consider the difference of the partition function of a CFT with an orbifolding of said CFT, i.e.
\bea
\cC(q, \bar q) = Z_{CFT}(q, \bar q)-Z_{CFT/H}(q, \bar q)~.
\eea  
The simplest example of this is again the cuspidal function obtained by applying the non-diagonal algorithm to the $\widehat{su(2)}_4$ WZW model. 

Here we consider a different example, namely the compact boson of radius $R$. We subtract from its partition function that of its $\Z_2$ orbifold, giving \cite{DiFrancesco:1997nk}
\es{}{\cC(q, \bar q;R) &= Z_R(q,\qb) - Z_{R/\Z_2}(q,\qb)\\ &= {1\o 2}\left(Z_R(q,\qb) - \left({|\vartheta_2\vartheta_3|\o |\eta|^2} + {|\vartheta_2\vartheta_4|\o |\eta|^2} + {|\vartheta_3\vartheta_4|\o |\eta|^2}\right)\right)\\
&= {1\o 2|\eta|^2}\left(\sum_{e,m\in\Z} q^{(e/R+mR/2)^2/2}\qb^{(e/R-mR/2)^2/2}- \left({|\vartheta_2\vartheta_3|} + {|\vartheta_2\vartheta_4|} + {|\vartheta_3\vartheta_4|}\right)\right)~,}
%
where $\vartheta_i$ are Jacobi theta functions. As $q \rar 0$ we have, using ${\sqrt{\vartheta_2(q)}} \sim \sqrt{2 q^{1/8}}$ and $\vartheta_{3,4}(q) \sim 1$, 
\es{}{\cC(q, \bar q;R)  &\sim \h{q^{-{1\o24}}\o \eta(\qb)}\left(1 - \sqrt{\vartheta_3(\qb)\vartheta_4(\qb)}\right)~\quad (q\rar 0~,~\qb~\text{fixed})~.}
Thus we read off 
\e{}{m_0 = -{1\o24}~.}
To determine $x_0$ we take $\qb=q$. In the convention $R\geq\sqrt{2}$, the lightest non-vacuum states in $Z_R$ are electric ($m=0$). Near $q=0$,
\e{}{\cC(q,q;R)\sim \h q^{-{1\o12}}((1+2q^{1\o R^2}+\ldots)-(1+4q^{1\o8}+\ldots))~.}
Thus we read off
\e{}{x_0 = -{1\o 12} + \text{min}\left({1\o R^2},{1\o 8}\right)~.}
This is cuspidal as long as $R^2< 12$. 


\sec{$\wh{su(N)_k}$ WZW data}
\label{appwzw}
Here we collect a few useful formulas about WZW models \c{DiFrancesco:1997nk}. For a model based on affine algebra $\wh{g}_k$, the central charge is
\e{}{c(\wh{g}_k)={k ~{\rm dim}\,g\over k+h^{\vee}}~,}
where $g$ is the Lie algebra generated by zero modes. For $g= su(N)$, the central charge, which we denote $c(N,k)$, is
\es{}{c(N,k) &= {k(N^2-1)\over (k+N)}~.}
The $\widehat{su(N)_k}$ representations that furnish primary states of WZW models are the highest-weight, or so-called ``integrable" representations. These are labeled by $su(N)$ Young diagrams with no more than $k$ boxes across. If the associated weight vector is $\vec \l$, then the left-moving conformal weight is
\e{}{h_{\vec\l}  = {C_2(\vec\l)\over k+N}~,}
where $C_2(\vec\l)$ is the quadratic Casimir of $su(N)$,
\e{}{C_2(\vec \l) = \sum_{i<j}\l_i\l_j{i(N-j)\over N}+\h\sum_{j=1}^{N-1}\l_j\left[\l_j{j(N-j)\over N}+{j}(N-j)\right]~.}
The representation of lowest conformal weight is the fundamental, $\vec\l = (1,0,0,\ldots,0)$, with
\e{}{h_\min=h_\square = {N^2-1\over 2N(k+N)}~.}
Note that
\e{hfnk}{h_\square= {12\o N k}{c\o 24}~.}
Thus all primaries are above threshold, i.e. $h > {c\o 24}$, when $Nk<12$.

\newpage
\sec{Modular systems}
\label{sec:modsystems}
\begin{figure}[!h]
\begin{center}
\includegraphics[scale=.11]{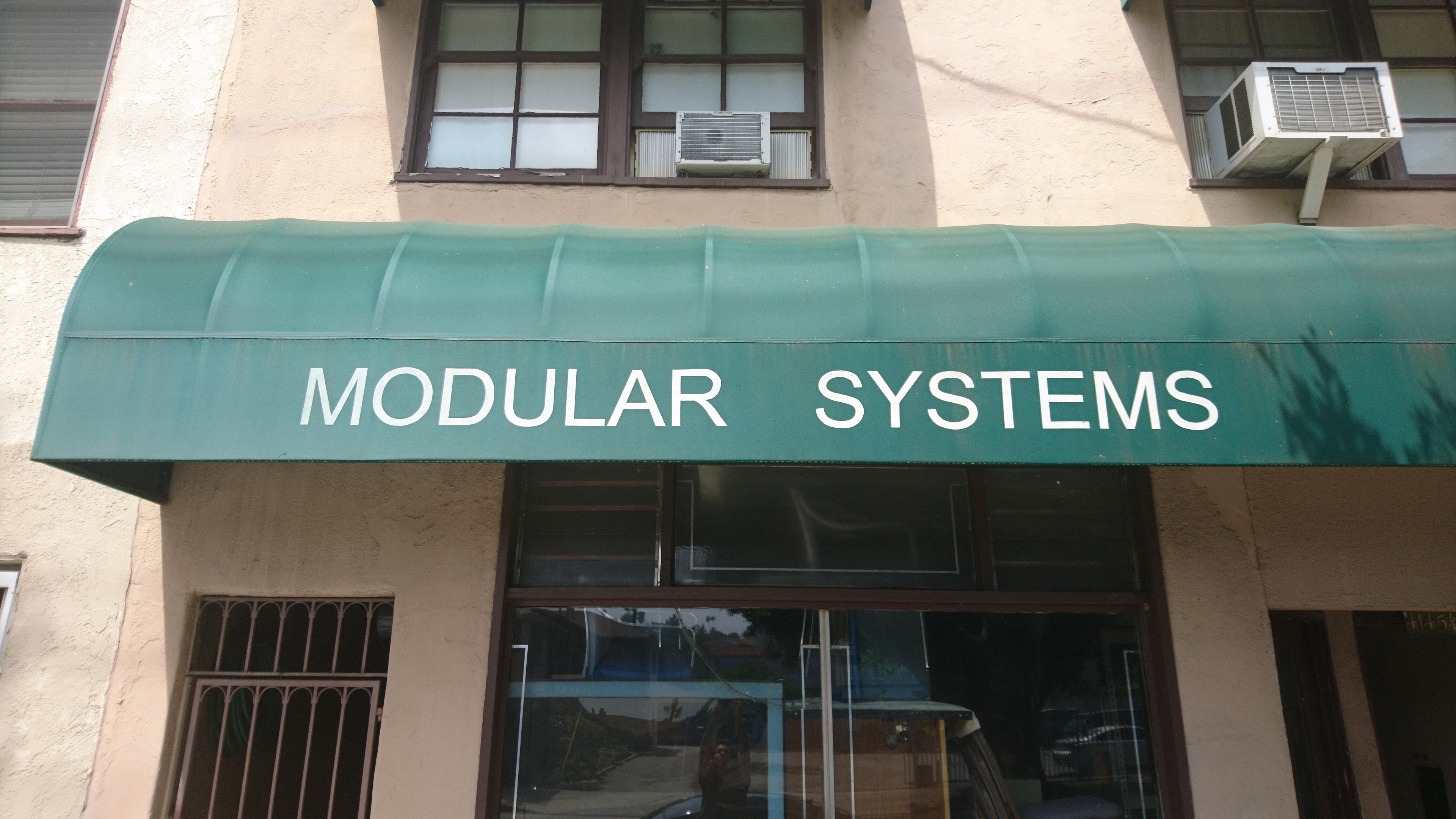}
\end{center}
\end{figure}

\bibliographystyle{JHEP}
\bibliography{bib}

\providecommand{\href}[2]{#2}\begingroup\raggedright\begin{thebibliography}{100}

\bibitem{Hellerman:2009bu}
S.~Hellerman, \emph{{A Universal Inequality for CFT and Quantum Gravity}},
  \href{http://dx.doi.org/10.1007/JHEP08(2011)130}{\emph{JHEP} {\bf 08} (2011)
  130}, [\href{http://arxiv.org/abs/0902.2790}{{\tt 0902.2790}}].

\bibitem{Friedan:2013cba}
D.~Friedan and C.~A. Keller, \emph{{Constraints on 2d CFT partition
  functions}}, \href{http://dx.doi.org/10.1007/JHEP10(2013)180}{\emph{JHEP}
  {\bf 10} (2013) 180}, [\href{http://arxiv.org/abs/1307.6562}{{\tt
  1307.6562}}].

\bibitem{Qualls:2013eha}
J.~D. Qualls and A.~D. Shapere, \emph{{Bounds on Operator Dimensions in 2D
  Conformal Field Theories}},
  \href{http://dx.doi.org/10.1007/JHEP05(2014)091}{\emph{JHEP} {\bf 05} (2014)
  091}, [\href{http://arxiv.org/abs/1312.0038}{{\tt 1312.0038}}].

\bibitem{Keller:2014xba}
C.~A. Keller and A.~Maloney, \emph{{Poincare Series, 3D Gravity and CFT
  Spectroscopy}}, \href{http://dx.doi.org/10.1007/JHEP02(2015)080}{\emph{JHEP}
  {\bf 02} (2015) 080}, [\href{http://arxiv.org/abs/1407.6008}{{\tt
  1407.6008}}].

\bibitem{Collier:2016cls}
S.~Collier, Y.-H. Lin and X.~Yin, \emph{{Modular Bootstrap Revisited}},
  \href{http://dx.doi.org/10.1007/JHEP09(2018)061}{\emph{JHEP} {\bf 09} (2018)
  061}, [\href{http://arxiv.org/abs/1608.06241}{{\tt 1608.06241}}].

\bibitem{Kraus:2016nwo}
P.~Kraus and A.~Maloney, \emph{{A cardy formula for three-point coefficients or
  how the black hole got its spots}},
  \href{http://dx.doi.org/10.1007/JHEP05(2017)160}{\emph{JHEP} {\bf 05} (2017)
  160}, [\href{http://arxiv.org/abs/1608.03284}{{\tt 1608.03284}}].

\bibitem{Collier:2017shs}
S.~Collier, P.~Kravchuk, Y.-H. Lin and X.~Yin, \emph{{Bootstrapping the
  Spectral Function: On the Uniqueness of Liouville and the Universality of
  BTZ}}, \href{http://dx.doi.org/10.1007/JHEP09(2018)150}{\emph{JHEP} {\bf 09}
  (2018) 150}, [\href{http://arxiv.org/abs/1702.00423}{{\tt 1702.00423}}].

\bibitem{Bae:2017kcl}
J.-B. Bae, S.~Lee and J.~Song, \emph{{Modular Constraints on Conformal Field
  Theories with Currents}},
  \href{http://dx.doi.org/10.1007/JHEP12(2017)045}{\emph{JHEP} {\bf 12} (2017)
  045}, [\href{http://arxiv.org/abs/1708.08815}{{\tt 1708.08815}}].

\bibitem{Anous:2018hjh}
T.~Anous, R.~Mahajan and E.~Shaghoulian, \emph{{Parity and the modular
  bootstrap}},
  \href{http://dx.doi.org/10.21468/SciPostPhys.5.3.022}{\emph{SciPost Phys.}
  {\bf 5} (2018) 022}, [\href{http://arxiv.org/abs/1803.04938}{{\tt
  1803.04938}}].

\bibitem{Bae:2018qym}
J.-B. Bae, S.~Lee and J.~Song, \emph{{Modular Constraints on Superconformal
  Field Theories}},
  \href{http://dx.doi.org/10.1007/JHEP01(2019)209}{\emph{JHEP} {\bf 01} (2019)
  209}, [\href{http://arxiv.org/abs/1811.00976}{{\tt 1811.00976}}].

\bibitem{Afkhami-Jeddi:2019zci}
N.~Afkhami-Jeddi, T.~Hartman and A.~Tajdini, \emph{{Fast Conformal Bootstrap
  and Constraints on 3d Gravity}},
  \href{http://dx.doi.org/10.1007/JHEP05(2019)087}{\emph{JHEP} {\bf 05} (2019)
  087}, [\href{http://arxiv.org/abs/1903.06272}{{\tt 1903.06272}}].

\bibitem{Ganguly:2019ksp}
S.~Ganguly and S.~Pal, \emph{{Bounds on the density of states and the spectral
  gap in CFT$_{2}$}},
  \href{http://dx.doi.org/10.1103/PhysRevD.101.106022}{\emph{Phys. Rev. D} {\bf
  101} (2020) 106022}, [\href{http://arxiv.org/abs/1905.12636}{{\tt
  1905.12636}}].

\bibitem{Mukhametzhanov:2019pzy}
B.~Mukhametzhanov and A.~Zhiboedov, \emph{{Modular invariance, tauberian
  theorems and microcanonical entropy}},
  \href{http://dx.doi.org/10.1007/JHEP10(2019)261}{\emph{JHEP} {\bf 10} (2019)
  261}, [\href{http://arxiv.org/abs/1904.06359}{{\tt 1904.06359}}].

\bibitem{Benjamin:2019stq}
N.~Benjamin, H.~Ooguri, S.-H. Shao and Y.~Wang, \emph{{Light-cone modular
  bootstrap and pure gravity}},
  \href{http://dx.doi.org/10.1103/PhysRevD.100.066029}{\emph{Phys. Rev. D} {\bf
  100} (2019) 066029}, [\href{http://arxiv.org/abs/1906.04184}{{\tt
  1906.04184}}].

\bibitem{Hartman:2019pcd}
T.~Hartman, D.~Maz\'{a}\v{c} and L.~Rastelli, \emph{{Sphere Packing and Quantum
  Gravity}}, \href{http://dx.doi.org/10.1007/JHEP12(2019)048}{\emph{JHEP} {\bf
  12} (2019) 048}, [\href{http://arxiv.org/abs/1905.01319}{{\tt 1905.01319}}].

\bibitem{Gliozzi:2019ewk}
F.~Gliozzi, \emph{{Modular Bootstrap, Elliptic Points, and Quantum Gravity}},
  \href{http://dx.doi.org/10.1103/PhysRevResearch.2.013327}{\emph{Phys. Rev.
  Res.} {\bf 2} (2020) 013327}, [\href{http://arxiv.org/abs/1908.00029}{{\tt
  1908.00029}}].

\bibitem{Brehm:2019pcx}
E.~M. Brehm and D.~Das, \emph{{Aspects of the S transformation Bootstrap}},
  \href{http://dx.doi.org/10.1088/1742-5468/ab7f36}{\emph{J. Stat. Mech.} {\bf
  2005} (2020) 053103}, [\href{http://arxiv.org/abs/1911.02309}{{\tt
  1911.02309}}].

\bibitem{Collier:2019weq}
S.~Collier, A.~Maloney, H.~Maxfield and I.~Tsiares, \emph{{Universal dynamics
  of heavy operators in CFT$_{2}$}},
  \href{http://dx.doi.org/10.1007/JHEP07(2020)074}{\emph{JHEP} {\bf 07} (2020)
  074}, [\href{http://arxiv.org/abs/1912.00222}{{\tt 1912.00222}}].

\bibitem{Mukhametzhanov:2020swe}
B.~Mukhametzhanov and S.~Pal, \emph{{Beurling-Selberg Extremization and Modular
  Bootstrap at High Energies}},
  \href{http://dx.doi.org/10.21468/SciPostPhys.8.6.088}{\emph{SciPost Phys.}
  {\bf 8} (2020) 088}, [\href{http://arxiv.org/abs/2003.14316}{{\tt
  2003.14316}}].

\bibitem{Benjamin:2020swg}
N.~Benjamin, H.~Ooguri, S.-H. Shao and Y.~Wang, \emph{{Twist gap and global
  symmetry in two dimensions}},
  \href{http://dx.doi.org/10.1103/PhysRevD.101.106026}{\emph{Phys. Rev. D} {\bf
  101} (2020) 106026}, [\href{http://arxiv.org/abs/2003.02844}{{\tt
  2003.02844}}].

\bibitem{Benjamin:2020zbs}
N.~Benjamin and Y.-H. Lin, \emph{{Lessons from the Ramond sector}},
  \href{http://arxiv.org/abs/2005.02394}{{\tt 2005.02394}}.

\bibitem{Pal:2020wwd}
S.~Pal and Z.~Sun, \emph{{High Energy Modular Bootstrap, Global Symmetries and
  Defects}},  \href{http://arxiv.org/abs/2004.12557}{{\tt 2004.12557}}.

\bibitem{Afkhami-Jeddi:2020ezh}
N.~Afkhami-Jeddi, H.~Cohn, T.~Hartman and A.~Tajdini, \emph{{Free partition
  functions and an averaged holographic duality}},
  \href{http://arxiv.org/abs/2006.04839}{{\tt 2006.04839}}.

\bibitem{Afkhami-Jeddi:2020hde}
N.~Afkhami-Jeddi, H.~Cohn, T.~Hartman, D.~de~Laat and A.~Tajdini,
  \emph{{High-dimensional sphere packing and the modular bootstrap}},
  \href{http://arxiv.org/abs/2006.02560}{{\tt 2006.02560}}.

\bibitem{Yin:2017yyn}
X.~Yin, \emph{{Aspects of Two-Dimensional Conformal Field Theories}},
  \href{http://dx.doi.org/10.22323/1.305.0003}{\emph{PoS} {\bf TASI2017} (2017)
  003}.

\bibitem{Benjamin:2020mfz}
N.~Benjamin, S.~Collier and A.~Maloney, \emph{{Pure Gravity and Conical
  Defects}},  \href{http://arxiv.org/abs/2004.14428}{{\tt 2004.14428}}.

\bibitem{mathur1988classification}
S.~D. Mathur, S.~Mukhi and A.~Sen, \emph{On the classification of rational
  conformal field theories}, {\emph{Physics Letters B} {\bf 213} (1988)
  303--308}.

\bibitem{Mathur:1988gt}
S.~D. Mathur, S.~Mukhi and A.~Sen, \emph{{Reconstruction of Conformal Field
  Theories From Modular Geometry on the Torus}},
  \href{http://dx.doi.org/10.1016/0550-3213(89)90615-9}{\emph{Nucl. Phys. B}
  {\bf 318} (1989) 483--540}.

\bibitem{Kiritsis:1988kq}
E.~B. Kiritsis, \emph{{Fuchsian Differential Equations for Characters on the
  Torus: A Classification}},
  \href{http://dx.doi.org/10.1016/0550-3213(89)90475-6}{\emph{Nucl. Phys. B}
  {\bf 324} (1989) 475--494}.

\bibitem{Naculich:1988xv}
S.~G. Naculich, \emph{{DIFFERENTIAL EQUATIONS FOR RATIONAL CONFORMAL
  CHARACTERS}},
  \href{http://dx.doi.org/10.1016/0550-3213(89)90150-8}{\emph{Nucl. Phys. B}
  {\bf 323} (1989) 423--440}.

\bibitem{Gaberdiel:2008ma}
M.~R. Gaberdiel and S.~Lang, \emph{{Modular differential equations for torus
  one-point functions}},
  \href{http://dx.doi.org/10.1088/1751-8113/42/4/045405}{\emph{J. Phys. A} {\bf
  42} (2009) 045405}, [\href{http://arxiv.org/abs/0810.0106}{{\tt 0810.0106}}].

\bibitem{Hampapura:2015cea}
H.~R. Hampapura and S.~Mukhi, \emph{{On 2d Conformal Field Theories with Two
  Characters}}, \href{http://dx.doi.org/10.1007/JHEP01(2016)005}{\emph{JHEP}
  {\bf 01} (2016) 005}, [\href{http://arxiv.org/abs/1510.04478}{{\tt
  1510.04478}}].

\bibitem{Gaberdiel:2016zke}
M.~R. Gaberdiel, H.~R. Hampapura and S.~Mukhi, \emph{{Cosets of Meromorphic
  CFTs and Modular Differential Equations}},
  \href{http://dx.doi.org/10.1007/JHEP04(2016)156}{\emph{JHEP} {\bf 04} (2016)
  156}, [\href{http://arxiv.org/abs/1602.01022}{{\tt 1602.01022}}].

\bibitem{Chandra:2018pjq}
A.~R. Chandra and S.~Mukhi, \emph{{Towards a Classification of Two-Character
  Rational Conformal Field Theories}},
  \href{http://dx.doi.org/10.1007/JHEP04(2019)153}{\emph{JHEP} {\bf 04} (2019)
  153}, [\href{http://arxiv.org/abs/1810.09472}{{\tt 1810.09472}}].

\bibitem{Chandra:2018ezv}
A.~R. Chandra and S.~Mukhi, \emph{{Curiosities above c = 24}},
  \href{http://dx.doi.org/10.21468/SciPostPhys.6.5.053}{\emph{SciPost Phys.}
  {\bf 6} (2019) 053}, [\href{http://arxiv.org/abs/1812.05109}{{\tt
  1812.05109}}].

\bibitem{Harvey:2018rdc}
J.~A. Harvey and Y.~Wu, \emph{{Hecke Relations in Rational Conformal Field
  Theory}}, \href{http://dx.doi.org/10.1007/JHEP09(2018)032}{\emph{JHEP} {\bf
  09} (2018) 032}, [\href{http://arxiv.org/abs/1804.06860}{{\tt 1804.06860}}].

\bibitem{Mukhi:2020gnj}
S.~Mukhi, R.~Poddar and P.~Singh, \emph{{Rational CFT with three characters:
  the quasi-character approach}},
  \href{http://dx.doi.org/10.1007/JHEP05(2020)003}{\emph{JHEP} {\bf 05} (2020)
  003}, [\href{http://arxiv.org/abs/2002.01949}{{\tt 2002.01949}}].

\bibitem{Cheng:2020srs}
M.~C. Cheng, T.~Gannon and G.~Lockhart, \emph{{Modular Exercises for Four-Point
  Blocks -- I}},  \href{http://arxiv.org/abs/2002.11125}{{\tt 2002.11125}}.

\bibitem{kaneko2002modular}
M.~Kaneko and M.~Koike, \emph{On modular forms arising from a differential
  equation of hypergeometric type},  2002.

\bibitem{knopp2004}
M.~Knopp and G.~Mason, \emph{Vector-valued modular forms and poincaré series},
  \href{http://dx.doi.org/10.1215/ijm/1258138515}{\emph{Illinois J. Math.} {\bf
  48} (10, 2004) 1345--1366}.

\bibitem{Bantay:2005vk}
P.~Bantay and T.~Gannon, \emph{{Conformal characters and the modular
  representation}},
  \href{http://dx.doi.org/10.1088/1126-6708/2006/02/005}{\emph{JHEP} {\bf 02}
  (2006) 005}, [\href{http://arxiv.org/abs/hep-th/0512011}{{\tt
  hep-th/0512011}}].

\bibitem{bantay2007vector}
P.~Bantay and T.~Gannon, \emph{Vector-valued modular functions for the modular
  group and the hypergeometric equation}, {\emph{arXiv preprint
  arXiv:0705.2467} (2007) }.

\bibitem{marks2010structure}
C.~Marks and G.~Mason, \emph{Structure of the module of vector-valued modular
  forms}, {\emph{Journal of the London Mathematical Society} {\bf 82} (2010)
  32--48}.

\bibitem{marks2012fourier}
C.~Marks, \emph{{Fourier coefficients of three-dimensional vector-valued
  modular forms}},  \href{http://arxiv.org/abs/1201.5165}{{\tt 1201.5165}}.

\bibitem{Marks:2010fm}
C.~Marks, \emph{{Irreducible vector-valued modular forms of dimension less than
  six}},  \href{http://arxiv.org/abs/1004.3019}{{\tt 1004.3019}}.

\bibitem{99dc2d57859749fcabfc151b9730f994}
M.~Kaneko, K.~Nagatomo and Y.~Sakai, \emph{Modular forms and second order
  ordinary differential equations: Applications to vertex operator algebras},
  \href{http://dx.doi.org/10.1007/s11005-012-0602-5}{\emph{Letters in
  Mathematical Physics} {\bf 103} (Jan., 2013) 439--453}.

\bibitem{franc2014fourier}
C.~Franc and G.~Mason, \emph{Fourier coefficients of vector-valued modular
  forms of dimension 2}, {\emph{Canadian Mathematical Bulletin} {\bf 57} (2014)
  485--494}.

\bibitem{franc2016hypergeometric}
C.~Franc and G.~Mason, \emph{Hypergeometric series, modular linear differential
  equations and vector-valued modular forms}, {\emph{The Ramanujan Journal}
  {\bf 41} (2016) 233--267}.

\bibitem{KANEKO2017332}
M.~Kaneko, K.~Nagatomo and Y.~Sakai, \emph{The third order modular linear
  differential equations},
  \href{http://dx.doi.org/https://doi.org/10.1016/j.jalgebra.2017.05.007}{\emph{Journal
  of Algebra} {\bf 485} (2017) 332 -- 352}.

\bibitem{franc2018constructions}
C.~Franc and G.~Mason, \emph{Constructions of vector-valued modular forms of
  rank four and level one}, {\emph{arXiv preprint arXiv:1810.09408} (2018) }.

\bibitem{mason2018vertex}
G.~Mason, K.~Nagatomo and Y.~Sakai, \emph{Vertex operator algebras with two
  simple modules - the mathur-mukhi-sen theorem revisited},  2018.

\bibitem{Franc2020}
C.~Franc and G.~Mason, \emph{Classification of some vertex operator algebras of
  rank 3}, \href{http://dx.doi.org/10.2140/ant.2020.14.1613}{\emph{Algebra \&
  Number Theory} {\bf 14} (Jul, 2020) 1613–1667}.

\bibitem{gannon2014theory}
T.~Gannon, \emph{The theory of vector-valued modular forms for the modular
  group},  in \emph{Conformal field theory, automorphic forms and related
  topics}, pp.~247--286.
\newblock Springer, 2014.

\bibitem{Mukhi:2019xjy}
S.~Mukhi, \emph{{Classification of RCFT from Holomorphic Modular Bootstrap: A
  Status Report}},  in \emph{{Pollica Summer Workshop 2019}: {Mathematical and
  Geometric Tools for Conformal Field Theories}}, 10, 2019.
\newblock \href{http://arxiv.org/abs/1910.02973}{{\tt 1910.02973}}.

\bibitem{Cardy:1987vr}
J.~L. Cardy, \emph{{Continuously Varying Exponents and the Value of the Central
  Charge}}, \href{http://dx.doi.org/10.1088/0305-4470/20/13/014}{\emph{J. Phys.
  A} {\bf 20} (1987) L891--L896}.

\bibitem{Komargodski:2016auf}
Z.~Komargodski and D.~Simmons-Duffin, \emph{{The Random-Bond Ising Model in
  2.01 and 3 Dimensions}},
  \href{http://dx.doi.org/10.1088/1751-8121/aa6087}{\emph{J. Phys. A} {\bf 50}
  (2017) 154001}, [\href{http://arxiv.org/abs/1603.04444}{{\tt 1603.04444}}].

\bibitem{Bashmakov:2017rko}
V.~Bashmakov, M.~Bertolini and H.~Raj, \emph{{On non-supersymmetric conformal
  manifolds: field theory and holography}},
  \href{http://dx.doi.org/10.1007/JHEP11(2017)167}{\emph{JHEP} {\bf 11} (2017)
  167}, [\href{http://arxiv.org/abs/1709.01749}{{\tt 1709.01749}}].

\bibitem{Behan:2017mwi}
C.~Behan, \emph{{Conformal manifolds: ODEs from OPEs}},
  \href{http://dx.doi.org/10.1007/JHEP03(2018)127}{\emph{JHEP} {\bf 03} (2018)
  127}, [\href{http://arxiv.org/abs/1709.03967}{{\tt 1709.03967}}].

\bibitem{Sen:2017gfr}
K.~Sen and Y.~Tachikawa, \emph{{First-order conformal perturbation theory by
  marginal operators}},  \href{http://arxiv.org/abs/1711.05947}{{\tt
  1711.05947}}.

\bibitem{Gaberdiel:2010pz}
M.~R. Gaberdiel and R.~Gopakumar, \emph{{An AdS\_3 Dual for Minimal Model
  CFTs}}, \href{http://dx.doi.org/10.1103/PhysRevD.83.066007}{\emph{Phys. Rev.
  D} {\bf 83} (2011) 066007}, [\href{http://arxiv.org/abs/1011.2986}{{\tt
  1011.2986}}].

\bibitem{Castro:2011zq}
A.~Castro, M.~R. Gaberdiel, T.~Hartman, A.~Maloney and R.~Volpato, \emph{{The
  Gravity Dual of the Ising Model}},
  \href{http://dx.doi.org/10.1103/PhysRevD.85.024032}{\emph{Phys. Rev. D} {\bf
  85} (2012) 024032}, [\href{http://arxiv.org/abs/1111.1987}{{\tt 1111.1987}}].

\bibitem{Jian:2019ubz}
C.-M. Jian, A.~W. Ludwig, Z.-X. Luo, H.-Y. Sun and Z.~Wang, \emph{{Establishing
  strongly-coupled 3D AdS quantum gravity with Ising dual using all-genus
  partition functions}},  \href{http://arxiv.org/abs/1907.06656}{{\tt
  1907.06656}}.

\bibitem{Maloney:2020nni}
A.~Maloney and E.~Witten, \emph{{Averaging Over Narain Moduli Space}},
  \href{http://arxiv.org/abs/2006.04855}{{\tt 2006.04855}}.

\bibitem{zamps}
A.~B. Zamolodchikov and A.~B. Zamolodchikov, \emph{{Liouville field theory on a
  pseudosphere}},  \href{http://arxiv.org/abs/hep-th/0101152}{{\tt
  hep-th/0101152}}.

\bibitem{Hartman:2014oaa}
T.~Hartman, C.~A. Keller and B.~Stoica, \emph{{Universal Spectrum of 2d
  Conformal Field Theory in the Large c Limit}},
  \href{http://dx.doi.org/10.1007/JHEP09(2014)118}{\emph{JHEP} {\bf 09} (2014)
  118}, [\href{http://arxiv.org/abs/1405.5137}{{\tt 1405.5137}}].

\bibitem{apostol2012modular}
T.~M. Apostol, \emph{Modular functions and Dirichlet series in number theory},
  vol.~41.
\newblock Springer Science \& Business Media, 2012.

\bibitem{conrad}
K.~Conrad, \emph{{$\text{SL}_2(\mathbb{Z})$}},
  \href{http://arxiv.org/abs/https://kconrad.math.uconn.edu/blurbs/grouptheory/SL(2,Z).pdf}{{\tt
  https://kconrad.math.uconn.edu/blurbs/grouptheory/SL(2,Z).pdf}}.

\bibitem{atkin1971modular}
A.~O.~L. Atkin and H.~P.~F. Swinnerton-Dyer, \emph{Modular forms on
  noncongruence subgroups}, {\emph{Combinatorics} {\bf 19} (1971) 1--25}.

\bibitem{mason2010fourier}
G.~Mason, \emph{On the fourier coefficients of 2-dimensional vector-valued
  modular forms},  2010.

\bibitem{FRANC2016186}
C.~Franc and G.~Mason, \emph{Three-dimensional imprimitive representations of
  the modular group and their associated modular forms},
  \href{http://dx.doi.org/https://doi.org/10.1016/j.jnt.2015.08.013}{\emph{Journal
  of Number Theory} {\bf 160} (2016) 186 -- 214}.

\bibitem{Diamond2005}
F.~Diamond and J.~Shurman, \emph{A First Course in Modular Forms}.
\newblock Springer New York, New York, NY, 2005.

\bibitem{schultz2015notes}
D.~Schultz, \emph{{Notes on modular forms}},
  \href{http://arxiv.org/abs/https://faculty.math.illinois.edu/~schult25/ModFormNotes.pdf}{{\tt
  https://faculty.math.illinois.edu/~schult25/ModFormNotes.pdf}}.

\bibitem{knopp2009logarithmic}
M.~Knopp and G.~Mason, \emph{Logarithmic vector-valued modular forms},  2009.

\bibitem{bantay2013dimension}
P.~Bantay, \emph{The dimension of spaces of vector-valued modular forms of
  integer weight}, {\emph{Letters in Mathematical Physics} {\bf 103} (2013)
  1243--1260}.

\bibitem{DiFrancesco:1997nk}
P.~Di~Francesco, P.~Mathieu and D.~Senechal, \emph{{Conformal Field Theory}}.
\newblock Graduate Texts in Contemporary Physics. Springer-Verlag, New York,
  1997,
  \href{http://dx.doi.org/10.1007/978-1-4612-2256-9}{10.1007/978-1-4612-2256-9}.

\bibitem{Headrick:2015gba}
M.~Headrick, A.~Maloney, E.~Perlmutter and I.~G. Zadeh, \emph{{Rényi
  entropies, the analytic bootstrap, and 3D quantum gravity at higher genus}},
  \href{http://dx.doi.org/10.1007/JHEP07(2015)059}{\emph{JHEP} {\bf 07} (2015)
  059}, [\href{http://arxiv.org/abs/1503.07111}{{\tt 1503.07111}}].

\bibitem{Maldacena:2015iua}
J.~Maldacena, D.~Simmons-Duffin and A.~Zhiboedov, \emph{{Looking for a bulk
  point}}, \href{http://dx.doi.org/10.1007/JHEP01(2017)013}{\emph{JHEP} {\bf
  01} (2017) 013}, [\href{http://arxiv.org/abs/1509.03612}{{\tt 1509.03612}}].

\bibitem{Poghossian:2009mk}
R.~Poghossian, \emph{{Recursion relations in CFT and N=2 SYM theory}},
  \href{http://dx.doi.org/10.1088/1126-6708/2009/12/038}{\emph{JHEP} {\bf 12}
  (2009) 038}, [\href{http://arxiv.org/abs/0909.3412}{{\tt 0909.3412}}].

\bibitem{Afkhami-Jeddi:2017idc}
N.~Afkhami-Jeddi, K.~Colville, T.~Hartman, A.~Maloney and E.~Perlmutter,
  \emph{{Constraints on higher spin CFT$_{2}$}},
  \href{http://dx.doi.org/10.1007/JHEP05(2018)092}{\emph{JHEP} {\bf 05} (2018)
  092}, [\href{http://arxiv.org/abs/1707.07717}{{\tt 1707.07717}}].

\bibitem{Hadasz:2009db}
L.~Hadasz, Z.~Jaskolski and P.~Suchanek, \emph{{Recursive representation of the
  torus 1-point conformal block}},
  \href{http://dx.doi.org/10.1007/JHEP01(2010)063}{\emph{JHEP} {\bf 01} (2010)
  063}, [\href{http://arxiv.org/abs/0911.2353}{{\tt 0911.2353}}].

\bibitem{Zamolodchikov:1985ie}
A.~Zamolodchikov, \emph{{CONFORMAL SYMMETRY IN TWO-DIMENSIONS: AN EXPLICIT
  RECURRENCE FORMULA FOR THE CONFORMAL PARTIAL WAVE AMPLITUDE}},
  \href{http://dx.doi.org/10.1007/BF01214585}{\emph{Commun. Math. Phys.} {\bf
  96} (1984) 419--422}.

\bibitem{zamo2}
A.~Zamolodchikov, \emph{Conformal symmetry in two-dimensional space: Recursion
  representation of conformal block},
  \href{http://dx.doi.org/10.1007/BF01022967}{\emph{Theoretical and
  Mathematical Physics} {\bf 73} (1987) 1088--1093}.

\bibitem{Alkalaev:2016fok}
K.~Alkalaev, R.~Geiko and V.~Rappoport, \emph{{Various semiclassical limits of
  torus conformal blocks}},
  \href{http://dx.doi.org/10.1007/JHEP04(2017)070}{\emph{JHEP} {\bf 04} (2017)
  070}, [\href{http://arxiv.org/abs/1612.05891}{{\tt 1612.05891}}].

\bibitem{Beem:2013sza}
C.~Beem, M.~Lemos, P.~Liendo, W.~Peelaers, L.~Rastelli and B.~C. van Rees,
  \emph{{Infinite Chiral Symmetry in Four Dimensions}},
  \href{http://dx.doi.org/10.1007/s00220-014-2272-x}{\emph{Commun. Math. Phys.}
  {\bf 336} (2015) 1359--1433}, [\href{http://arxiv.org/abs/1312.5344}{{\tt
  1312.5344}}].

\bibitem{Beem:2017ooy}
C.~Beem and L.~Rastelli, \emph{{Vertex operator algebras, Higgs branches, and
  modular differential equations}},
  \href{http://dx.doi.org/10.1007/JHEP08(2018)114}{\emph{JHEP} {\bf 08} (2018)
  114}, [\href{http://arxiv.org/abs/1707.07679}{{\tt 1707.07679}}].

\bibitem{Ardehali:2015bla}
A.~Arabi~Ardehali, \emph{{High-temperature asymptotics of supersymmetric
  partition functions}},
  \href{http://dx.doi.org/10.1007/JHEP07(2016)025}{\emph{JHEP} {\bf 07} (2016)
  025}, [\href{http://arxiv.org/abs/1512.03376}{{\tt 1512.03376}}].

\bibitem{beem}
C.~Beem, encrypted communication.

\bibitem{Bianchi:2019sxz}
L.~Bianchi and M.~Lemos, \emph{{Superconformal surfaces in four dimensions}},
  \href{http://dx.doi.org/10.1007/JHEP06(2020)056}{\emph{JHEP} {\bf 06} (2020)
  056}, [\href{http://arxiv.org/abs/1911.05082}{{\tt 1911.05082}}].

\bibitem{Runkel:2001ng}
I.~Runkel and G.~Watts, \emph{{A Nonrational CFT with c = 1 as a limit of
  minimal models}},
  \href{http://dx.doi.org/10.1088/1126-6708/2001/09/006}{\emph{JHEP} {\bf 09}
  (2001) 006}, [\href{http://arxiv.org/abs/hep-th/0107118}{{\tt
  hep-th/0107118}}].

\bibitem{Ribault:2014hia}
S.~Ribault, \emph{{Conformal field theory on the plane}},
  \href{http://arxiv.org/abs/1406.4290}{{\tt 1406.4290}}.

\bibitem{Benjamin:2018kre}
N.~Benjamin, E.~Dyer, A.~L. Fitzpatrick and Y.~Xin, \emph{{The Most Irrational
  Rational Theories}},
  \href{http://dx.doi.org/10.1007/JHEP04(2019)025}{\emph{JHEP} {\bf 04} (2019)
  025}, [\href{http://arxiv.org/abs/1812.07579}{{\tt 1812.07579}}].

\bibitem{Brown:2017}
F.~Brown, \emph{{A class of non-holomorphic modular forms III: Real Analytic
  Cusp Forms for $SL(2, \mathbb{Z})$}},  10, 2017.
\newblock \href{http://arxiv.org/abs/1710.07912}{{\tt 1710.07912}}.

\bibitem{DHoker:2019txf}
E.~D'Hoker and J.~Kaidi, \emph{{Modular graph functions and odd cuspidal
  functions. Fourier and Poincar\'e series}},
  \href{http://dx.doi.org/10.1007/JHEP04(2019)136}{\emph{JHEP} {\bf 04} (2019)
  136}, [\href{http://arxiv.org/abs/1902.04180}{{\tt 1902.04180}}].

\bibitem{DHoker:2015wxz}
E.~D'Hoker, M.~B. Green, O.~Gürdo{\u g}an and P.~Vanhove, \emph{{Modular Graph
  Functions}},
  \href{http://dx.doi.org/10.4310/CNTP.2017.v11.n1.a4}{\emph{Commun. Num.
  Theor. Phys.} {\bf 11} (2017) 165--218},
  [\href{http://arxiv.org/abs/1512.06779}{{\tt 1512.06779}}].

\bibitem{DHoker:2016mwo}
E.~D'Hoker and M.~B. Green, \emph{{Identities between Modular Graph Forms}},
  {\emph{J. Number Theor.} {\bf 189} (2018) 25--88},
  [\href{http://arxiv.org/abs/1603.00839}{{\tt 1603.00839}}].

\bibitem{DHoker:2016quv}
E.~D'Hoker and J.~Kaidi, \emph{{Hierarchy of Modular Graph Identities}},
  \href{http://dx.doi.org/10.1007/JHEP11(2016)051}{\emph{JHEP} {\bf 11} (2016)
  051}, [\href{http://arxiv.org/abs/1608.04393}{{\tt 1608.04393}}].

\bibitem{DHoker:2017zhq}
E.~D'Hoker and W.~Duke, \emph{{Fourier series of modular graph functions}},
  \href{http://arxiv.org/abs/1708.07998}{{\tt 1708.07998}}.

\bibitem{Gerken:2018zcy}
J.~E. Gerken and J.~Kaidi, \emph{{Holomorphic subgraph reduction of
  higher-point modular graph forms}},
  \href{http://dx.doi.org/10.1007/JHEP01(2019)131}{\emph{JHEP} {\bf 01} (2019)
  131}, [\href{http://arxiv.org/abs/1809.05122}{{\tt 1809.05122}}].

\bibitem{Gerken:2018jrq}
J.~E. Gerken, A.~Kleinschmidt and O.~Schlotterer, \emph{{Heterotic-string
  amplitudes at one loop: modular graph forms and relations to open strings}},
  \href{http://dx.doi.org/10.1007/JHEP01(2019)052}{\emph{JHEP} {\bf 01} (2019)
  052}, [\href{http://arxiv.org/abs/1811.02548}{{\tt 1811.02548}}].

\bibitem{Gerken:2019cxz}
J.~E. Gerken, A.~Kleinschmidt and O.~Schlotterer, \emph{{All-order differential
  equations for one-loop closed-string integrals and modular graph forms}},
  \href{http://dx.doi.org/10.1007/JHEP01(2020)064}{\emph{JHEP} {\bf 01} (2020)
  064}, [\href{http://arxiv.org/abs/1911.03476}{{\tt 1911.03476}}].

\bibitem{Datta:2019jeo}
S.~Datta, P.~Kraus and B.~Michel, \emph{{Typicality and thermality in 2d CFT}},
  \href{http://dx.doi.org/10.1007/JHEP07(2019)143}{\emph{JHEP} {\bf 07} (2019)
  143}, [\href{http://arxiv.org/abs/1904.00668}{{\tt 1904.00668}}].

\bibitem{Dyer:2016pou}
E.~Dyer and G.~Gur-Ari, \emph{{2D CFT Partition Functions at Late Times}},
  \href{http://dx.doi.org/10.1007/JHEP08(2017)075}{\emph{JHEP} {\bf 08} (2017)
  075}, [\href{http://arxiv.org/abs/1611.04592}{{\tt 1611.04592}}].

\bibitem{Fitzpatrick:2016ive}
A.~L. Fitzpatrick, J.~Kaplan, D.~Li and J.~Wang, \emph{{On information loss in
  AdS$_{3}$/CFT$_{2}$}},
  \href{http://dx.doi.org/10.1007/JHEP05(2016)109}{\emph{JHEP} {\bf 05} (2016)
  109}, [\href{http://arxiv.org/abs/1603.08925}{{\tt 1603.08925}}].

\bibitem{Alday:2019vdr}
L.~F. Alday and J.-B. Bae, \emph{{Rademacher Expansions and the Spectrum of 2d
  CFT}},  \href{http://arxiv.org/abs/2001.00022}{{\tt 2001.00022}}.

\bibitem{Moore:1988ss}
G.~W. Moore and N.~Seiberg, \emph{{Naturality in Conformal Field Theory}},
  \href{http://dx.doi.org/10.1016/0550-3213(89)90511-7}{\emph{Nucl. Phys. B}
  {\bf 313} (1989) 16--40}.

\bibitem{Iles:2014gra}
N.~J. Iles and G.~M. Watts, \emph{{Modular properties of characters of the
  W$_{3}$ algebra}},
  \href{http://dx.doi.org/10.1007/JHEP01(2016)089}{\emph{JHEP} {\bf 01} (2016)
  089}, [\href{http://arxiv.org/abs/1411.4039}{{\tt 1411.4039}}].

\bibitem{Saad:2019lba}
P.~Saad, S.~H. Shenker and D.~Stanford, \emph{{JT gravity as a matrix
  integral}},  \href{http://arxiv.org/abs/1903.11115}{{\tt 1903.11115}}.

\bibitem{Cotler:2020ugk}
J.~Cotler and K.~Jensen, \emph{{AdS$_3$ gravity and random CFT}},
  \href{http://arxiv.org/abs/2006.08648}{{\tt 2006.08648}}.

\bibitem{Perez:2020klz}
A.~Pérez and R.~Troncoso, \emph{{Gravitational dual of averaged free CFT's
  over the Narain lattice}},  \href{http://arxiv.org/abs/2006.08216}{{\tt
  2006.08216}}.

\bibitem{Belin:2020hea}
A.~Belin and J.~de~Boer, \emph{{Random Statistics of OPE Coefficients and
  Euclidean Wormholes}},  \href{http://arxiv.org/abs/2006.05499}{{\tt
  2006.05499}}.

\bibitem{Cotler:2020hgz}
J.~Cotler and K.~Jensen, \emph{{AdS$_3$ wormholes from a modular bootstrap}},
  \href{http://arxiv.org/abs/2007.15653}{{\tt 2007.15653}}.

\bibitem{alex}
A.~Maloney, Caltech hep-th informal discussion.

\bibitem{Benjamin:2016aww}
N.~Benjamin, E.~Dyer, A.~L. Fitzpatrick, A.~Maloney and E.~Perlmutter,
  \emph{{Small Black Holes and Near-Extremal CFTs}},
  \href{http://dx.doi.org/10.1007/JHEP08(2016)023}{\emph{JHEP} {\bf 08} (2016)
  023}, [\href{http://arxiv.org/abs/1603.08524}{{\tt 1603.08524}}].

\bibitem{Maxfield:2020ale}
H.~Maxfield and G.~J. Turiaci, \emph{{The path integral of 3D gravity near
  extremality; or, JT gravity with defects as a matrix integral}},
  \href{http://arxiv.org/abs/2006.11317}{{\tt 2006.11317}}.

\bibitem{Komargodski:2012ek}
Z.~Komargodski and A.~Zhiboedov, \emph{{Convexity and Liberation at Large
  Spin}}, \href{http://dx.doi.org/10.1007/JHEP11(2013)140}{\emph{JHEP} {\bf 11}
  (2013) 140}, [\href{http://arxiv.org/abs/1212.4103}{{\tt 1212.4103}}].

\bibitem{Fitzpatrick:2012yx}
A.~Fitzpatrick, J.~Kaplan, D.~Poland and D.~Simmons-Duffin, \emph{{The Analytic
  Bootstrap and AdS Superhorizon Locality}},
  \href{http://dx.doi.org/10.1007/JHEP12(2013)004}{\emph{JHEP} {\bf 12} (2013)
  004}, [\href{http://arxiv.org/abs/1212.3616}{{\tt 1212.3616}}].

\bibitem{Collier:2018exn}
S.~Collier, Y.~Gobeil, H.~Maxfield and E.~Perlmutter, \emph{{Quantum Regge
  Trajectories and the Virasoro Analytic Bootstrap}},
  \href{http://dx.doi.org/10.1007/JHEP05(2019)212}{\emph{JHEP} {\bf 05} (2019)
  212}, [\href{http://arxiv.org/abs/1811.05710}{{\tt 1811.05710}}].

\bibitem{Dijkgraaf:2000fq}
R.~Dijkgraaf, J.~M. Maldacena, G.~W. Moore and E.~P. Verlinde, \emph{{A Black
  hole Farey tail}},  \href{http://arxiv.org/abs/hep-th/0005003}{{\tt
  hep-th/0005003}}.

\bibitem{Denef:2007vg}
F.~Denef and G.~W. Moore, \emph{{Split states, entropy enigmas, holes and
  halos}}, \href{http://dx.doi.org/10.1007/JHEP11(2011)129}{\emph{JHEP} {\bf
  11} (2011) 129}, [\href{http://arxiv.org/abs/hep-th/0702146}{{\tt
  hep-th/0702146}}].

\bibitem{deBoer:2008fk}
J.~de~Boer, F.~Denef, S.~El-Showk, I.~Messamah and D.~Van~den Bleeken,
  \emph{{Black hole bound states in AdS(3) x S**2}},
  \href{http://dx.doi.org/10.1088/1126-6708/2008/11/050}{\emph{JHEP} {\bf 11}
  (2008) 050}, [\href{http://arxiv.org/abs/0802.2257}{{\tt 0802.2257}}].

\bibitem{Bena:2011zw}
I.~Bena, B.~D. Chowdhury, J.~de~Boer, S.~El-Showk and M.~Shigemori,
  \emph{{Moulting Black Holes}},
  \href{http://dx.doi.org/10.1007/JHEP03(2012)094}{\emph{JHEP} {\bf 03} (2012)
  094}, [\href{http://arxiv.org/abs/1108.0411}{{\tt 1108.0411}}].

\bibitem{Coste:1999yc}
A.~Coste and T.~Gannon, \emph{{Congruence subgroups and rational conformal
  field theory}},  \href{http://arxiv.org/abs/math/9909080}{{\tt
  math/9909080}}.

\bibitem{Maloney:2016kee}
A.~Maloney, H.~Maxfield and G.~S. Ng, \emph{{A conformal block Farey tail}},
  \href{http://dx.doi.org/10.1007/JHEP06(2017)117}{\emph{JHEP} {\bf 06} (2017)
  117}, [\href{http://arxiv.org/abs/1609.02165}{{\tt 1609.02165}}].

\bibitem{iwaniecspectral}
H.~Iwaniec, \emph{Spectral Methods of Automorphic Forms}.
\newblock Graduate studies in mathematics. American Mathematical Soc.

\end{thebibliography}\endgroup

\end{document}